\newif\ifarxiv
\arxivtrue

\ifarxiv
\documentclass[a4paper,10pt,abstract=true]{scrartcl}
\usepackage{libertine}
\usepackage{natbib}
\newcommand\Description[1]{}
\usepackage[includeheadfoot, head=13pt, foot=2pc, top=58pt, bottom=44pt, inner=60pt, outer=60pt, marginparwidth=2pc, heightrounded]{geometry}
\else
\documentclass[format=acmsmall,natbib]{acmart}
\usepackage{libertine}
\fi

\usepackage{mathpazo}
\usepackage[T1]{fontenc}
\usepackage[utf8]{inputenc}
\usepackage[scaled=0.81]{beramono}  

\usepackage[multiple]{footmisc} \usepackage{tabularx}
\usepackage[yyyymmdd]{datetime}

\ifarxiv\else
\makeatletter
\def\@copyrightpermission{}
\def\@copyrightowner{}
\acmPrice{}
\makeatother
\fi

\setcitestyle{square,numbers,sort}

\usepackage{graphicx}
\graphicspath{{.}{figs/}}
\DeclareGraphicsExtensions{.png,.pdf}

\usepackage{xparse,xstring,xspace}
\usepackage{dcolumn}
\usepackage{caption,subcaption}

\DeclareRobustCommand{\var}[1]{\ensuremath{\textnormal{\textsl{#1}}}}

\DeclareRobustCommand{\dist}[1]{\ensuremath{\textsf{#1}}}

\usepackage{tikz}
\usetikzlibrary{positioning,shapes.geometric,arrows,calc}
\tikzset{unobserved/.style={draw,circle,thick,inner sep=1pt}}

\usepackage{pgfkeys,pgfplots,numprint,relsize}
\xspaceaddexceptions{\%}  \def\keyfamily{/cj37/}    \def\NAN{??}              

\usepackage{fmtcount}     

\DeclareDocumentCommand{\spellout}{m O{13}}{\IfStrEq{#1}{ }{\NAN}{\IfInteger{#1}{\expandafter\ifnum#1<#2{\numberstringnum{#1}}\else#1\fi }{\NAN}}}

\newcommand{\cj}{\textsf{Code Jam}\xspace}

\DeclareDocumentCommand{\n}{t. t: o m o O{} t| t!}{\begingroup \pgfkeys{/pgf/fpu=true}\IfBooleanTF{#2}{\pgfkeyssetvalue{/tmp/value}{#4}\pgfkeyssetvalue{/tmp/found}{found}}{\pgfkeysifdefined{\keyfamily#4}{\pgfkeyssetvalue{/tmp/value}{\pgfkeysvalueof{\keyfamily#4}}\pgfkeyssetvalue{/tmp/found}{found}}{}}\pgfkeysifdefined{/tmp/found}{\IfNoValueF{#5}{\pgfkeyssetvalue{/tmp/multiplier}{#5}\pgfmathparse{\pgfkeysvalueof{/tmp/multiplier} * \pgfkeysvalueof{/tmp/value}}\pgfkeyslet{/tmp/value}\pgfmathresult }\IfBooleanTF{#1}{\IfBooleanTF{#8}{\spellout{\pgfkeysvalueof{/tmp/value}}}{\pgfkeysvalueof{/tmp/value}}}{\IfNoValueTF{#3}{\pgfmathprintnumber [set thousands separator={\,},int detect,#6]{\pgfkeysvalueof{/tmp/value}}}{{\pgfmathprintnumber [precision=#3,fixed,zerofill,set thousands separator={\,},#6]{\pgfkeysvalueof{/tmp/value}}}}}\IfBooleanT{#7}{{\smaller[1.2]\%}}}{\NAN }\pgfkeys{/pgf/fpu=false}\endgroup }

\DeclareDocumentCommand{\absmell}{m}{``\textsf{#1}''}

\pgfkeyssetvalue{/cj37/rounds2010/qualification round:nicks}{8459}
\pgfkeyssetvalue{/cj37/rounds2010/qualification round:langs}{37}
\pgfkeyssetvalue{/cj37/rounds2010/round 1A:nicks}{2075}
\pgfkeyssetvalue{/cj37/rounds2010/round 1A:langs}{22}
\pgfkeyssetvalue{/cj37/rounds2010/round 1B:nicks}{2885}
\pgfkeyssetvalue{/cj37/rounds2010/round 1B:langs}{29}
\pgfkeyssetvalue{/cj37/rounds2010/round 1C:nicks}{2863}
\pgfkeyssetvalue{/cj37/rounds2010/round 1C:langs}{27}
\pgfkeyssetvalue{/cj37/rounds2010/round 1:nicks}{5553}
\pgfkeyssetvalue{/cj37/rounds2010/round 1:langs}{34}
\pgfkeyssetvalue{/cj37/rounds2010/round 2:nicks}{1924}
\pgfkeyssetvalue{/cj37/rounds2010/round 2:langs}{19}
\pgfkeyssetvalue{/cj37/rounds2010/round 3:nicks}{361}
\pgfkeyssetvalue{/cj37/rounds2010/round 3:langs}{10}
\pgfkeyssetvalue{/cj37/rounds2010/finals:nicks}{24}
\pgfkeyssetvalue{/cj37/rounds2010/finals:langs}{3}
\pgfkeyssetvalue{/cj37/num languages}{75}
\pgfkeyssetvalue{/cj37/used languages count/1}{90682}
\pgfkeyssetvalue{/cj37/used languages count/2}{36012}
\pgfkeyssetvalue{/cj37/used languages count/3}{23845}
\pgfkeyssetvalue{/cj37/used languages count/4}{8398}
\pgfkeyssetvalue{/cj37/used languages count/5}{8210}
\pgfkeyssetvalue{/cj37/used languages count/6}{3070}
\pgfkeyssetvalue{/cj37/used languages count/7}{1915}
\pgfkeyssetvalue{/cj37/used languages count/8}{1580}
\pgfkeyssetvalue{/cj37/used languages count/9}{1291}
\pgfkeyssetvalue{/cj37/used languages count/10}{1288}
\pgfkeyssetvalue{/cj37/used languages count/11}{597}
\pgfkeyssetvalue{/cj37/used languages count/12}{474}
\pgfkeyssetvalue{/cj37/used languages count/13}{361}
\pgfkeyssetvalue{/cj37/used languages count/14}{335}
\pgfkeyssetvalue{/cj37/used languages count/15}{309}
\pgfkeyssetvalue{/cj37/used languages count/16}{256}
\pgfkeyssetvalue{/cj37/used languages count/17}{201}
\pgfkeyssetvalue{/cj37/used languages count/18}{180}
\pgfkeyssetvalue{/cj37/used languages count/19}{179}
\pgfkeyssetvalue{/cj37/used languages count/20}{135}
\pgfkeyssetvalue{/cj37/used languages count/21}{114}
\pgfkeyssetvalue{/cj37/used languages count/22}{110}
\pgfkeyssetvalue{/cj37/used languages count/23}{80}
\pgfkeyssetvalue{/cj37/used languages count/24}{71}
\pgfkeyssetvalue{/cj37/used languages count/25}{47}
\pgfkeyssetvalue{/cj37/used languages count/26}{46}
\pgfkeyssetvalue{/cj37/used languages count/27}{45}
\pgfkeyssetvalue{/cj37/used languages count/28}{43}
\pgfkeyssetvalue{/cj37/used languages count/29}{42}
\pgfkeyssetvalue{/cj37/used languages count/30}{40}
\pgfkeyssetvalue{/cj37/used languages count/31}{26}
\pgfkeyssetvalue{/cj37/used languages count/32}{25}
\pgfkeyssetvalue{/cj37/used languages count/33}{23}
\pgfkeyssetvalue{/cj37/used languages count/34}{21}
\pgfkeyssetvalue{/cj37/used languages count/35}{16}
\pgfkeyssetvalue{/cj37/used languages count/36}{16}
\pgfkeyssetvalue{/cj37/used languages count/37}{14}
\pgfkeyssetvalue{/cj37/used languages count/38}{13}
\pgfkeyssetvalue{/cj37/used languages count/39}{12}
\pgfkeyssetvalue{/cj37/used languages count/40}{11}
\pgfkeyssetvalue{/cj37/used languages count/41}{10}
\pgfkeyssetvalue{/cj37/used languages count/42}{9}
\pgfkeyssetvalue{/cj37/used languages count/43}{8}
\pgfkeyssetvalue{/cj37/used languages count/44}{8}
\pgfkeyssetvalue{/cj37/used languages count/45}{7}
\pgfkeyssetvalue{/cj37/used languages count/46}{7}
\pgfkeyssetvalue{/cj37/used languages count/47}{6}
\pgfkeyssetvalue{/cj37/used languages count/48}{5}
\pgfkeyssetvalue{/cj37/used languages count/49}{5}
\pgfkeyssetvalue{/cj37/used languages count/50}{5}
\pgfkeyssetvalue{/cj37/used languages count/51}{5}
\pgfkeyssetvalue{/cj37/used languages count/52}{4}
\pgfkeyssetvalue{/cj37/used languages count/53}{4}
\pgfkeyssetvalue{/cj37/used languages count/54}{3}
\pgfkeyssetvalue{/cj37/used languages count/55}{3}
\pgfkeyssetvalue{/cj37/used languages count/56}{2}
\pgfkeyssetvalue{/cj37/used languages count/57}{2}
\pgfkeyssetvalue{/cj37/used languages count/58}{2}
\pgfkeyssetvalue{/cj37/used languages count/59}{2}
\pgfkeyssetvalue{/cj37/used languages count/60}{2}
\pgfkeyssetvalue{/cj37/used languages count/61}{2}
\pgfkeyssetvalue{/cj37/used languages count/62}{2}
\pgfkeyssetvalue{/cj37/used languages count/63}{2}
\pgfkeyssetvalue{/cj37/used languages count/64}{2}
\pgfkeyssetvalue{/cj37/used languages count/65}{2}
\pgfkeyssetvalue{/cj37/used languages count/66}{1}
\pgfkeyssetvalue{/cj37/used languages count/67}{1}
\pgfkeyssetvalue{/cj37/used languages count/68}{1}
\pgfkeyssetvalue{/cj37/used languages count/69}{1}
\pgfkeyssetvalue{/cj37/used languages count/70}{1}
\pgfkeyssetvalue{/cj37/used languages count/71}{1}
\pgfkeyssetvalue{/cj37/used languages count/72}{1}
\pgfkeyssetvalue{/cj37/used languages count/73}{1}
\pgfkeyssetvalue{/cj37/used languages count/74}{1}
\pgfkeyssetvalue{/cj37/used languages count/75}{1}
\pgfkeyssetvalue{/cj37/used languages count/76}{0}
\pgfkeyssetvalue{/cj37/used languages count/77}{0}
\pgfkeyssetvalue{/cj37/used languages count/78}{0}
\pgfkeyssetvalue{/cj37/used languages count/79}{0}
\pgfkeyssetvalue{/cj37/used languages count/80}{0}
\pgfkeyssetvalue{/cj37/used languages count/81}{0}
\pgfkeyssetvalue{/cj37/used languages count/82}{0}
\pgfkeyssetvalue{/cj37/used languages count/83}{0}
\pgfkeyssetvalue{/cj37/used languages count/84}{0}
\pgfkeyssetvalue{/cj37/used languages count/85}{0}
\pgfkeyssetvalue{/cj37/used languages count/86}{0}
\pgfkeyssetvalue{/cj37/used languages count/87}{0}
\pgfkeyssetvalue{/cj37/used languages count/88}{0}
\pgfkeyssetvalue{/cj37/used languages count/89}{0}
\pgfkeyssetvalue{/cj37/used languages count/90}{0}
\pgfkeyssetvalue{/cj37/used languages count/91}{0}
\pgfkeyssetvalue{/cj37/used languages count/92}{0}
\pgfkeyssetvalue{/cj37/used languages count/93}{0}
\pgfkeyssetvalue{/cj37/used languages count/94}{0}
\pgfkeyssetvalue{/cj37/used languages count/95}{0}
\pgfkeyssetvalue{/cj37/used languages count/96}{0}
\pgfkeyssetvalue{/cj37/used languages count/97}{0}
\pgfkeyssetvalue{/cj37/used languages count/98}{0}
\pgfkeyssetvalue{/cj37/used languages count/99}{0}
\pgfkeyssetvalue{/cj37/used languages count/100}{0}
\pgfkeyssetvalue{/cj37/used languages percent/1}{0.503084570490203}
\pgfkeyssetvalue{/cj37/used languages percent/2}{0.199786964915785}
\pgfkeyssetvalue{/cj37/used languages percent/3}{0.132287020393671}
\pgfkeyssetvalue{/cj37/used languages percent/4}{0.0465903290948228}
\pgfkeyssetvalue{/cj37/used languages percent/5}{0.0455473448283514}
\pgfkeyssetvalue{/cj37/used languages percent/6}{0.0170317111599317}
\pgfkeyssetvalue{/cj37/used languages percent/7}{0.0106240152675144}
\pgfkeyssetvalue{/cj37/used languages percent/8}{0.00876550606928079}
\pgfkeyssetvalue{/cj37/used languages percent/9}{0.0071621951490136}
\pgfkeyssetvalue{/cj37/used languages percent/10}{0.00714555178305927}
\pgfkeyssetvalue{/cj37/used languages percent/11}{0.00331202982491179}
\pgfkeyssetvalue{/cj37/used languages percent/12}{0.00262965182078424}
\pgfkeyssetvalue{/cj37/used languages percent/13}{0.00200275170317112}
\pgfkeyssetvalue{/cj37/used languages percent/14}{0.00185850919823358}
\pgfkeyssetvalue{/cj37/used languages percent/15}{0.00171426669329605}
\pgfkeyssetvalue{/cj37/used languages percent/16}{0.00142023389476954}
\pgfkeyssetvalue{/cj37/used languages percent/17}{0.00111510551894015}
\pgfkeyssetvalue{/cj37/used languages percent/18}{0.000998601957259836}
\pgfkeyssetvalue{/cj37/used languages percent/19}{0.000993054168608393}
\pgfkeyssetvalue{/cj37/used languages percent/20}{0.000748951467944877}
\pgfkeyssetvalue{/cj37/used languages percent/21}{0.000632447906264563}
\pgfkeyssetvalue{/cj37/used languages percent/22}{0.000610256751658789}
\pgfkeyssetvalue{/cj37/used languages percent/23}{0.000443823092115483}
\pgfkeyssetvalue{/cj37/used languages percent/24}{0.000393892994252491}
\pgfkeyssetvalue{/cj37/used languages percent/25}{0.000260746066617846}
\pgfkeyssetvalue{/cj37/used languages percent/26}{0.000255198277966403}
\pgfkeyssetvalue{/cj37/used languages percent/27}{0.000249650489314959}
\pgfkeyssetvalue{/cj37/used languages percent/28}{0.000238554912012072}
\pgfkeyssetvalue{/cj37/used languages percent/29}{0.000233007123360628}
\pgfkeyssetvalue{/cj37/used languages percent/30}{0.000221911546057741}
\pgfkeyssetvalue{/cj37/used languages percent/31}{0.000144242504937532}
\pgfkeyssetvalue{/cj37/used languages percent/32}{0.000138694716286088}
\pgfkeyssetvalue{/cj37/used languages percent/33}{0.000127599138983201}
\pgfkeyssetvalue{/cj37/used languages percent/34}{0.000116503561680314}
\pgfkeyssetvalue{/cj37/used languages percent/35}{8.87646184230966e-05}
\pgfkeyssetvalue{/cj37/used languages percent/36}{8.87646184230966e-05}
\pgfkeyssetvalue{/cj37/used languages percent/37}{7.76690411202095e-05}
\pgfkeyssetvalue{/cj37/used languages percent/38}{7.2121252468766e-05}
\pgfkeyssetvalue{/cj37/used languages percent/39}{6.65734638173224e-05}
\pgfkeyssetvalue{/cj37/used languages percent/40}{6.10256751658789e-05}
\pgfkeyssetvalue{/cj37/used languages percent/41}{5.54778865144353e-05}
\pgfkeyssetvalue{/cj37/used languages percent/42}{4.99300978629918e-05}
\pgfkeyssetvalue{/cj37/used languages percent/43}{4.43823092115483e-05}
\pgfkeyssetvalue{/cj37/used languages percent/44}{4.43823092115483e-05}
\pgfkeyssetvalue{/cj37/used languages percent/45}{3.88345205601047e-05}
\pgfkeyssetvalue{/cj37/used languages percent/46}{3.88345205601047e-05}
\pgfkeyssetvalue{/cj37/used languages percent/47}{3.32867319086612e-05}
\pgfkeyssetvalue{/cj37/used languages percent/48}{2.77389432572177e-05}
\pgfkeyssetvalue{/cj37/used languages percent/49}{2.77389432572177e-05}
\pgfkeyssetvalue{/cj37/used languages percent/50}{2.77389432572177e-05}
\pgfkeyssetvalue{/cj37/used languages percent/51}{2.77389432572177e-05}
\pgfkeyssetvalue{/cj37/used languages percent/52}{2.21911546057741e-05}
\pgfkeyssetvalue{/cj37/used languages percent/53}{2.21911546057741e-05}
\pgfkeyssetvalue{/cj37/used languages percent/54}{1.66433659543306e-05}
\pgfkeyssetvalue{/cj37/used languages percent/55}{1.66433659543306e-05}
\pgfkeyssetvalue{/cj37/used languages percent/56}{1.10955773028871e-05}
\pgfkeyssetvalue{/cj37/used languages percent/57}{1.10955773028871e-05}
\pgfkeyssetvalue{/cj37/used languages percent/58}{1.10955773028871e-05}
\pgfkeyssetvalue{/cj37/used languages percent/59}{1.10955773028871e-05}
\pgfkeyssetvalue{/cj37/used languages percent/60}{1.10955773028871e-05}
\pgfkeyssetvalue{/cj37/used languages percent/61}{1.10955773028871e-05}
\pgfkeyssetvalue{/cj37/used languages percent/62}{1.10955773028871e-05}
\pgfkeyssetvalue{/cj37/used languages percent/63}{1.10955773028871e-05}
\pgfkeyssetvalue{/cj37/used languages percent/64}{1.10955773028871e-05}
\pgfkeyssetvalue{/cj37/used languages percent/65}{1.10955773028871e-05}
\pgfkeyssetvalue{/cj37/used languages percent/66}{5.54778865144353e-06}
\pgfkeyssetvalue{/cj37/used languages percent/67}{5.54778865144353e-06}
\pgfkeyssetvalue{/cj37/used languages percent/68}{5.54778865144353e-06}
\pgfkeyssetvalue{/cj37/used languages percent/69}{5.54778865144353e-06}
\pgfkeyssetvalue{/cj37/used languages percent/70}{5.54778865144353e-06}
\pgfkeyssetvalue{/cj37/used languages percent/71}{5.54778865144353e-06}
\pgfkeyssetvalue{/cj37/used languages percent/72}{5.54778865144353e-06}
\pgfkeyssetvalue{/cj37/used languages percent/73}{5.54778865144353e-06}
\pgfkeyssetvalue{/cj37/used languages percent/74}{5.54778865144353e-06}
\pgfkeyssetvalue{/cj37/used languages percent/75}{5.54778865144353e-06}
\pgfkeyssetvalue{/cj37/used languages percent/76}{0}
\pgfkeyssetvalue{/cj37/used languages percent/77}{0}
\pgfkeyssetvalue{/cj37/used languages percent/78}{0}
\pgfkeyssetvalue{/cj37/used languages percent/79}{0}
\pgfkeyssetvalue{/cj37/used languages percent/80}{0}
\pgfkeyssetvalue{/cj37/used languages percent/81}{0}
\pgfkeyssetvalue{/cj37/used languages percent/82}{0}
\pgfkeyssetvalue{/cj37/used languages percent/83}{0}
\pgfkeyssetvalue{/cj37/used languages percent/84}{0}
\pgfkeyssetvalue{/cj37/used languages percent/85}{0}
\pgfkeyssetvalue{/cj37/used languages percent/86}{0}
\pgfkeyssetvalue{/cj37/used languages percent/87}{0}
\pgfkeyssetvalue{/cj37/used languages percent/88}{0}
\pgfkeyssetvalue{/cj37/used languages percent/89}{0}
\pgfkeyssetvalue{/cj37/used languages percent/90}{0}
\pgfkeyssetvalue{/cj37/used languages percent/91}{0}
\pgfkeyssetvalue{/cj37/used languages percent/92}{0}
\pgfkeyssetvalue{/cj37/used languages percent/93}{0}
\pgfkeyssetvalue{/cj37/used languages percent/94}{0}
\pgfkeyssetvalue{/cj37/used languages percent/95}{0}
\pgfkeyssetvalue{/cj37/used languages percent/96}{0}
\pgfkeyssetvalue{/cj37/used languages percent/97}{0}
\pgfkeyssetvalue{/cj37/used languages percent/98}{0}
\pgfkeyssetvalue{/cj37/used languages percent/99}{0}
\pgfkeyssetvalue{/cj37/used languages percent/100}{0}
\pgfkeyssetvalue{/cj37/num all rounds}{150539}
\pgfkeyssetvalue{/cj37/num all participants}{49808}
\pgfkeyssetvalue{/cj37/num editions}{7}
\pgfkeyssetvalue{/cj37/first edition}{2008}
\pgfkeyssetvalue{/cj37/last edition}{2014}
\pgfkeyssetvalue{/cj37/cutoffs/num.years}{2}
\pgfkeyssetvalue{/cj37/cutoffs/num.rounds}{6}
\pgfkeyssetvalue{/cj37/cutoffs/geq.years.rounds}{1896}
\pgfkeyssetvalue{/cj37/cutoffs/percent.rounds}{1}
\pgfkeyssetvalue{/cj37/cutoffs/n.nicknames}{105}
\pgfkeyssetvalue{/cj37/cutoffs/percent.nicknames}{0}
\pgfkeyssetvalue{/cj37/cutoffs/gid}{37}
\pgfkeyssetvalue{/cj37/min num rounds per year}{7}
\pgfkeyssetvalue{/cj37/max num rounds per year}{10}
\pgfkeyssetvalue{/cj37/rank variance}{3962574.95177404}
\pgfkeyssetvalue{/cj37/rank mean}{928.877637130802}
\pgfkeyssetvalue{/cj37/size min}{457}
\pgfkeyssetvalue{/cj37/size max}{202965}
\pgfkeyssetvalue{/cj37/loo m1m2m3m4/m4:elpd_diff}{0}
\pgfkeyssetvalue{/cj37/loo m1m2m3m4/m4:se_diff}{0}
\pgfkeyssetvalue{/cj37/loo m1m2m3m4/m4:elpd_loo}{-12906.4882459334}
\pgfkeyssetvalue{/cj37/loo m1m2m3m4/m4:se_elpd_loo}{81.0703710808259}
\pgfkeyssetvalue{/cj37/loo m1m2m3m4/m4:p_loo}{134.547324619432}
\pgfkeyssetvalue{/cj37/loo m1m2m3m4/m4:se_p_loo}{9.02270850693688}
\pgfkeyssetvalue{/cj37/loo m1m2m3m4/m4:looic}{25812.9764918667}
\pgfkeyssetvalue{/cj37/loo m1m2m3m4/m4:se_looic}{162.140742161652}
\pgfkeyssetvalue{/cj37/loo m1m2m3m4/m4:stdiffs}{NaN}
\pgfkeyssetvalue{/cj37/loo m1m2m3m4/m4:reldiffs}{0}
\pgfkeyssetvalue{/cj37/loo m1m2m3m4/m4:relstdiffs}{NaN}
\pgfkeyssetvalue{/cj37/loo m1m2m3m4/m3:elpd_diff}{-387.634100160419}
\pgfkeyssetvalue{/cj37/loo m1m2m3m4/m3:se_diff}{26.5263780380487}
\pgfkeyssetvalue{/cj37/loo m1m2m3m4/m3:elpd_loo}{-13294.1223460938}
\pgfkeyssetvalue{/cj37/loo m1m2m3m4/m3:se_elpd_loo}{79.6205649231556}
\pgfkeyssetvalue{/cj37/loo m1m2m3m4/m3:p_loo}{126.934319943186}
\pgfkeyssetvalue{/cj37/loo m1m2m3m4/m3:se_p_loo}{8.66257094934146}
\pgfkeyssetvalue{/cj37/loo m1m2m3m4/m3:looic}{26588.2446921875}
\pgfkeyssetvalue{/cj37/loo m1m2m3m4/m3:se_looic}{159.241129846311}
\pgfkeyssetvalue{/cj37/loo m1m2m3m4/m3:stdiffs}{-14.6131559915344}
\pgfkeyssetvalue{/cj37/loo m1m2m3m4/m3:reldiffs}{-387.634100160419}
\pgfkeyssetvalue{/cj37/loo m1m2m3m4/m3:relstdiffs}{-14.6131559915344}
\pgfkeyssetvalue{/cj37/loo m1m2m3m4/m2:elpd_diff}{-1173.0316818518}
\pgfkeyssetvalue{/cj37/loo m1m2m3m4/m2:se_diff}{42.3664701621349}
\pgfkeyssetvalue{/cj37/loo m1m2m3m4/m2:elpd_loo}{-14079.5199277852}
\pgfkeyssetvalue{/cj37/loo m1m2m3m4/m2:se_elpd_loo}{85.2102398966393}
\pgfkeyssetvalue{/cj37/loo m1m2m3m4/m2:p_loo}{117.353713142715}
\pgfkeyssetvalue{/cj37/loo m1m2m3m4/m2:se_p_loo}{12.8967348214721}
\pgfkeyssetvalue{/cj37/loo m1m2m3m4/m2:looic}{28159.0398555703}
\pgfkeyssetvalue{/cj37/loo m1m2m3m4/m2:se_looic}{170.420479793279}
\pgfkeyssetvalue{/cj37/loo m1m2m3m4/m2:stdiffs}{-27.6877369618628}
\pgfkeyssetvalue{/cj37/loo m1m2m3m4/m2:reldiffs}{-785.397581691383}
\pgfkeyssetvalue{/cj37/loo m1m2m3m4/m2:relstdiffs}{-18.5381878331071}
\pgfkeyssetvalue{/cj37/loo m1m2m3m4/m1:elpd_diff}{-1258.80002570593}
\pgfkeyssetvalue{/cj37/loo m1m2m3m4/m1:se_diff}{41.3460520563897}
\pgfkeyssetvalue{/cj37/loo m1m2m3m4/m1:elpd_loo}{-14165.2882716393}
\pgfkeyssetvalue{/cj37/loo m1m2m3m4/m1:se_elpd_loo}{80.2835525967569}
\pgfkeyssetvalue{/cj37/loo m1m2m3m4/m1:p_loo}{5.66047694655346}
\pgfkeyssetvalue{/cj37/loo m1m2m3m4/m1:se_p_loo}{0.685200064154968}
\pgfkeyssetvalue{/cj37/loo m1m2m3m4/m1:looic}{28330.5765432786}
\pgfkeyssetvalue{/cj37/loo m1m2m3m4/m1:se_looic}{160.567105193514}
\pgfkeyssetvalue{/cj37/loo m1m2m3m4/m1:stdiffs}{-30.4454709240223}
\pgfkeyssetvalue{/cj37/loo m1m2m3m4/m1:reldiffs}{-85.7683438541253}
\pgfkeyssetvalue{/cj37/loo m1m2m3m4/m1:relstdiffs}{-2.07440226063544}
\pgfkeyssetvalue{/cj37/mid1 alpha num}{1}
\pgfkeyssetvalue{/cj37/mid1 alpha perc}{0.0196078431372549}
\pgfkeyssetvalue{/cj37/mid1 alpha length}{51}
\pgfkeyssetvalue{/cj37/m3 beta low 50}{-1.01614656730683}
\pgfkeyssetvalue{/cj37/m3 beta high 50}{-0.945305268193174}
\pgfkeyssetvalue{/cj37/mid2 alpha num}{36}
\pgfkeyssetvalue{/cj37/mid2 alpha perc}{0.705882352941177}
\pgfkeyssetvalue{/cj37/mid2 alpha length}{51}
\pgfkeyssetvalue{/cj37/m1/lower 95:C++}{-0.509484099666667}
\pgfkeyssetvalue{/cj37/m1/lower 95:Java}{-0.462434099666666}
\pgfkeyssetvalue{/cj37/m1/lower 95:Python}{0.387225900333334}
\pgfkeyssetvalue{/cj37/m1/lower 90:C++}{-0.497434099666666}
\pgfkeyssetvalue{/cj37/m1/lower 90:Java}{-0.439774099666667}
\pgfkeyssetvalue{/cj37/m1/lower 90:Python}{0.435325900333334}
\pgfkeyssetvalue{/cj37/m1/lower 70:C++}{-0.470254099666667}
\pgfkeyssetvalue{/cj37/m1/lower 70:Java}{-0.387354099666666}
\pgfkeyssetvalue{/cj37/m1/lower 70:Python}{0.528265900333333}
\pgfkeyssetvalue{/cj37/m1/lower 50:C++}{-0.461514099666666}
\pgfkeyssetvalue{/cj37/m1/lower 50:Java}{-0.363104099666667}
\pgfkeyssetvalue{/cj37/m1/lower 50:Python}{0.599305900333333}
\pgfkeyssetvalue{/cj37/m1/upper 50:C++}{-0.408704099666666}
\pgfkeyssetvalue{/cj37/m1/upper 50:Java}{-0.255734099666666}
\pgfkeyssetvalue{/cj37/m1/upper 50:Python}{0.843405900333333}
\pgfkeyssetvalue{/cj37/m1/upper 70:C++}{-0.388344099666666}
\pgfkeyssetvalue{/cj37/m1/upper 70:Java}{-0.225624099666667}
\pgfkeyssetvalue{/cj37/m1/upper 70:Python}{0.909015900333334}
\pgfkeyssetvalue{/cj37/m1/upper 90:C++}{-0.366804099666666}
\pgfkeyssetvalue{/cj37/m1/upper 90:Java}{-0.177384099666666}
\pgfkeyssetvalue{/cj37/m1/upper 90:Python}{1.03648590033333}
\pgfkeyssetvalue{/cj37/m1/upper 95:C++}{-0.354654099666666}
\pgfkeyssetvalue{/cj37/m1/upper 95:Java}{-0.148364099666666}
\pgfkeyssetvalue{/cj37/m1/upper 95:Python}{1.10009590033333}
\pgfkeyssetvalue{/cj37/m2/lower 95:C++}{-0.603612788833333}
\pgfkeyssetvalue{/cj37/m2/lower 95:Java}{-0.636472788833333}
\pgfkeyssetvalue{/cj37/m2/lower 95:Python}{0.440247211166667}
\pgfkeyssetvalue{/cj37/m2/lower 90:C++}{-0.585962788833333}
\pgfkeyssetvalue{/cj37/m2/lower 90:Java}{-0.599112788833334}
\pgfkeyssetvalue{/cj37/m2/lower 90:Python}{0.501457211166667}
\pgfkeyssetvalue{/cj37/m2/lower 70:C++}{-0.541742788833333}
\pgfkeyssetvalue{/cj37/m2/lower 70:Java}{-0.520572788833333}
\pgfkeyssetvalue{/cj37/m2/lower 70:Python}{0.625927211166667}
\pgfkeyssetvalue{/cj37/m2/lower 50:C++}{-0.523822788833333}
\pgfkeyssetvalue{/cj37/m2/lower 50:Java}{-0.488092788833333}
\pgfkeyssetvalue{/cj37/m2/lower 50:Python}{0.687737211166667}
\pgfkeyssetvalue{/cj37/m2/upper 50:C++}{-0.433652788833333}
\pgfkeyssetvalue{/cj37/m2/upper 50:Java}{-0.312652788833333}
\pgfkeyssetvalue{/cj37/m2/upper 50:Python}{0.981097211166667}
\pgfkeyssetvalue{/cj37/m2/upper 70:C++}{-0.404082788833334}
\pgfkeyssetvalue{/cj37/m2/upper 70:Java}{-0.249552788833333}
\pgfkeyssetvalue{/cj37/m2/upper 70:Python}{1.07179721116667}
\pgfkeyssetvalue{/cj37/m2/upper 90:C++}{-0.366732788833334}
\pgfkeyssetvalue{/cj37/m2/upper 90:Java}{-0.167652788833333}
\pgfkeyssetvalue{/cj37/m2/upper 90:Python}{1.20876721116667}
\pgfkeyssetvalue{/cj37/m2/upper 95:C++}{-0.344342788833333}
\pgfkeyssetvalue{/cj37/m2/upper 95:Java}{-0.122982788833333}
\pgfkeyssetvalue{/cj37/m2/upper 95:Python}{1.28459721116667}
\pgfkeyssetvalue{/cj37/m3/lower 95:C++}{-0.322146012}
\pgfkeyssetvalue{/cj37/m3/lower 95:Java}{-0.270516012}
\pgfkeyssetvalue{/cj37/m3/lower 95:Python}{-0.164206012}
\pgfkeyssetvalue{/cj37/m3/lower 90:C++}{-0.293376012}
\pgfkeyssetvalue{/cj37/m3/lower 90:Java}{-0.233936012}
\pgfkeyssetvalue{/cj37/m3/lower 90:Python}{-0.103966012}
\pgfkeyssetvalue{/cj37/m3/lower 70:C++}{-0.247106012000001}
\pgfkeyssetvalue{/cj37/m3/lower 70:Java}{-0.161256012}
\pgfkeyssetvalue{/cj37/m3/lower 70:Python}{0.0192339879999999}
\pgfkeyssetvalue{/cj37/m3/lower 50:C++}{-0.215826012}
\pgfkeyssetvalue{/cj37/m3/lower 50:Java}{-0.122346012}
\pgfkeyssetvalue{/cj37/m3/lower 50:Python}{0.0705839880000001}
\pgfkeyssetvalue{/cj37/m3/upper 50:C++}{-0.111676012}
\pgfkeyssetvalue{/cj37/m3/upper 50:Java}{0.044803988}
\pgfkeyssetvalue{/cj37/m3/upper 50:Python}{0.312443988}
\pgfkeyssetvalue{/cj37/m3/upper 70:C++}{-0.0857360119999999}
\pgfkeyssetvalue{/cj37/m3/upper 70:Java}{0.0966239880000002}
\pgfkeyssetvalue{/cj37/m3/upper 70:Python}{0.392293988}
\pgfkeyssetvalue{/cj37/m3/upper 90:C++}{-0.0342560120000002}
\pgfkeyssetvalue{/cj37/m3/upper 90:Java}{0.175433988}
\pgfkeyssetvalue{/cj37/m3/upper 90:Python}{0.486983988}
\pgfkeyssetvalue{/cj37/m3/upper 95:C++}{-0.0122760120000001}
\pgfkeyssetvalue{/cj37/m3/upper 95:Java}{0.218943987999999}
\pgfkeyssetvalue{/cj37/m3/upper 95:Python}{0.542163988}
\pgfkeyssetvalue{/cj37/m4/lower 95:C++}{-0.661867688333334}
\pgfkeyssetvalue{/cj37/m4/lower 95:Java}{-0.378167688333335}
\pgfkeyssetvalue{/cj37/m4/lower 95:Python}{-1.07116768833333}
\pgfkeyssetvalue{/cj37/m4/lower 90:C++}{-0.566767688333334}
\pgfkeyssetvalue{/cj37/m4/lower 90:Java}{-0.218067688333335}
\pgfkeyssetvalue{/cj37/m4/lower 90:Python}{-0.987267688333334}
\pgfkeyssetvalue{/cj37/m4/lower 70:C++}{-0.344267688333334}
\pgfkeyssetvalue{/cj37/m4/lower 70:Java}{0.00403231166666629}
\pgfkeyssetvalue{/cj37/m4/lower 70:Python}{-0.767367688333334}
\pgfkeyssetvalue{/cj37/m4/lower 50:C++}{-0.192067688333333}
\pgfkeyssetvalue{/cj37/m4/lower 50:Java}{0.133232311666667}
\pgfkeyssetvalue{/cj37/m4/lower 50:Python}{-0.629567688333333}
\pgfkeyssetvalue{/cj37/m4/upper 50:C++}{0.272732311666665}
\pgfkeyssetvalue{/cj37/m4/upper 50:Java}{0.629532311666665}
\pgfkeyssetvalue{/cj37/m4/upper 50:Python}{-0.151767688333335}
\pgfkeyssetvalue{/cj37/m4/upper 70:C++}{0.379632311666665}
\pgfkeyssetvalue{/cj37/m4/upper 70:Java}{0.769532311666666}
\pgfkeyssetvalue{/cj37/m4/upper 70:Python}{-0.0321676883333346}
\pgfkeyssetvalue{/cj37/m4/upper 90:C++}{0.587632311666667}
\pgfkeyssetvalue{/cj37/m4/upper 90:Java}{0.993432311666666}
\pgfkeyssetvalue{/cj37/m4/upper 90:Python}{0.175432311666665}
\pgfkeyssetvalue{/cj37/m4/upper 95:C++}{0.711532311666666}
\pgfkeyssetvalue{/cj37/m4/upper 95:Java}{1.07463231166667}
\pgfkeyssetvalue{/cj37/m4/upper 95:Python}{0.310532311666666}
\pgfkeyssetvalue{/cj37/m5/lower 95:C++}{-0.569805975}
\pgfkeyssetvalue{/cj37/m5/lower 95:Java}{-0.4956504}
\pgfkeyssetvalue{/cj37/m5/lower 95:Python}{-0.3239186}
\pgfkeyssetvalue{/cj37/m5/lower 90:C++}{-0.4679974}
\pgfkeyssetvalue{/cj37/m5/lower 90:Java}{-0.40193045}
\pgfkeyssetvalue{/cj37/m5/lower 90:Python}{-0.233107}
\pgfkeyssetvalue{/cj37/m5/lower 70:C++}{-0.2972614}
\pgfkeyssetvalue{/cj37/m5/lower 70:Java}{-0.2331624}
\pgfkeyssetvalue{/cj37/m5/lower 70:Python}{-0.0801759199999999}
\pgfkeyssetvalue{/cj37/m5/lower 50:C++}{-0.214709}
\pgfkeyssetvalue{/cj37/m5/lower 50:Java}{-0.1526115}
\pgfkeyssetvalue{/cj37/m5/lower 50:Python}{-0.0166407}
\pgfkeyssetvalue{/cj37/m5/upper 50:C++}{-0.0014327975}
\pgfkeyssetvalue{/cj37/m5/upper 50:Java}{0.060945225}
\pgfkeyssetvalue{/cj37/m5/upper 50:Python}{0.202038}
\pgfkeyssetvalue{/cj37/m5/upper 70:C++}{0.0509331049999999}
\pgfkeyssetvalue{/cj37/m5/upper 70:Java}{0.12072675}
\pgfkeyssetvalue{/cj37/m5/upper 70:Python}{0.28273185}
\pgfkeyssetvalue{/cj37/m5/upper 90:C++}{0.1707204}
\pgfkeyssetvalue{/cj37/m5/upper 90:Java}{0.24746225}
\pgfkeyssetvalue{/cj37/m5/upper 90:Python}{0.4402325}
\pgfkeyssetvalue{/cj37/m5/upper 95:C++}{0.2483222}
\pgfkeyssetvalue{/cj37/m5/upper 95:Java}{0.333365224999999}
\pgfkeyssetvalue{/cj37/m5/upper 95:Python}{0.528225649999999}
\pgfkeyssetvalue{/cj37/m6/lower 95:C++}{-1.05101875}
\pgfkeyssetvalue{/cj37/m6/lower 95:Java}{-0.947024675}
\pgfkeyssetvalue{/cj37/m6/lower 95:Python}{-0.1662468}
\pgfkeyssetvalue{/cj37/m6/lower 90:C++}{-0.9226428}
\pgfkeyssetvalue{/cj37/m6/lower 90:Java}{-0.80557635}
\pgfkeyssetvalue{/cj37/m6/lower 90:Python}{-0.042790375}
\pgfkeyssetvalue{/cj37/m6/lower 70:C++}{-0.6930595}
\pgfkeyssetvalue{/cj37/m6/lower 70:Java}{-0.56500355}
\pgfkeyssetvalue{/cj37/m6/lower 70:Python}{0.1528808}
\pgfkeyssetvalue{/cj37/m6/lower 50:C++}{-0.56900275}
\pgfkeyssetvalue{/cj37/m6/lower 50:Java}{-0.43302275}
\pgfkeyssetvalue{/cj37/m6/lower 50:Python}{0.2624835}
\pgfkeyssetvalue{/cj37/m6/upper 50:C++}{-0.17905475}
\pgfkeyssetvalue{/cj37/m6/upper 50:Java}{-0.02082055}
\pgfkeyssetvalue{/cj37/m6/upper 50:Python}{0.691057}
\pgfkeyssetvalue{/cj37/m6/upper 70:C++}{-0.0801780650000002}
\pgfkeyssetvalue{/cj37/m6/upper 70:Java}{0.0840433349999996}
\pgfkeyssetvalue{/cj37/m6/upper 70:Python}{0.822606849999999}
\pgfkeyssetvalue{/cj37/m6/upper 90:C++}{0.11053425}
\pgfkeyssetvalue{/cj37/m6/upper 90:Java}{0.28000605}
\pgfkeyssetvalue{/cj37/m6/upper 90:Python}{1.061062}
\pgfkeyssetvalue{/cj37/m6/upper 95:C++}{0.222711125}
\pgfkeyssetvalue{/cj37/m6/upper 95:Java}{0.387904474999999}
\pgfkeyssetvalue{/cj37/m6/upper 95:Python}{1.20405675}
\pgfkeyssetvalue{/cj37/max diff/language:diff}{1.2591}
\pgfkeyssetvalue{/cj37/max diff/language:exp}{3.52225003539421}
\pgfkeyssetvalue{/cj37/max diff/nickname:diff}{3.11099}
\pgfkeyssetvalue{/cj37/max diff/nickname:exp}{22.4432522257626}
\pgfkeyssetvalue{/cj37/max diff/challenge:diff}{5.03176}
\pgfkeyssetvalue{/cj37/max diff/challenge:exp}{153.20241179963}
\pgfkeyssetvalue{/cj37/max diff/size:diff}{6.62367803180834}
\pgfkeyssetvalue{/cj37/max diff/size:exp}{752.708497856233}
\pgfkeyssetvalue{/cj37/main size index}{6}
\pgfkeyssetvalue{/cj37/other datasets/1:size}{1}
\pgfkeyssetvalue{/cj37/other datasets/1:num.years}{7}
\pgfkeyssetvalue{/cj37/other datasets/1:num.rounds}{6}
\pgfkeyssetvalue{/cj37/other datasets/1:geq.years.rounds}{721}
\pgfkeyssetvalue{/cj37/other datasets/1:n.nicknames}{27}
\pgfkeyssetvalue{/cj37/other datasets/2:size}{2}
\pgfkeyssetvalue{/cj37/other datasets/2:num.years}{6}
\pgfkeyssetvalue{/cj37/other datasets/2:num.rounds}{6}
\pgfkeyssetvalue{/cj37/other datasets/2:geq.years.rounds}{960}
\pgfkeyssetvalue{/cj37/other datasets/2:n.nicknames}{39}
\pgfkeyssetvalue{/cj37/other datasets/3:size}{3}
\pgfkeyssetvalue{/cj37/other datasets/3:num.years}{5}
\pgfkeyssetvalue{/cj37/other datasets/3:num.rounds}{6}
\pgfkeyssetvalue{/cj37/other datasets/3:geq.years.rounds}{1230}
\pgfkeyssetvalue{/cj37/other datasets/3:n.nicknames}{53}
\pgfkeyssetvalue{/cj37/other datasets/4:size}{4}
\pgfkeyssetvalue{/cj37/other datasets/4:num.years}{4}
\pgfkeyssetvalue{/cj37/other datasets/4:num.rounds}{6}
\pgfkeyssetvalue{/cj37/other datasets/4:geq.years.rounds}{1564}
\pgfkeyssetvalue{/cj37/other datasets/4:n.nicknames}{75}
\pgfkeyssetvalue{/cj37/other datasets/5:size}{5}
\pgfkeyssetvalue{/cj37/other datasets/5:num.years}{3}
\pgfkeyssetvalue{/cj37/other datasets/5:num.rounds}{6}
\pgfkeyssetvalue{/cj37/other datasets/5:geq.years.rounds}{1807}
\pgfkeyssetvalue{/cj37/other datasets/5:n.nicknames}{95}
\pgfkeyssetvalue{/cj37/other datasets/6:size}{6}
\pgfkeyssetvalue{/cj37/other datasets/6:num.years}{2}
\pgfkeyssetvalue{/cj37/other datasets/6:num.rounds}{6}
\pgfkeyssetvalue{/cj37/other datasets/6:geq.years.rounds}{1896}
\pgfkeyssetvalue{/cj37/other datasets/6:n.nicknames}{105}
\pgfkeyssetvalue{/cj37/other datasets/7:size}{7}
\pgfkeyssetvalue{/cj37/other datasets/7:num.years}{1}
\pgfkeyssetvalue{/cj37/other datasets/7:num.rounds}{6}
\pgfkeyssetvalue{/cj37/other datasets/7:geq.years.rounds}{2010}
\pgfkeyssetvalue{/cj37/other datasets/7:n.nicknames}{124}
\pgfkeyssetvalue{/cj37/other datasets/8:size}{8}
\pgfkeyssetvalue{/cj37/other datasets/8:num.years}{7}
\pgfkeyssetvalue{/cj37/other datasets/8:num.rounds}{5}
\pgfkeyssetvalue{/cj37/other datasets/8:geq.years.rounds}{2762}
\pgfkeyssetvalue{/cj37/other datasets/8:n.nicknames}{115}
\pgfkeyssetvalue{/cj37/other datasets/9:size}{9}
\pgfkeyssetvalue{/cj37/other datasets/9:num.years}{7}
\pgfkeyssetvalue{/cj37/other datasets/9:num.rounds}{4}
\pgfkeyssetvalue{/cj37/other datasets/9:geq.years.rounds}{4889}
\pgfkeyssetvalue{/cj37/other datasets/9:n.nicknames}{221}
\pgfkeyssetvalue{/cj37/other datasets/10:size}{10}
\pgfkeyssetvalue{/cj37/other datasets/10:num.years}{6}
\pgfkeyssetvalue{/cj37/other datasets/10:num.rounds}{5}
\pgfkeyssetvalue{/cj37/other datasets/10:geq.years.rounds}{5106}
\pgfkeyssetvalue{/cj37/other datasets/10:n.nicknames}{233}
\pgfkeyssetvalue{/cj37/other datasets/11:size}{11}
\pgfkeyssetvalue{/cj37/other datasets/11:num.years}{7}
\pgfkeyssetvalue{/cj37/other datasets/11:num.rounds}{3}
\pgfkeyssetvalue{/cj37/other datasets/11:geq.years.rounds}{5707}
\pgfkeyssetvalue{/cj37/other datasets/11:n.nicknames}{272}
\pgfkeyssetvalue{/cj37/other datasets/12:size}{12}
\pgfkeyssetvalue{/cj37/other datasets/12:num.years}{7}
\pgfkeyssetvalue{/cj37/other datasets/12:num.rounds}{1}
\pgfkeyssetvalue{/cj37/other datasets/12:geq.years.rounds}{5776}
\pgfkeyssetvalue{/cj37/other datasets/12:n.nicknames}{279}
\pgfkeyssetvalue{/cj37/other datasets/13:size}{13}
\pgfkeyssetvalue{/cj37/other datasets/13:num.years}{7}
\pgfkeyssetvalue{/cj37/other datasets/13:num.rounds}{2}
\pgfkeyssetvalue{/cj37/other datasets/13:geq.years.rounds}{5776}
\pgfkeyssetvalue{/cj37/other datasets/13:n.nicknames}{279}
\pgfkeyssetvalue{/cj37/other datasets/14:size}{14}
\pgfkeyssetvalue{/cj37/other datasets/14:num.years}{5}
\pgfkeyssetvalue{/cj37/other datasets/14:num.rounds}{5}
\pgfkeyssetvalue{/cj37/other datasets/14:geq.years.rounds}{7359}
\pgfkeyssetvalue{/cj37/other datasets/14:n.nicknames}{366}
\pgfkeyssetvalue{/cj37/other datasets/15:size}{15}
\pgfkeyssetvalue{/cj37/other datasets/15:num.years}{4}
\pgfkeyssetvalue{/cj37/other datasets/15:num.rounds}{5}
\pgfkeyssetvalue{/cj37/other datasets/15:geq.years.rounds}{9667}
\pgfkeyssetvalue{/cj37/other datasets/15:n.nicknames}{536}
\pgfkeyssetvalue{/cj37/other datasets/16:size}{16}
\pgfkeyssetvalue{/cj37/other datasets/16:num.years}{6}
\pgfkeyssetvalue{/cj37/other datasets/16:num.rounds}{4}
\pgfkeyssetvalue{/cj37/other datasets/16:geq.years.rounds}{11242}
\pgfkeyssetvalue{/cj37/other datasets/16:n.nicknames}{576}
\pgfkeyssetvalue{/cj37/other datasets/17:size}{17}
\pgfkeyssetvalue{/cj37/other datasets/17:num.years}{3}
\pgfkeyssetvalue{/cj37/other datasets/17:num.rounds}{5}
\pgfkeyssetvalue{/cj37/other datasets/17:geq.years.rounds}{11814}
\pgfkeyssetvalue{/cj37/other datasets/17:n.nicknames}{739}
\pgfkeyssetvalue{/cj37/other datasets/18:size}{18}
\pgfkeyssetvalue{/cj37/other datasets/18:num.years}{2}
\pgfkeyssetvalue{/cj37/other datasets/18:num.rounds}{5}
\pgfkeyssetvalue{/cj37/other datasets/18:geq.years.rounds}{13478}
\pgfkeyssetvalue{/cj37/other datasets/18:n.nicknames}{957}
\pgfkeyssetvalue{/cj37/other datasets/19:size}{19}
\pgfkeyssetvalue{/cj37/other datasets/19:num.years}{6}
\pgfkeyssetvalue{/cj37/other datasets/19:num.rounds}{3}
\pgfkeyssetvalue{/cj37/other datasets/19:geq.years.rounds}{13652}
\pgfkeyssetvalue{/cj37/other datasets/19:n.nicknames}{749}
\pgfkeyssetvalue{/cj37/other datasets/20:size}{20}
\pgfkeyssetvalue{/cj37/other datasets/20:num.years}{6}
\pgfkeyssetvalue{/cj37/other datasets/20:num.rounds}{2}
\pgfkeyssetvalue{/cj37/other datasets/20:geq.years.rounds}{13884}
\pgfkeyssetvalue{/cj37/other datasets/20:n.nicknames}{775}
\pgfkeyssetvalue{/cj37/other datasets/21:size}{21}
\pgfkeyssetvalue{/cj37/other datasets/21:num.years}{6}
\pgfkeyssetvalue{/cj37/other datasets/21:num.rounds}{1}
\pgfkeyssetvalue{/cj37/other datasets/21:geq.years.rounds}{13908}
\pgfkeyssetvalue{/cj37/other datasets/21:n.nicknames}{779}
\pgfkeyssetvalue{/cj37/used languages/1}{C++}
\pgfkeyssetvalue{/cj37/used languages/2}{Java}
\pgfkeyssetvalue{/cj37/used languages/3}{Python}
\pgfkeyssetvalue{/cj37/used languages/4}{C\#}
\pgfkeyssetvalue{/cj37/used languages/5}{C}
\pgfkeyssetvalue{/cj37/used languages/6}{Ruby}
\pgfkeyssetvalue{/cj37/used languages/7}{PHP}
\pgfkeyssetvalue{/cj37/used languages/8}{Perl}
\pgfkeyssetvalue{/cj37/used languages/9}{Pascal}
\pgfkeyssetvalue{/cj37/used languages/10}{Haskell}
\pgfkeyssetvalue{/cj37/used languages/11}{Scala}
\pgfkeyssetvalue{/cj37/used languages/12}{JavaScript}
\pgfkeyssetvalue{/cj37/used languages/13}{Visual Basic}
\pgfkeyssetvalue{/cj37/used languages/14}{Go}
\pgfkeyssetvalue{/cj37/used languages/15}{MATLAB}
\pgfkeyssetvalue{/cj37/used languages/16}{OCaml}
\pgfkeyssetvalue{/cj37/used languages/17}{Lisp}
\pgfkeyssetvalue{/cj37/used languages/18}{Groovy}
\pgfkeyssetvalue{/cj37/used languages/19}{Clojure}
\pgfkeyssetvalue{/cj37/used languages/20}{D}
\pgfkeyssetvalue{/cj37/used languages/21}{Lua}
\pgfkeyssetvalue{/cj37/used languages/22}{F\#}
\pgfkeyssetvalue{/cj37/used languages/23}{Scheme}
\pgfkeyssetvalue{/cj37/used languages/24}{Objective-C}
\pgfkeyssetvalue{/cj37/used languages/25}{CoffeeScript}
\pgfkeyssetvalue{/cj37/used languages/26}{ActionScript}
\pgfkeyssetvalue{/cj37/used languages/27}{Racket}
\pgfkeyssetvalue{/cj37/used languages/28}{Shell}
\pgfkeyssetvalue{/cj37/used languages/29}{R}
\pgfkeyssetvalue{/cj37/used languages/30}{Erlang}
\pgfkeyssetvalue{/cj37/used languages/31}{J}
\pgfkeyssetvalue{/cj37/used languages/32}{TCL}
\pgfkeyssetvalue{/cj37/used languages/33}{AWK}
\pgfkeyssetvalue{/cj37/used languages/34}{Dart}
\pgfkeyssetvalue{/cj37/used languages/35}{Julia}
\pgfkeyssetvalue{/cj37/used languages/36}{WRAPL}
\pgfkeyssetvalue{/cj37/used languages/37}{PowerShell}
\pgfkeyssetvalue{/cj37/used languages/38}{Fortran}
\pgfkeyssetvalue{/cj37/used languages/39}{PL/SQL}
\pgfkeyssetvalue{/cj37/used languages/40}{Processing}
\pgfkeyssetvalue{/cj37/used languages/41}{Prolog}
\pgfkeyssetvalue{/cj37/used languages/42}{Standard ML}
\pgfkeyssetvalue{/cj37/used languages/43}{Gosu}
\pgfkeyssetvalue{/cj37/used languages/44}{Kotlin}
\pgfkeyssetvalue{/cj37/used languages/45}{AutoIt}
\pgfkeyssetvalue{/cj37/used languages/46}{Basic}
\pgfkeyssetvalue{/cj37/used languages/47}{Rust}
\pgfkeyssetvalue{/cj37/used languages/48}{Ada}
\pgfkeyssetvalue{/cj37/used languages/49}{Factor}
\pgfkeyssetvalue{/cj37/used languages/50}{SQL}
\pgfkeyssetvalue{/cj37/used languages/51}{Xtend}
\pgfkeyssetvalue{/cj37/used languages/52}{Cache}
\pgfkeyssetvalue{/cj37/used languages/53}{Octave}
\pgfkeyssetvalue{/cj37/used languages/54}{GAP}
\pgfkeyssetvalue{/cj37/used languages/55}{Pike}
\pgfkeyssetvalue{/cj37/used languages/56}{AutoHotkey}
\pgfkeyssetvalue{/cj37/used languages/57}{Boo}
\pgfkeyssetvalue{/cj37/used languages/58}{Clay}
\pgfkeyssetvalue{/cj37/used languages/59}{COBOL}
\pgfkeyssetvalue{/cj37/used languages/60}{Cobra}
\pgfkeyssetvalue{/cj37/used languages/61}{Maple}
\pgfkeyssetvalue{/cj37/used languages/62}{Mathematica}
\pgfkeyssetvalue{/cj37/used languages/63}{Nemerle}
\pgfkeyssetvalue{/cj37/used languages/64}{Rexx}
\pgfkeyssetvalue{/cj37/used languages/65}{Smalltalk}
\pgfkeyssetvalue{/cj37/used languages/66}{Eiffel}
\pgfkeyssetvalue{/cj37/used languages/67}{FRACTRAN}
\pgfkeyssetvalue{/cj37/used languages/68}{Golfscript}
\pgfkeyssetvalue{/cj37/used languages/69}{Haxe}
\pgfkeyssetvalue{/cj37/used languages/70}{HSP}
\pgfkeyssetvalue{/cj37/used languages/71}{Idris}
\pgfkeyssetvalue{/cj37/used languages/72}{Mercury}
\pgfkeyssetvalue{/cj37/used languages/73}{mIRC}
\pgfkeyssetvalue{/cj37/used languages/74}{MoonScript}
\pgfkeyssetvalue{/cj37/used languages/75}{Whitespace}
\pgfkeyssetvalue{/cj37/used languages/76}{Agda}
\pgfkeyssetvalue{/cj37/used languages/77}{ALGOL}
\pgfkeyssetvalue{/cj37/used languages/78}{Alice}
\pgfkeyssetvalue{/cj37/used languages/79}{Assembly}
\pgfkeyssetvalue{/cj37/used languages/80}{Asymptote}
\pgfkeyssetvalue{/cj37/used languages/81}{B}
\pgfkeyssetvalue{/cj37/used languages/82}{bc}
\pgfkeyssetvalue{/cj37/used languages/83}{Befunge}
\pgfkeyssetvalue{/cj37/used languages/84}{BibTeX}
\pgfkeyssetvalue{/cj37/used languages/85}{Brainfuck}
\pgfkeyssetvalue{/cj37/used languages/86}{CJam}
\pgfkeyssetvalue{/cj37/used languages/87}{Claire}
\pgfkeyssetvalue{/cj37/used languages/88}{Clasp}
\pgfkeyssetvalue{/cj37/used languages/89}{Clean}
\pgfkeyssetvalue{/cj37/used languages/90}{CMake}
\pgfkeyssetvalue{/cj37/used languages/91}{Coco}
\pgfkeyssetvalue{/cj37/used languages/92}{Cool}
\pgfkeyssetvalue{/cj37/used languages/93}{Coq}
\pgfkeyssetvalue{/cj37/used languages/94}{dc}
\pgfkeyssetvalue{/cj37/used languages/95}{Dylan}
\pgfkeyssetvalue{/cj37/used languages/96}{ECLiPSe}
\pgfkeyssetvalue{/cj37/used languages/97}{Event Related Model}
\pgfkeyssetvalue{/cj37/used languages/98}{FALSE}
\pgfkeyssetvalue{/cj37/used languages/99}{FreeMat}
\pgfkeyssetvalue{/cj37/used languages/100}{(Other)}

\title{Towards Causal Analysis of\\ Empirical Software Engineering Data}
\subtitle{The Impact of Programming Languages on Coding Competitions}

\ifarxiv
\author{Carlo A.\ Furia$^1$
        $\quad\cdot\quad$
        Richard Torkar$^{2,3}$
        $\quad\cdot\quad$
        Robert Feldt$^2$ \\[2mm]
        \normalsize
        $^1$ Software Institute, USI Universit\`a della Svizzera italiana, Switzerland \\
        \normalsize
        $^2$ Chalmers and the University of Gothenburg, Sweden \\
        \normalsize
        $^3$ Stellenbosch Institute for Advanced Study (STIAS), South Africa
}
\else
\author{Carlo A.\ Furia}
\orcid{http://orcid.org/0000-0003-1040-3201}
\affiliation{\institution{Software Institute, USI Universit\`a della Svizzera italiana}
  \streetaddress{Via G.\ Buffi 13}
  \postcode{CH-6900}
  \city{Lugano}
  \country{Switzerland}}
\email{https://bugcounting.net}

\author{Richard Torkar}
\orcid{http://orcid.org/0000-0002-0118-8143}
\affiliation{\institution{University of Gothenburg}
  \streetaddress{Chalmersplaten 4}
  \postcode{SE-41296}
  \city{Gothenburg}
  \country{Sweden}}
\affiliation{\institution{Stellenbosch Institute for Advanced Study (STIAS)}
  \city{Stellenbosch}
  \country{South Africa}}
\email{https://www.torkar.se}

\author{Robert Feldt}
\orcid{http://orcid.org/0000-0002-5179-4205}
\affiliation{\institution{Chalmers and University of Gothenburg}
  \streetaddress{Chalmersplaten 4}
  \postcode{SE-41296}
  \city{Gothenburg}
  \country{Sweden}}
\email{https://www.roberfeldt.net}
\fi

\ifarxiv\else
\authorsaddresses{}

\acmBooktitle{}

\acmJournal{TOSEM}


\fi

\makeatletter
\let\Title\@title
\makeatother

\ifarxiv\else
\fi

\date{First version: 2023-01-18. Current version: 2023-06-27.}

\usepackage{amsmath,amsthm,amsfonts,dsfont}
\usepackage{mathtools}

\newcommand{\bigCI}{\mathrel{\text{\scalebox{1.07}{$\perp\mkern-10mu\perp$}}}}

\DeclareDocumentCommand{\bracketcol}{s m m O{black} O{black} O{black}}
{\IfBooleanTF{#1}{\begingroup
    \color{#4}
    \overbracket{\color{#6}#2}^{\color{#5}#3}
    \endgroup
  }{\begingroup
    \color{#4}
    \underbracket{\color{#6}#2}_{\color{#5}#3}
    \endgroup
  }}

\newcommand{\idm}[1]{\ensuremath{\mathit{mid}_{#1}}}

\DeclareDocumentCommand{\effect}{s o}{\ensuremath{\delta \IfBooleanTF{#1}{\IfValueT{#2}{(#2)}}{\IfValueT{#2}_{#2}}}}
\newcommand{\mean}[1]{\ensuremath{\mathds{E}\!\left[#1\right]}}

\usepackage{booktabs}
\usepackage{xcolor}
\definecolor{m1col}{HTML}{1B9E77}
\definecolor{m2col}{HTML}{D95F02}
\definecolor{m3col}{HTML}{7570B3}
\definecolor{m4col}{HTML}{E7298A}

\usepackage{rotating}
\usepackage{multirow}

\usepackage{enumerate}
 \usepackage[inline]{enumitem}
\setlist[description]{leftmargin=\parindent,labelindent=\parindent,align=left}

\usepackage{changepage}

\usepackage{array}

\usepackage{ifthen}

\usepackage{boxedminipage}

\usepackage[colorinlistoftodos,textsize=tiny,obeyFinal,textwidth=16mm]{todonotes}
\NewDocumentCommand{\ColorComment}{O{color=white} m}{\todo[size=\scriptsize,#1]{#2}}

\usepackage{fontawesome}

\pgfplotsset{compat=1.17}

\usepackage{silence}
\usepackage{hyperref}

\hypersetup{
pdftitle={\Title{}},
pdfauthor={Carlo A. Furia, Richard Torkar, Robert Feldt},
pdfsubject={},
pdfkeywords={},
pdfcreator={pdfTeX}
}

\begin{document}

\ifarxiv
  \maketitle
\fi

\begin{abstract}
There is abundant observational data in the software engineering domain,
  whereas running large-scale controlled experiments is often practically impossible.
  Thus, most empirical studies can only report statistical \emph{correlations}---instead
  of potentially more insightful and robust \emph{causal} relations.

To support analyzing purely observational data for causal relations,
  and to assess any differences between purely predictive and causal models
  of the same data, 
  this paper discusses some novel techniques based on structural
  causal models (such as directed acyclic graphs of causal Bayesian networks).
  Using these techniques, one can rigorously
  express, and partially validate, causal hypotheses; and then use the
  causal information to guide the construction of a statistical model
  that captures genuine causal relations---such that
  correlation \emph{does imply} causation.
  
We apply these ideas to analyzing public data about programmer
  performance in Code Jam, a large world-wide coding contest organized
  by Google every year. Specifically, we look at the impact of
  different programming languages on a participant's performance in
  the contest.
While the overall effect associated with programming
  languages is weak compared to other variables---regardless of
  whether we consider correlational or causal links---we found
  considerable differences between a purely associational and a causal
  analysis of the very same data.

The takeaway message is that even an
  imperfect causal analysis of observational data can help answer the
  salient research questions more precisely and more robustly than
  with just purely predictive techniques---where genuine causal effects may be confounded.
\end{abstract}

\ifarxiv\else
\maketitle
\fi

\section{Introduction}

It is commonplace that ``correlation does not imply causation'':
just because two variables appear to change together,
it does not mean that one \emph{makes} the other change.
This fundamental limitation 
is especially damning for fields like
empirical software engineering, 
where there is plenty of detailed observational data---for example by mining software repositories---but fully controlled experiments are challenging to design and to run.
As a result, most (quantitative) empirical studies in software engineering
apply statistical analysis to 
detect \emph{associations} (such as correlations),
which may suggest and be consistent with---but do not necessarily establish---causal effects.

On the bright side,
if our only, or primary, goal is \emph{predicting}
an outcome from measured variables,
we do not necessarily need to understand causal relations,
and a more traditional approach to statistical analysis may be enough.
Practical solutions for engineering can often be built on prediction alone:
the spectacular success of machine learning---which often learns from and predicts observational data---indicates that lots of applications can go a long way with correlational data alone.
In contrast,
if we want to \emph{understand} the specific contributions
of some variables on some outcomes, and to \emph{generalize} 
their effects in a different scenario,
we must be able to distinguish between mere correlations
and true causal effects.
After all,
an understanding grounded in causal relations has not only practical value,
but should be a fundamental goal of any rigorous scientific discipline.

In this paper, we demonstrate on a case study
(outlined in \autoref{sec:langs})
how 
thinking about causality can make a difference---even when all we have is observational data---to reliably isolate the impact of specific factors in software engineering empirical data, 
and to develop a statistical analysis that
precisely answers our research questions.
In a nutshell, we will apply some basic tools
of causal analysis
(in particular, causal directed acyclic graphs)
to model and assess possible causal relations between variables~\cite{pearl09a@reason}.\footnote{As we further discuss in \autoref{sec:rw:causality},
  the over-arching goal of modern causal analysis is
  to formally capture the intuitive notion of causality
  with mathematical\slash statistical constructs, so that
  one can rigorously reason about causal links
  in experimental or observational data.
  }
This, in turn, will help us
choose a statistical model that reliably identifies the magnitude of causal effects.
We will also demonstrate that \emph{not}
accounting for causality leads to choosing different statistical models
for our case study,
which make accurate predictions ``in the small''
but spuriously misrepresent the impact of
the variables of interest.

The main lesson following this exercise
will be that
software engineering empirical research
needs conceptual and technical tools
to model and reason about causal relations.
Failing to do so---sticking to purely associational interpretations of empirical data---can severely limit the robustness and generalizability of
any empirical results,
and would ultimately stifle the long-term standing
of software engineering's scientific progress.

\subsection{Programming: Competitions and Languages}
\label{sec:langs}

As a case study
of applying causal modeling to empirical software engineering data,
we consider programmers that take part in \emph{programming competitions}.
More precisely, we analyze data collected during \n.{num editions}! past editions
of the \cj
coding competition---organized every year by Google.
\autoref{sec:code-jam-data} describes how the \cj contest works,
and what public data we analyzed.

As they grow in popularity and accessibility,
programming competitions
help promote programming and its applications
in informal and fun settings.
From the point of view of empirical software engineering,
programming competitions 
provide a setting where
one can easily collect data about the behavior and performance of devoted programmers.
Of course, competitions usually take place
under somewhat artificial settings with numerous constraints,
which do not necessarily reflect software development as done in a professional environment.
On the flip side, 
the same constraints also make them a partially controlled environment,
where it's easier
to zero in on some specific aspects of the behavior of programmers
without too many confounding factors.
In particular, a programmer's work in a programming competition
usually has very clear goals and constraints,
and there are objective criteria to determine to what extent a submission
meets those goals and satisfies those constraints.

In this paper, we use \cj data to study, in a specific setting, a
long-standing question in computer science: the impact of using
different programming languages.
This evergreen topic featured in countless empirical studies---including some recent hot-button work~\cite{FSE,TOPLAS,TOSEM-Bayes-guidelines}---that usually analyzed observational data by purely associational means.
As summarized in \autoref{sec:related-work},
different studies do not always agree on their findings;
but they tend to suggest that the association to using different programming languages
is generally small compared to other correlational factors.
In this paper, we take yet another look at the same topic
but through the lens of causal relations.
Does this stance affect the conclusions that we can draw from an empirical analysis?

\cj accepts submissions in a broad variety of programming languages;
in practice, as we discuss in \autoref{sec:code-jam-data},
only a handful of programming languages are overwhelmingly chosen
by participants to \cj.
In order to focus our analysis on data that is extensive and consistent,
we only consider
\begin{enumerate*}[label=\textit{\roman*})]
\item submissions in the most \emph{popular languages} (namely, C++, Java, and Python),
  and
\item\emph{experienced} participants---that is, participants who successfully took part
  in several editions of the \cj contest.
\end{enumerate*}
Thus, we investigate the following fundamental research question.
\begin{center}
  \textbf{RQ}: For experienced participants to the \cj contest,
  how does \\using different programming languages relate to their results in the contest?
\end{center}

\subsection{Correlation vs.\ Causation}
\label{sec:intro:corr-vs-caus}

The phrasing of question RQ is deliberately vague about what ``relate to'' means---in other words, what kind of relations we are looking for.
Two common interpretations are:

\begin{description}
\item[correlational:] are some programming languages \emph{associated with}
  a certain performance in \cj?

\item[causal:] does using some programming languages \emph{affect}
  the performance in \cj?
\end{description}
The first interpretation only looks for correlations among the programming language
and other variables (in particular, the outcome in the competition).
The second interpretation is stricter, in a way, as it demands evidence that the programming language
tends to \emph{cause} a change in the competition outcome.

In this paper, we first carry out a statistical analysis
following the \emph{correlational\slash associational interpretation}.
This represents a type of data analysis that is typical in present-day
empirical software engineering.
The analysis, described in \autoref{sec:correlation},
indicates that:
\begin{enumerate*}[label=\textit{\roman*})]
\item using Java is associated with worse-than-average results in the \cj competition;
\item using Python is associated with better-than-average results; and
\item using C++ has no consistent association with better or worse results.
\end{enumerate*}

\autoref{sec:causation},
revisits the correlational analysis
under a \emph{causal} lens.
This represents a type of data analysis that could add value
if used more frequently in empirical software engineering.
In particular,
\autoref{sec:corr-doesnt-imply}
explains in what sense
``correlation does not imply causation'':
the \cj contest is obviously not a randomized experiment,
and hence we cannot distinguish a causal relation from a mere correlation
based on its data alone.
However, 
we can still build, under some reasonable assumptions,
a different statistical model than
the one used in the correlational analysis
that is poised to measure not mere correlations but causal effects.
Remarkably, the analysis of this ``causally-consistent'' model,
presented in \autoref{sec:analysis-effects},
leads to results that are nearly the opposite of
the correlational analysis's:
\begin{enumerate*}[label=\textit{\roman*})]
\item using Python is associated with worse-than-average results;
\item using C++ is associated with better-than-average results; and
\item using Java has no consistent association with better or worse results.
\end{enumerate*}
 
This difference in results between associational and causal analysis
is especially troubling given that 
the former clearly outperforms the latter 
in terms of purely \emph{predictive} accuracy;
yet, it fails to properly isolate causal effects.

As we further discuss in \autoref{sec:effects-comparison},
an underlying issue is that, in our data, the \emph{magnitude} of
the impact of choosing different programming languages is anyway small
compared to other factors (such as a programmer's individual skills).
Thus,
the main contribution of our work lies not so much in picking
the ``right'' programming language but rather in demonstrating---on a case study of empirical programming data---that thinking about causality can be crucial to choose
the statistical models that most accurately answer our research questions,
and to identify confounding factors that may affect the reliability of
a study's results.

\subsection{Contributions}
\label{sec:contributions}

This paper makes the following contributions:

\begin{itemize}
\item Highlights the importance, for empirical software engineering, of modeling and reasoning about causal relations to fully understand, and properly analyze, observational data;

\item Demonstrates using some basic tools of causal analysis on a case study,
  where the main findings differ from those that one would draw based on a purely correlational (associational)
  statistical analysis;

\item Argues, based on the case study, that a causal analysis's capability of identifying confounders (sources of non-causal, spurious correlations)
  provides a basis to more clearly identify the relative
  importance of different factors, and more easily disentangle them;

\item For reproducibility, all data and analysis scripts are available online---together with additional results and detailed data visualization:
    \begin{center}
        {\small \textsc{replication package:}}$\quad$ \url{https://doi.org/10.5281/zenodo.7541480}$\quad$ \cite{replication-package}.
    \end{center}
  
\end{itemize}

\subsection{Organization}

The rest of the paper is organized as follows.
\autoref{sec:related-work} summarizes some fundamental work on causal modeling and analysis,
recalls some (recent) empirical work on the analysis of programming languages on software development practices,
and discusses the few other applications of causal models to software engineering data analysis to date.
\autoref{sec:code-jam-data} presents the \cj programming contest,
the available data from several of its editions, and how we selected a subset of this data for analysis.
\autoref{sec:correlation} follows a ``canonical'', purely correlational,
statistical analysis of the \cj data---primarily using Bayesian techniques.
\autoref{sec:causation} peruses the statistical models developed in the previous section
according to a model of causal relations among observed variables,
which leads to revising some key study results.
\autoref{sec:effects-comparison} explains how
to reconcile the two conflicting outcomes of \autoref{sec:correlation}'s and \autoref{sec:causation}'s analysis, and corroborates previous evidence
that the absolute impact of programming languages is often small compared to other
factors.
\autoref{sec:robustness-and-threats} discusses further threats to the validity of the
case study's findings; and \autoref{sec:conclusions} concludes with a high-level discussion
of the paper's ideas and results.

\section{Related Work}
\label{sec:related-work}

This section reviews related work in the main areas that are relevant for this paper.
\autoref{sec:rw:causality} outlines fundamental concepts of modern causal inference---applicable, in principle, to every branch of science and engineering.
\autoref{sec:rw:se-causal} summarizes studies that have applied causal analysis techniques
to empirical data in the software engineering domain.
\autoref{sec:rw:pl-studies} reviews studies of the impact of using different programming languages, including on data emerging from programming contests of various kind---this paper's area of application.

\subsection{Causal Inference}
\label{sec:rw:causality}

Causal inference denotes a process, usually built on top of mathematical\slash statistical concepts, for identifying and analyzing which factors lead to specific outcomes~\cite{Pearl:2009aa,Pearl:2009causalOverview}.
It has grown in popularity as a tool for data analysis, and can be used in conjunction with, or in place of, traditional statistical and data analysis methods~\cite{Peters:2017elements}.
In a nutshell, while plain statistical methods identify \emph{associations},
causal analysis determines which factors \emph{lead to} (cause) changes in others,
and \emph{quantifies} such causal effects.

The so-called \emph{ladder of causality} illustrates the crucial differences between various types of analysis~\cite{Pearl:2019seven}.
Classical, associative statistical analysis is the bottom level (``seeing''),
addressing questions of the form 
``How does seeing $X$ change my belief in $Y$?''.
The analysis of interventions (``doing'')
is one step higher:
``What would happen to $Y$ if I do $X$?''.
Counterfactual analysis
is the highest level (``imagining''):
``What would have happened to $Y$ if $X$ had not occurred?''.

In principle, the benefits of causal inference in empirical research are clear:
finding the causes of studied effects is a key goal of science.
However, it can also have practical benefits, leading to more robust and accurate estimates. 
For instance, a re-analysis of medical data on hip fractures among elderly patients
found that causal analysis could identify which factors contribute to the occurrence of hip fractures and to what extent, and also provide predictions that were comparable to traditional methods~\cite{Caillet:2015hip}.
Another study~\cite{Richens:2020improving}
demonstrated that medical diagnosis based on causal inference was almost twice as effective
(25th percentile vs.\ 48th percentile of human doctors) as those based on traditional, associational statistical methods.

One key ingredient of modern-day causal inference is the use of directed acyclic graphs (DAGs) to model the observed factors' dependence \emph{structure}. Nodes in a DAG denote factors, and edges denote which nodes (edge source\slash parent)
cause changes in other nodes (edge target\slash child).
Thus, the DAG of a causal model makes the causal dependence structure explicit.
In addition to DAGs, a causal analysis framework normally offers means of \emph{estimating} 
the value of effect nodes from the values of their cause nodes.
In a so-called structural equation model, this is achieved via equations---typically linear, but more general forms have also been used---that relate each child node to its parents in the DAG.
Another kind of causal model, the probabilistic causal model, uses instead a single probability distribution over all factors~\cite{Pearl:2009aa}.

The causal structure (the DAG) underlying some empirical data
may be known a priori or be deduced from domain theory.
In some cases, one can apply so-called structural learning (also called causal discovery)
to identify causal structures from data~\cite{Spirtes:2016causal}.
Structural learning methods may combine observational data with interventional data---that is,
data measured after taking actions that cause changes in some of a DAG's factors.
A key benefit of structural causal models is that DAGs
can be used both to guide which factors to intervene on (change),
and to calculate the causal effects from the observations.

There are frameworks for causal inference that are not based on the modeling of causal structure with DAGs.
A notable example is the potential outcomes framework~\cite{rubin1974estimating}, also referred to as the Neyman-Rubin causal model. This framework estimates the causal effect of an intervention by comparing the outcome that would have occurred under the intervention to the outcome that would have occurred in the absence of the intervention.
Pearl has criticized this approach by pointing out concrete mistakes in the estimation of causal effects
that may occur if failing to consider the causal structure~\cite{pearl11NR}.
However, the potential outcomes framework may still have advantages in some fields such as econometrics~\cite{imbens2020potential}, where it fits better their typical assumptions and restrictions.
In this paper,
we focus on DAG-based causal inference, and leave the evaluation of other causal inference approaches to future work.

\subsection{Causal Inference and Analysis in Software Engineering}
\label{sec:rw:se-causal}

While causal analysis is certainly not (yet) a common practice in software engineering research, there is a growing number of studies that tried to apply it to specific problems. In a recent survey~\cite{Siebert:2021causal},
Siebert reviews a total of 31 studies that applied some kind of causal analysis to software engineering data. All these papers were published after 2010, and are considered ``software engineering'' because they focus on at least one of the activities of the SWEBOK (Software Engineering Body of Knowledge)~\cite{abran2004software}. 
The papers belong to \spellout{4} groups depending on their main activity: fault localization (17 papers), testing (4), performance analysis (5), and others (5).

The \spellout{17} fault localization papers build on, or at least cite, Baah et al.'s work~\cite{baah2010causal}.
Most of them model the causal effect that program elements (such as variables and statements)
have on values and, ultimately, failures in an executing program;
identifying these effects can help locate and debug actual faults.
One paper in this group~\cite{li2019thinking}, targets a different unit of analysis, as it is concerned with
fault localization in services composed of multiple, interacting programs.

Among the \spellout{4} testing papers,
one builds on the authors' fault localization results to identify elements to mutate~\cite{lee2021causal}.
Clark et al.~\cite{clark2022testing} test simulation models to determine whether they exhibit the expected causal effects. Liu et al.~\cite{liu2022bayesian} apply Bayesian causal modeling to improve the A/B testing used to evaluate changes in automotive software; they use the potential outcomes framework as their model of causality. 

Most of the \spellout{5} papers on performance analysis
focus on identifying components that determine a system's observed performance. An exception is the paper by Scholz et al.~\cite{scholz2021empirical}, which focuses instead on the performance of fault prediction algorithms.

The remaining \spellout{5} papers surveyed by Siebert~\cite{Siebert:2021causal}
are on miscellaneous topics that do not fit any of the other main categories:
the impact of open-source license choice, newsletter strategies, the effect of gender on pull request acceptance,
and productivity analysis.

Not all papers considered by Siebert apply a full set of causal modeling tools. Most of them
deploy some specific statistical methods that can help detect causal effects, but 
do not use structural\slash graphical models (DAGs).
The few papers that did consider graphical causal models
built them by applying causal discovery algorithms---possibly complementing their output with expert, manual modeling. 

A few additional recent papers applying causal analysis to software engineering data did not make it into Siebert's survey.
Heyn and Knauss~\cite{heyn2022structural} use causal relations and DAGs to clarify, communicate, and apply human knowledge to artificial intelligence system development.
Fang et al.~\cite{fang2022slick} use an econometric causal inference technique (based on the aforementioned potential outcomes framework) to study if tweeting about open-source development projects can affect their popularity.
Finally, Dubslaff et al.~\cite{causal-configs} apply Halpern and Pearl's concept of actual causality~\cite{halpern2015modification} to understand and analyze dependencies in configurable software systems.

\subsection{Empirical Studies of Programming Languages}
\label{sec:rw:pl-studies}

Empirical studies of programming languages fall into three main categories
according to what kind of data they collect and analyze.
\emph{Controlled experiments}
usually compare a small set of language features on a specific task;
for example, static vs.\ dynamic type systems~\cite{DBLP:conf/oopsla/Hanenberg10},
or different concurrency primitives~\cite{rossbach-et-al:2010:transactional,nanz-et-al:2011:design}.
Their precise, controlled experimental setups work well to investigate fine-grained questions;
for the same reason, their results may have limited generalizability.
\emph{Surveys}, interviews, and other ways of
methodically collecting individual observations
are the tool of choice to investigate qualitative questions
and to assess programmer perceptions and subjective preferences;
for example, to understand trends behind language preferences and adoption~\cite{Meyerovich:2013:EAP:2509136.2509515}.
\emph{Repository} data, obtained by mining GitHub or other artifact repositories
(such as the \cj contest data we analyze in this paper),
offer plenty of information, which can support large-scale studies~\cite{SeoSEAB14,FSE,NF-ICSE15};
the flip-side is that the data may be noisy, or too heterogeneous to discern robust, general trends within it.

In fact, empirical studies of programming languages based on mining repositories
can lead to clear, solid results when they target characteristics that are easy to measure reliably
(such as the \emph{conciseness} of programs written in different programming languages);
in contrast, many confounding factors may arise
that complicate discerning the actual, practical impact of programming languages on 
features lacking a universally accepted, easy-to-measure operationalization 
(such as \emph{error proneness}).
As an example, take
Nanz and Furia's analysis of the Rosetta Code repository~\cite{rosettacode-extended},
which compared solutions to programming tasks written in \spellout{8} programming languages.
Their comparison of conciseness found strong effect-size differences (Cohen's $d > 0.7$)
for \spellout{14} out of \spellout{28} language pairs,
corroborating other studies of conciseness~\cite{Prechelt:2000:ECS:619056.621567}.
In contrast, their comparison of failure proneness
(simplistically measured as how many programs terminate with a runtime failure)
found no strong effect-size differences, and only \spellout{5} medium effect-size differences ($0.3 \leq d < 0.5$),
among the same \spellout{28} language pairs.

Fault proneness is an especially tricky measure to associate with the chosen programming language.
Ray et al.'s study of ``code quality'' in GitHub projects~\cite{FSE}
also failed to find \emph{strong} associations between programming languages and presence of bugs,
concluding that, while ``some languages have a greater association with defects than other languages [...]
the effect is small'';
Berger at al.'s critical reanalysis of~\cite{FSE}
suggested to further ``reduce the number of languages with an association with defects''
and found that ``the practical effect size is exceedingly small''~\cite{TOPLAS}.
Our Bayesian reanalysis of the same data~\cite{TOSEM-Bayes-guidelines}
also found that
``the fault proneness of a language over another strongly depends on the conditions in which the languages are to be used'';
namely, any ``disproportionate differences [...] are project-specific rather than language-specific''.
Therefore, it is not surprising that the present paper will also conclude (\autoref{sec:effects-comparison})
that the choice of programming language has a modest effect
on predicting the success of a person participating in the \cj competition,
and it is largely secondary compared to other dominant factors---most important, the intrinsic skills and experience of each participant.

\paragraph{Empirical studies of programming contests.}
Programming contest empirical data have also been the subject of empirical studies---usually with a focus on using such contests to foster educational programs.
Often, such studies are mainly qualitative,
aiming at surveying the state of the art as broadly as possible~\cite{informatics-competitions,sysrev-feedback-generation},
and at suggesting ways to improve the impact of programming contests in education~\cite{competition-in-ed}.

To our knowledge,
Back and Westman's master's thesis~\cite{AE-thesis}\footnote{This thesis was supervised by the first author of the present paper.}
is the only other empirical study of Google \cj data to date.
It analyzes solutions to the contest's problems written in C, C++, C\#, Java, and Python,
and compares them for attributes such as size, running time, and memory consumption
(along the lines of~\cite{NF-ICSE15}).

\section{The Google \cj Data}
\label{sec:code-jam-data}

\cj is a worldwide programming contest organized by Google every year since \n.{first edition}.
Each edition consists of several elimination rounds; all rounds---except possibly the final one---take place online.

Each round begins and ends at predefined times set by the organizers.
As long as a round is open, registered participants can download the coding problems
for that round and submit their solutions.
Solutions can be written in a wide variety of programming languages,
provided they are runnable on the platform Google sets up for the contest.

Participants can submit as many solution attempts as they want for each problem,
in any order.
A submission gets a number of points that depends on the number of tests it passes
in the test suite that accompanies each problem.
Participants are ranked based on the points
of their best submission in each problem,
using the submission time as a tie breaker.
At the end of each round, the top participants in the ranking advance to the next round.

\begin{figure}[!bth]
  \centering
  \resizebox{0.9\textwidth}{!}{
    \begin{tikzpicture}[node distance=14mm and 12mm,
      stage/.style={align=center,draw,very thick,font=\sffamily,minimum width=20mm,minimum height=10mm}]
    \node (entry) {};
    \begin{scope}
      \node[stage,right=10mm of entry] (quali) {Qualification\\Round};
      \node[right=10mm of quali] (out-quali) {};
      \coordinate[right=20mm of out-quali] (r1b);
      \coordinate[above=of r1b] (r1a);
      \coordinate[below=of r1b] (r1c);
      \node[stage,xshift=3mm] (R1B) at (r1b) {Round \\ 1B};
      \node[stage] (R1A) at (r1a) {Round \\ 1A};
      \node[stage,xshift=6mm] (R1C) at (r1c) {Round \\ 1C};
      \node[stage,right=19mm of R1B] (R2) {Round \\ 2};
      \node[stage,right=of R2] (R3) {Round \\ 3};
      \node[stage,right=of R3] (final) {Finals};
    \end{scope}
    \node[right=10mm of final] (exit) {};

    \draw[black,very thick,fill=black] (out-quali) circle (1mm);

    \begin{scope}[-latex,very thick]
      \draw (entry) to node[above] {\n.{rounds2010/qualification round:nicks}} node[below] {\n.{rounds2010/qualification round:langs}} (quali);

      \draw (quali) to node[above] {\n.{rounds2010/round 1:nicks}} node[below] {\n.{rounds2010/round 1:langs}} (out-quali);
      
     \draw (out-quali) |- node[above,sloped,near end] {} (R1A);
      \draw (out-quali) to node[above] {} node[below] {} (R1B);
      \draw (out-quali) |- node[above,sloped,near end] {} node[below,sloped,near end] {} (R1C);
      
      \draw (R1B) to node[above,pos=0.6] {\n.{rounds2010/round 2:nicks}} node[below,pos=0.6] {\n.{rounds2010/round 2:langs}} (R2);

      \draw (R2) to node[above] {\n.{rounds2010/round 3:nicks}} node[below] {\n.{rounds2010/round 3:langs}} (R3);

      \draw (R3) to node[above] {\n.{rounds2010/finals:nicks}} node[below] {\n.{rounds2010/finals:langs}} (final);
    \end{scope}
      
    \draw[very thick] (R1A) -| ($(R1B)!0.4!(R2)$);
    \draw[very thick] (R1C) -| ($(R1B)!0.4!(R2)$);
    
  \end{tikzpicture}
  }
  \Description{Diagram describing the 2010 edition of Google CodeJam. From left to right, it shows the progression of rounds and participants: 8459 participants using 37 languages enter block ``Qualification Round''; 5553 participants using 34 languages enter any of the blocks ``Round 1A'', ``Round 1B'', and ``Round 1C''; 1924 participants using 19 languages enter block ``Round 2''; 361 participants using 10 languages enter block ``Round 3''; 24 participants using 3 languages enter the final block ``Finals''.}
  \caption{The structure of the 2010 edition of Google \cj. Each arrow indicates the number of participants entering each round of the contest (above the arrow) and the number of different programming languages used by these participants (below the arrow). Rounds 1A, 1B, and 1C are accessible to all participants who passed the qualification round.}
  \label{fig:cj2010}
\end{figure}
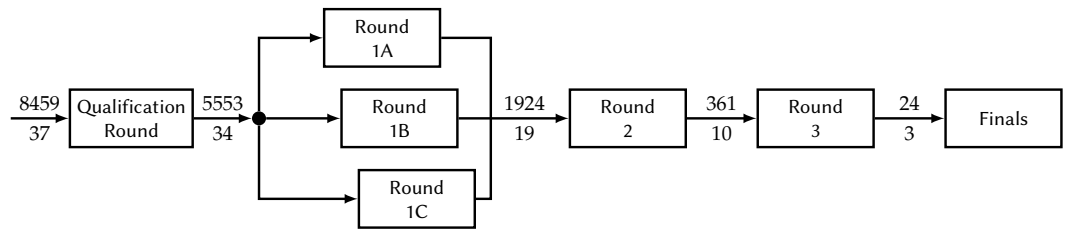

\autoref{fig:cj2010} illustrates the progression of the 2010 edition of \cj,
which included seven rounds.
Initially,
\n.{rounds2010/qualification round:nicks} programmers entered the preliminary qualification round;
\n.{rounds2010/round 1:nicks} were admitted to the next stage,
and allowed to submit to one or more of the three parallel rounds 1A, 1B, and 1C;
\n.{rounds2010/round 2:nicks} participants were ranked high enough to progress
to round~2; \n.{rounds2010/round 3:nicks} further progressed to round~3; and \n.{rounds2010/finals:nicks} took part in the world finals that determined the overall contest winners.

While some details of the contest
(such as the number of rounds, how many advance to the next round, or some details of how points are counted)
may change from year to year, the general contest structure
has remained largely unchanged.
Thus, we can merge the data about submissions to \cj in several yearly editions of the contest
and analyze them consistently.
Furthermore, we do not study the progression of participants within a contest,
but treat each round as a separate ranker of the participants' performance
based exclusively on their submissions to that round.

\subsection{The \cj Dataset}
\label{sec:cj-dataset}

Data about all submissions to \cj are publicly available on the contest's website.\footnote{\url{https://codingcompetitions.withgoogle.com/codejam}}
We analyzed data for the first \n.{num editions}! yearly editions
of the contest, \n.{first edition}--\n.{last edition},
which we got from an unofficial database dump of the contest data.\footnote{\url{https://www.go-hero.net/jam/}}

A \emph{datapoint} in this dataset reports information about a participant's results
in a single \emph{round} of the \cj contest, consisting of values for following variables:

\begin{description}[style=multiline,leftmargin=22mm]
\item[\var{challenge}:] an identifier of the yearly edition of \cj and the round number within that edition
\item[\var{nickname}:] the unique identifier of the programmer who submitted in this challenge
\item[\var{language}:] the programming language used by these submissions
\item[\var{size}:] the submissions' total size (in bytes)
\item[\var{rank}:] the final rank of the participant's submissions within the challenge (1 is the top rank)
\end{description}

As we discussed above,
each round includes one or more problems;
for each problem, a participant's
submission with the most points is taken as their final submission.
Since a participant's rank depends, in general,
on the points that all their (final) submissions collected in the round,
\var{size} is simply the sum of the sizes
of all their final submissions in that round, to give an idea of their overall programming effort.
In the following, we sometimes refer to any such datapoints as ``round\slash challenge participation''.

A clarification about terminology:
a \emph{high} rank denotes a rank
that is close to the top rank (the first rank), 
whereas a \emph{low} rank denotes a rank that is further away from the top rank.
Thus, a high-rank submission is one with a small ordinal number;
and a low-rank submission is one with a large ordinal number.
Consistently with this standard usage,
we will talk about
\emph{better}-than-average ranks
to denote smaller-than-average rank ordinals,
and \emph{worse}-than-average ranks
to denote larger-than-average rank ordinals.

\subsection{Selecting Programming Languages}
\label{sec:language-selection}

The \cj dataset includes submission in \n.{num languages} different programming languages,
including esoterica such as Whitespace\footnote{\url{https://en.wikipedia.org/wiki/Whitespace_(programming_language)}} and GolfScript.\footnote{\url{https://en.wikipedia.org/wiki/Code_golf\#Dedicated_golfing_languages}}
As we might expect, though, the popularity of languages in \cj is not uniform but highly skewed,
with the usual suspects dominating.
More precisely, the top 10 most used languages are:
\begin{enumerate*}
\item \n.{used languages/1} (\n[0]{used languages percent/1}[100]| of all datapoints);
\item \n.{used languages/2}~\mbox{(\n[0]{used languages percent/2}[100]|)};
\item \n.{used languages/3}~\mbox{(\n[0]{used languages percent/3}[100]|)};
\item \n.{used languages/4}~\mbox{(\n[0]{used languages percent/4}[100]|)};
\item \n.{used languages/5}~\mbox{(\n[0]{used languages percent/5}[100]|)};
\item \n.{used languages/6}~\mbox{(\n[0]{used languages percent/6}[100]|)};
\item \n.{used languages/7}~\mbox{(\n[0]{used languages percent/7}[100]|)};
\item \n.{used languages/8}~\mbox{(\n[0]{used languages percent/8}[100]|)};
\item \n.{used languages/9}~\mbox{(\n[0]{used languages percent/9}[100]|)};
\item \n.{used languages/10}~\mbox{(\n[0]{used languages percent/10}[100]|)}.
\end{enumerate*}

We want to focus our analysis on a small number of languages that are widely used in \cj,
so that there is abundant data to reliably estimate correlations and effects for those languages.
Therefore, we only consider submissions in the top-3 most used languages (\n.{used languages/1}, \n.{used languages/2}, and \n.{used languages/3}), which account for an order of magnitude more submissions than all other languages combined.
In particular,
these languages are disproportionately used by the programmers that advance to the
later rounds of the contest; for example, they are the \spellout{3} languages used in the 2010 final round (rightmost arrow in~\autoref{fig:cj2010}).

\subsection{Selecting Experienced Participants}
\label{sec:expert-selection}

\begin{figure}[!bt]
  \centering
\begin{tikzpicture}
    \node[inner sep=0] (pstats) at (0,0)
    {\includegraphics[width=0.8\textwidth]{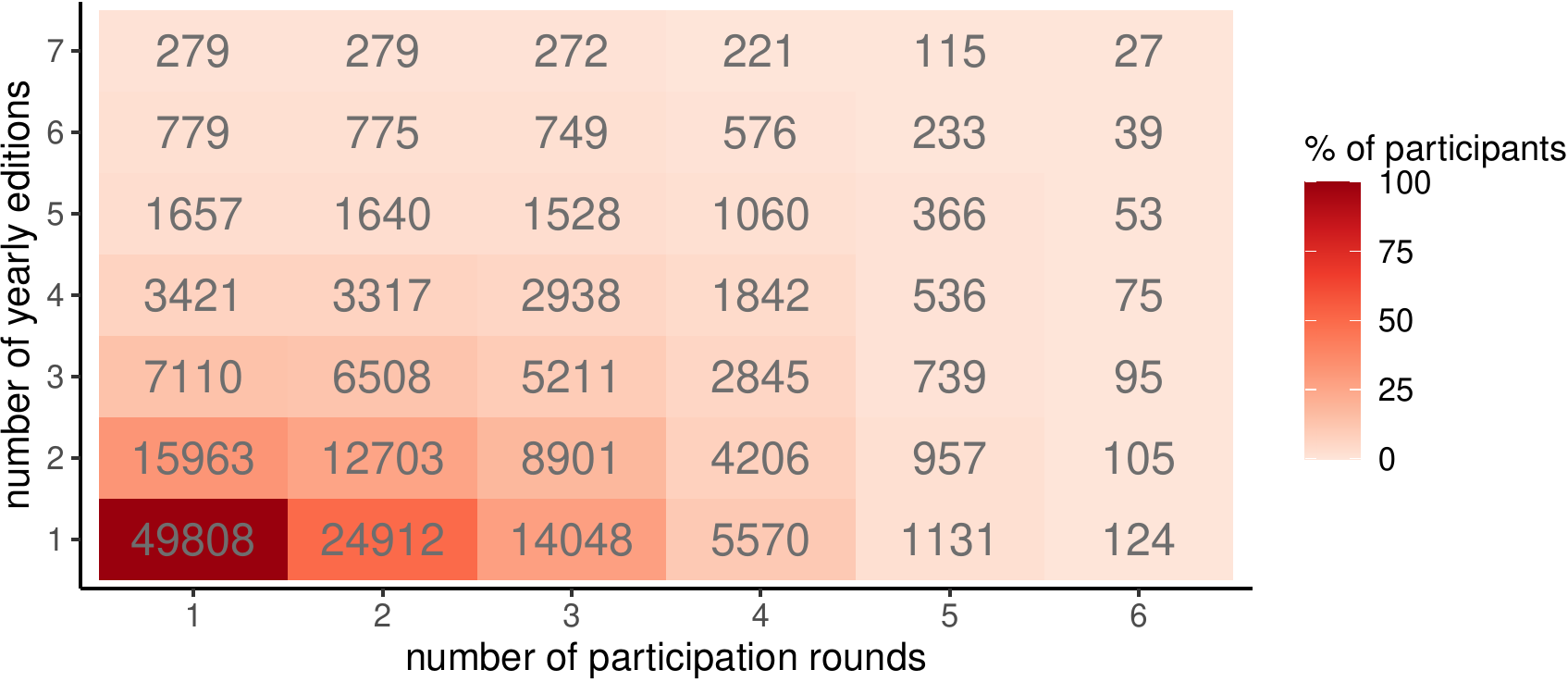}};

    \ifarxiv
      \node[inner sep=0,outer sep=0] (ecenter) at (30.5mm, -10.0mm) {};
    \else
      \node[inner sep=0,outer sep=0] (ecenter) at (25.5mm, -8.2mm) {};
    \fi

    \draw[blue,very thick] (ecenter) ellipse [x radius=5mm, y radius=2.5mm];
  \end{tikzpicture}
  \Description{A grid with 7 rows and 6 columns.
    The cell at row $r$ and column $c$ is the number of participants who took part in at least $r$ yearly editions and at least $c$ participation rounds in any one edition. In particular, 105 participants took part in at least $r = 2$ editions and at least $c = 6$ rounds in either of those editions.}
  \caption{Number of participants to \cj who entered at least $y$ yearly editions of \cj and at least $x$ participation rounds in any one edition. The background color denotes the corresponding percentage of all \n{num all participants} participants to \cj. An ellipse marks the subset of participants featuring in the main analysis.}
  \label{fig:participants-stats}
\end{figure}
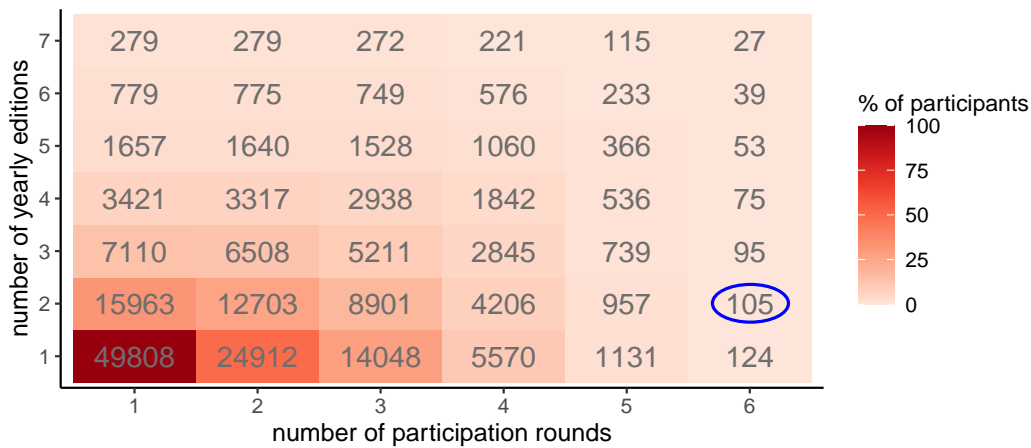

Even after ignoring all but the \spellout{3} most used programming languages,
the \cj dataset still includes \n{num all rounds} datapoints
for the submissions by \n{num all participants} unique participants.
Besides being an impractically large dataset to analyze, it dilutes a smaller number of \emph{experienced} participants within a far greater number of casual participants.
As we can see more clearly in \autoref{fig:participants-stats},
the overwhelming majority of participants took part in only one or two yearly editions of \cj,
where they advanced at most to the second round.

To select experienced participants, we further filter the dataset and only retain data about participants who
took part in at least \n.{cutoffs/num.years}! yearly editions of \cj and submitted,
in any one edition, to at least \n.{cutoffs/num.rounds}! rounds.
Since each edition of \cj in our dataset includes \n.{min num rounds per year}! rounds or more,
the latter requirement is met by who qualifies for the penultimate round at least once---clearly, an above-average result.
This selection criterion leaves us with \n{cutoffs/geq.years.rounds} datapoints
corresponding to the submissions of \n{cutoffs/n.nicknames} experienced participants to \cj.

We found that this data selection reasonably balances 
dataset size and experience of participants:
it's not so small as to be insignificant,
nor so large as to be impractical to analyze and including too many casual participants.
However, there is nothing special about it:
other data selections could also be reasonable and feasible.
To demonstrate this claim,
\autoref{sec:robustness} discusses the results of running the same analysis on significantly larger and smaller samples of the \cj dataset.
While some detailed results may change, the overall answer to our research question
mostly does not, and remains largely independent of the data selection criteria
within a broad range.

\section{Correlational Analysis of Programming Language Effects}
\label{sec:correlation}

This section presents a statistical analysis of the \cj dataset.
We build four Bayesian regression models $m_1, m_2, m_3, m_4$ 
of increasing complexity, and we compare them quantitatively according to their predictive capabilities.
More precisely, we use generalized linear (regression) models:
a broad, widely-used family of flexible statistical models~\cite{gelman2020regression}.
The analysis follows the guidelines we discussed in previous work~\cite{TOSEM-Bayes-guidelines};
for brevity, we do not discuss here the application of the guidelines, but refer the interested readers
to the replication package for details.

By following the recommended guidelines to build and evaluate regression models,
this section's analysis is arguably similar to any standard, up-to-date, rigorous 
statistical analysis of the same data made by an informed researcher.
Even with some latitude to accommodate their favorite statistical techniques,
the big picture that emerges should still be consistent with this section's.
In particular, \autoref{sec:frequentist} explains that using frequentist rather than Bayesian
techniques would lead to similar overall conclusions.

\begin{figure}[!tb]
  \centering
  \begin{subfigure}[t]{1.0\linewidth}
    \begin{equation*}
      \begin{aligned}
        \var{rank}_i &\sim\ \dist{NegativeBinomial}(\lambda_i, \phi) 
        \\
        \log(\lambda_i) &=\
        \bracketcol*{
          \bracketcol{
            \bracketcol*{
              \bracketcol{
                \alpha_{\var{language}[i]}}{m_1}[m1col][m1col]
              + \alpha_{\var{nickname}[i]}}{m_2}[m2col][m2col]
            + \alpha_{\var{challenge}[i]}}{m_3}[m3col][m3col]
          + \beta \cdot \log(\var{size}_i)}{m_4}[m4col][m4col]
      \end{aligned}
    \end{equation*}
    \Description{Equations defining the likelihoods of models $m_1$, $m_2$, $m_3$, and $m_4$, all using a negative binomial distribution and a logarithmic link function.}
    \caption{Likelihoods of models $m_1$, $m_2$, $m_3$, and $m_4$.}
    \label{eq:likelihoods-m14}
  \end{subfigure}\\[4mm]
  \begin{subfigure}[t]{1.0\linewidth}
    \begin{equation*}
      \begin{aligned}
  \alpha_\ell &\sim\ \dist{Normal}(5, 3.5)  & \text{for every }\var{language}\ \ell
  \\
  \alpha_n &\sim\ \dist{Normal}(0, 0.5)  & \text{for every }\var{nickname}\ n
  \\
  \alpha_c &\sim\ \dist{Normal}(0, 0.5)  & \text{for every }\var{challenge}\ c
  \\
  \beta &\sim\ \dist{Normal}(0, 1)
  \\
  \phi &\sim\ \dist{Gamma}(0.01, 0.01)
      \end{aligned}
    \end{equation*}
    \Description{Equations defining the priors of the predictors' coefficients in models $m_1$, $m_2$, $m_3$, and $m_4$.}
    \caption{Priors for the coefficients $\alpha$, $\beta$, and $\phi$ of models $m_1$, $m_2$, $m_3$, and $m_4$.}
    \label{eq:priors-m14}
  \end{subfigure}
  \Description{This figure consists of two subfigures with the equations of statistical models $m_1$, $m_2$, $m_3$, and $m_4$.}
  \caption{Definitions of regression models $m_1$, $m_2$, $m_3$, and $m_4$. All these models have the same negative binomial likelihood with log-linear parameter $\lambda$; each model $m_k$ includes exactly $k$ predictors.}
  \label{fig:models-m14}
\end{figure}

\subsection{Modeling}
\label{sec:stat-modeling}

\autoref{fig:models-m14} shows the four models,
whose components we now describe and justify.
These four models follow standard guidelines,
and are serviceable to support the analysis of the paper;
but we do not imply that they are the definitive or only models
for the data.
The replication package includes a few model variants,
as well as the data that can be fitted on any other kind of statistical model
to explore different research questions.

\paragraph{Variables.}
First of all, we should choose the \emph{variables} of our models.
Since our goal is to analyze the impact of programming languages and other factors
on the results in rounds of the \cj contest,
we pick \var{rank} as \emph{outcome} variable in all our models.

The other variables in the dataset can be used as \emph{predictors} (also: \emph{treatments}).
Since we are especially interested in the effect of programming languages,
we include variable \var{language} as a predictor in all our models.
Then, we introduce progressively more complex models by adding the other variables as predictors:
model $m_1$ only uses \var{language};
model $m_2$ adds \var{nickname},
since different participants are likely to have different results that reflect their skills;
model $m_3$ adds \var{challenge},
since each year\slash round of the competition is different, and arguably has a different intrinsic complexity;
and model $m_4$ adds \var{size},
which partially indicates the amount of work done by a participant in each round.

Plausibly, some of these four ``independent'' variables interact
(for example, participants may have their own favorite programming language),
and hence they are not really independent.
To account for this, \autoref{sec:model-comparison}
compares the four models using a kind of regularization
that weighs off predictive accuracy against complexity:
any model with redundant variables will be penalized
under this criterion.
In a frequentist setting (\autoref{sec:frequentist}),
a variable selection process provides a similar safeguard against
including strongly interacting variables.
Later, \autoref{sec:causation} will revisit variable interactions
and interpret them under a causal lens.

\paragraph{Likelihood.}
The \emph{likelihood} is a probability distribution
of the outcome given some parameters;
in a regression model,
the likelihood's parameters
are (generalized) linear functions of the predictors.

Since the outcome variable \var{rank} is a positive integer,
the Poisson distribution family is the most appropriate choice
since it has the highest information entropy for this kind of data~\cite{jaynes03prob};
in other words, it does not encode any assumption other than that \var{rank} must be integer and not negative~\cite{mcelreath2020statistical}.
Specifically, we pick the \emph{negative binomial} distribution within the Poisson family,
which is better suited for data that is overdispersed.
This is the case of variable \var{rank}
whose mean \n[0]{rank mean} is much smaller than its variance \n[0]{rank variance}.
Mean and variance coincide in a Poisson distribution,
which would fail to accurately capture the distribution of \var{rank};
in contrast, a negative binomial distribution $\dist{NegativeBinomial}(\lambda, \phi)$
has two parameters $\lambda$ and $\phi$,
so that its mean $\lambda$ can differ from its variance $\lambda + \lambda^2/\phi$.

\paragraph{Parameters.}
The \emph{mean} $\lambda$ of the negative binomial distribution in our models
is a linear function of the predictors after applying a customary logarithmic \emph{link function},
which ensures that the mean remains nonnegative.

\begin{description}
\item[Model $m_1$.] In model $m_1$,
  $\log(\lambda_i) = \alpha_{\var{language}[i]}$: in each datapoint
  $i$, the logarithm of parameter $\lambda$ equals a
  language-dependent parameter $\alpha_{\var{language}[i]}$.  In other
  words, model $m_1$ estimates a different mean for each programming
  language, and uses that to predict a challenge's rank.

\item[Model $m_2$.] In model $m_2$,
  $\log(\lambda_i) = \alpha_{\var{language}[i]} +
  \alpha_{\var{nickname}[i]}$, where $\alpha_{\var{nickname}[i]}$ is
  another parameter, which depends on \var{nickname}.  In other words,
  model $m_2$ estimates the mean rank by combining mean estimates for
  each programming language and for each participant (\var{nickname}).

\item[Model $m_3$.] In model $m_3$,
  $\log(\lambda_i) = \alpha_{\var{language}[i]} +
  \alpha_{\var{nickname}[i]} + \alpha_{\var{challenge}[i]}$,
  where $\alpha_{\var{challenge}[i]}$ is
  another parameter, which depends on \var{challenge}.  In other
  words, model $m_3$ estimates the mean rank by also combining mean
  estimates for each challenge (a round in a specific yearly edition of the contest).

\item[Model $m_4$.] In model $m_4$,
  $\log(\lambda_i) = \alpha_{\var{language}[i]} +
  \alpha_{\var{nickname}[i]} + \alpha_{\var{challenge}[i]} + \beta \cdot \log(\var{size}_i)$,
  where $\beta$ is a new linear parameter, which multiplies the logarithm of \var{size}.
  In other words, model $m_4$ estimates the mean rank by
  also estimating a contribution proportional to a submission's size.  We
  use the \emph{logarithm} of \var{size} because this variable varies
  greatly in the dataset (from \n{size min} to \n{size max} bytes);
  taking the logarithm is a standard practice to smoothen a wide
  variability range and focus on its ``orders of magnitude'' changes.
\end{description}

Parameters $\alpha_{\ell}$ for every language $\ell$,
$\alpha_n$ for every nickname $n$,
and $\alpha_c$ for every challenge $c$
are also called ``intercepts'', since they correspond to constant terms in the linear model.
For a similar reason, parameter $\beta$ is also called ``slope''.

\paragraph{Priors.}
The last components of a Bayesian model are \emph{priors}, that is
``initial'' probability distributions on the parameters 
$\alpha_{\ell}$ for every language $\ell$,
$\alpha_n$ for every nickname $n$,
$\beta$,
$\alpha_c$ for every challenge $c$,
and $\phi$.
The rest of this section explains and justifies our choice of priors;
however, there is a good deal of latitude in how we precisely
select them---as long as they pass the validation that we outline next.

We use standard weakly informative priors for all parameters:
normal distributions for the $\alpha$s and $\beta$,
and a gamma distribution for $\phi$.
As you can see in \autoref{eq:priors-m14},
the priors of the language intercepts $\alpha_\ell$
are not centered (their mean is 5);
in contrast, 
the priors of $\alpha_{n}$, $\alpha_c$, and $\beta$ 
are centered (their mean is 0)
and have smaller standard deviations than the prior of the language intercepts $\alpha_{\ell}$.
This reflects the analysis's focus on the effect of languages:
every model includes the language intercepts,
which are thus the primary determinant of the overall expected rank and are comparable between models;
then, each other predictor may move the overall estimate higher or lower, starting from the language's baseline.
Since the rank is a positive integer, the mean rank must also be positive;
hence, a prior with a positive mean is more likely to be consistent with the data.
However, the language intercepts' broad prior standard deviation makes the prior weak; 
this means that the data can freely sway the estimate of the actual impact of languages in either direction. 
As shown by the validation step, these priors make for an efficient fitting of the data;
but there's nothing ``special'' about them, since they are broad,
and in fact different choices for the priors
would lead to very similar overall estimates.

\paragraph{Model validation.}

Before we fit the models and use them for analyzing the data and answering our RQ,
we should \emph{validate} them to ensure that they are well-formed and their predictions can be trusted~\cite{TOSEM-Bayes-guidelines}.
Here, we summarize the outcome of validation; all details are in the replication package.

First, we check that the priors are \emph{plausible},
that is they are not unnecessarily constraining compared to the observed data.
The prior predictive simulation plots confirm that this is the case 
for all four models.

Second, we check that the models are \emph{workable},
that is they can be fitted without computational problems such as divergences.
To this end, we inspect the fitting diagnostic metrics
(divergent transitions,
$\widehat{R}$ ratio of within-to-between chain variance,
effective sample size,
and trace plots)
and confirm that they are within the expected ranges
for all four models.

Third, we check that the models are \emph{adequate},
that is they can generate data that resembles the analyzed empirical data.
This is an important property to ensure that the model does not merely pass
some diagnostic checks but can actually be trusted to capture (at least some) aspects
of the data.
Inspecting the posterior predictive checks plots for
all four models 
confirms that they are all reasonably adequate.\footnote{As expected, the adequacy of the simplistic model $m_1$ is visibly lower than the other, more complex models; however, they all are capable of reflecting at least the general shape and trend of the empirical data.}

\subsection{Model Comparison}
\label{sec:model-comparison}

Given that all four models 
pass the fundamental well-formedness checks,
which model should we use for our analysis?
Our current analysis goal is making good predictions---that is, estimating correlations between predictors and outcome.
Thus, we can use an information-theoretic criterion to assess
whether some model offers better out-of-sample predictions than some other models.

In particular, we use the widely used PSIS-LOO criterion~\cite{vehtariGG17loo},
which estimates the \emph{relative} adequacy of a number of competing models
using a form of leave-one-out validation.
Applying PSIS-LOO ranks the four models 
according to a score that estimates the relative predictive capabilities of
one model compared to the others;
each score difference has a standard error, which quantifies the uncertainty in the score difference.

\begin{table}[!bth]
\centering
\begin{tabular}{crrr}
  \toprule
    \multicolumn{1}{c}{\textsc{model}} & \multicolumn{1}{c}{\textsc{ranking}} & \multicolumn{2}{c}{\textsc{score difference}}
    \\
                                       & & \multicolumn{1}{c}{\textsc{absolute}} & \multicolumn{1}{c}{\textsc{standardized}}
    \\
    \midrule
    $m_4$ & 1 & -- & -- 
    \\
    $m_3$ & 2 & \n[1]{loo m1m2m3m4/m3:reldiffs} & \n[1]{loo m1m2m3m4/m3:relstdiffs}
    \\
    $m_2$ & 3 & \n[1]{loo m1m2m3m4/m2:reldiffs} & \n[1]{loo m1m2m3m4/m2:relstdiffs}
    \\
    $m_1$ & 4 & \n[1]{loo m1m2m3m4/m1:reldiffs} & \n[1]{loo m1m2m3m4/m1:relstdiffs}
  \\
  \bottomrule
  \end{tabular}
  \caption{Comparison of four \textsc{model}s according to the PSIS-LOO validation criterion.
    The models are ranked from best to worst according to the criterion (the ``best'' model has rank 1).
    The table also reports, 
    the relative \textsc{difference} in predictive capability score
    between each model and the one that precedes it in the ranking,
    both in \textsc{absolute} value and \textsc{standardized} (i.e., absolute value difference divided by standard error).}
  \label{tab:comparison-m14}
\end{table}

\autoref{tab:comparison-m14} shows the results of applying PSIS-LOO to the models $m_1, m_2, m_3, m_4$.
Model $m_4$ clearly outperforms all others in terms of predictive capabilities:
the score of $m_3$---the next best model in the ranking---is a massive \n[1]{loo m1m2m3m4/m3:relstdiffs}[-1.0]
standard errors lower.
In turn, model $m_3$ outperforms model $m_2$: their difference is still massive (\n[1]{loo m1m2m3m4/m2:relstdiffs}[-1.0] standard errors).
Model $m_1$ is the worst, although it's closer to $m_2$ (\n[1]{loo m1m2m3m4/m1:relstdiffs}[-1.0] standard errors).

The conclusion of this comparison is indisputable:
if our goal is \emph{predicting} with accuracy
the statistical associations between the predictors (including \var{language})
and the outcome \var{rank},
we should definitely use model $m_4$, which offers far better out-of-sample predictive accuracy
than the other, simpler models that incorporate less information.\footnote{The replication package shows that $m_4$ even outperforms considerably more sophisticated multilevel models that explicitly model variable interactions but do not include variable \var{size}.}

\begin{figure}[!tb]
  \centering
  \includegraphics[width=\textwidth]{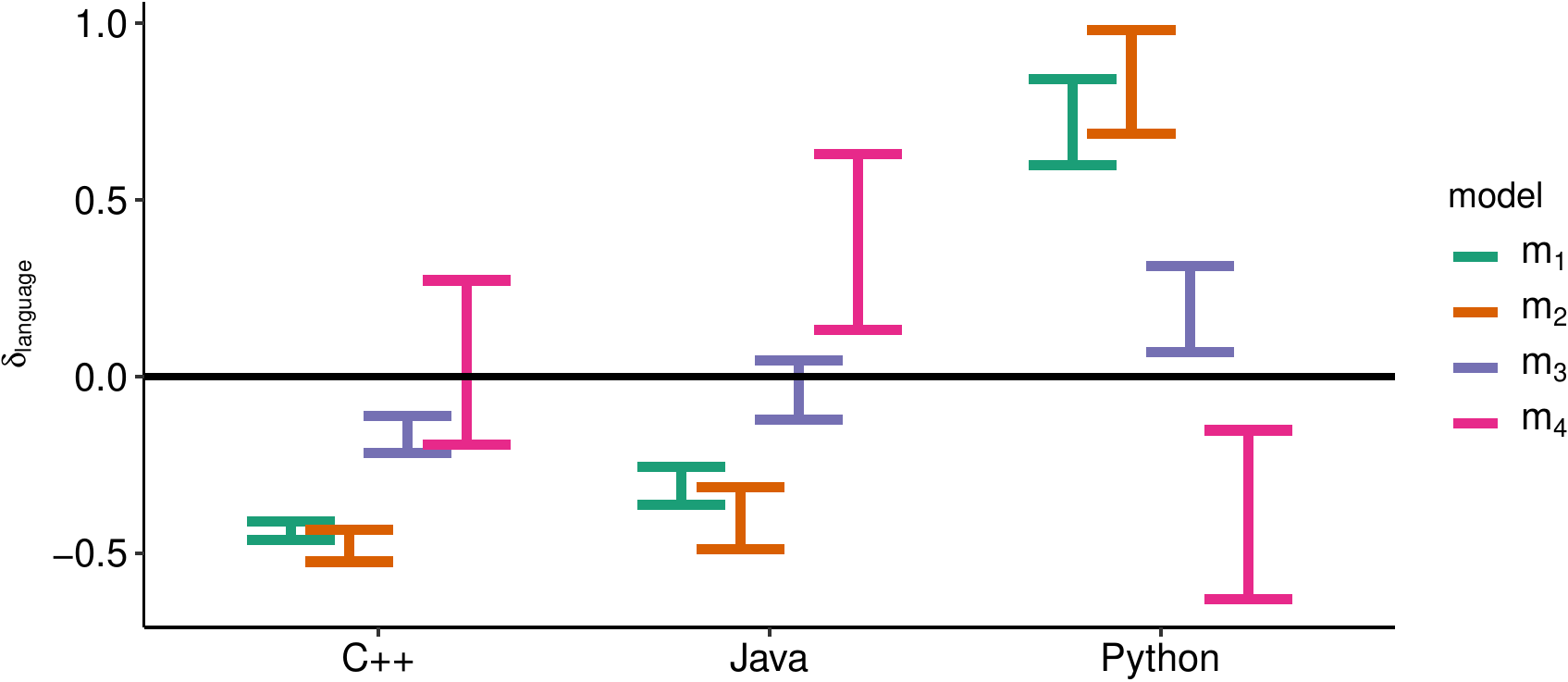}
  \Description{Twelve intervals, corresponding to the 50\% probability intervals of each language coefficient in the regression models $m_1, m_2, m_3, m_4$. In models $m_1$ and $m_2$, the intervals for C++ and Java are completely below zero (upper bounds below $-0.25$), and the intervals for Python are completely above zero (lower bounds above $+0.6$). In model $m_3$, the interval for C++ is completely below zero (upper bound around $-0.1$), the interval for Java barely overlaps zero (upper bound around $0.04$), and the interval for Python is just completely above zero (lower bound around $+0.07$). In model $m_4$, the interval for C++ is nearly centered around zero, the interval for Java is completely above zero (lower bound around $+0.13$), and the interval for Python is completely below zero (upper bound around $-0.15$).}
  \caption{For each language $\ell$, in each model $m_1, m_2, m_3, m_4$,
    the 50\% highest posterior density intervals of the distribution
    of the difference $\effect[\ell]$ between $\ell$'s intercept $\alpha_\ell$ and the average intercept $\overline{\alpha_L}$
    for all languages in the model. See \autoref{sec:analysis-correlation} for details on how these intervals are computed.}
  \label{fig:alphas-models-14}
\end{figure}

\subsection{Analysis: Programming Language Associations}
\label{sec:analysis-correlation}

Fitting a model on the \cj data
gives a \emph{posterior distribution} of the parameters
$\alpha_{\ell}$ for each language $\ell = \textrm{C++}, \text{Java}, \text{Python}$---as well as distributions for the other parameters 
$\alpha_n$, $\beta$, and $\alpha_c$,
but our RQ focuses on the analysis of languages.

To answer our RQ, we compute \emph{contrasts}
that quantify the difference between the absolute value of the $\alpha_{\ell}$ for different languages.
For each $\ell$, we derive the distribution\footnote{More precisely, we numerically estimate it on the posterior samples.}
of $\effect[\ell] = \alpha_\ell - \overline{\alpha_L}$,
where $\overline{\alpha_L}$ is the mean of the distribution of all coefficients $\alpha_{\text{C++}}, \alpha_{\text{Java}}, \alpha_{\text{Python}}$ taken together.
In other words, $\effect[\ell]$ is the difference between language $\ell$'s contribution to the rank
and the average language contribution to the rank in the model.

\autoref{fig:alphas-models-14} plots the 50\% highest-posterior density intervals of the distributions of $\effect[\ell]$
for each language $\ell$ and in each model.
If an interval $I_\ell$ for language $\ell$ is entirely below the zero line,
it means that, with 50\% probability, submissions written in language $\ell$ are associated with better-than-average ranks
(that is, smaller rank ordinals in \cj).
If $I_\ell$ is entirely above the zero line,
it means that, with 50\% probability, submissions written in language $\ell$ are associated with worse-than-average ranks
(that is, larger rank ordinals in \cj).
If $I_\ell$ includes the zero line, there is no consistent association with better- or worse-than-average
results for language $\ell$ at 50\% probability.

Why do we focus on 50\% probability intervals, instead of a higher, more common probability such as 95\%?
As we demonstrate in \autoref{sec:effects-comparison},
the absolute magnitude of the impact of programming languages is modest compared to
other variables in our data.
Therefore, there is quite some uncertainty about the precise effects associated with languages, and
we are unlikely to find consistent effects at high probability levels.
The 50\% probability intervals are still meaningful and give a ``sense of where the parameters and predicted values
will be'' instead of aiming for ``an unrealistic near-certainty''~\cite{gelman-50p-intervals}.
In the replication package, we also compute the 70\%, 90\%, and 95\% probability intervals,
showing that they still largely exhibit the same \emph{tendencies} as the 50\% intervals,
but with larger uncertainty
(for example, an interval that is completely above zero at 50\% probability
is \emph{mostly} above zero at 95\% probability---but also includes it).
In other words, higher-probability intervals
display the \emph{same} tendency as the 50\% intervals---just with more uncertainty about the precise interval spans.
In summary, 50\% probability intervals provide robust, valuable information,
that clearly conveys the main qualitative findings of our analysis.

\paragraph{Results: model $m_4$.}
According to model $m_4$
(which model comparison identified as the one delivering the most accurate predictions),
at the 50\% probability level:
\begin{itemize}
\item Python is associated with smaller-than-average rank ordinals, that is \emph{better} contest results;
\item Java is associated with larger-than-average rank ordinals, that is \emph{worse} contest results;
\item there is no consistent association for C++, as its 50\% interval is nearly centered around zero.
\end{itemize}
Based on this, we can answer the paper's main RQ under the correlational interpretation:

\begin{center}
\begin{boxedminipage}{0.8\columnwidth}
  \centering
  \textbf{AQ (correlation)}: For experienced participants to the \cj contest,\\
  using Python is associated with better contest results,\\
  using Java is associated with worse contest results,\\
  and using C++ has no consistent association.
\end{boxedminipage}
\end{center}

\begin{table}[!bt]
  \centering
  \begin{tabular}{clll}
  \toprule
  \multicolumn{1}{c}{\textsc{model}} &
                                                         \multicolumn{1}{c}{\textsc{better}} & \multicolumn{1}{c}{\textsc{worse}} & \multicolumn{1}{c}{\textsc{no association}}
  \\
  \midrule
  $m_1$ & C++, Java & Python
  \\
  $m_2$ & C++, Java & Python
  \\
  $m_3$ & C++ & Python & Java
  \\
  $m_4$ & Python & Java & C++
  \\
  \bottomrule
\end{tabular}
\caption{For each \textsc{model}, which languages are associated with \textsc{better} and \textsc{worse} contest results, or have \textsc{no} consistent \textsc{association}, at the 50\% probability level.}
\label{tab:betterworse-models-14}
\end{table}

\paragraph{Results: models $m_1$, $m_2$, $m_3$.}
Further inspecting \autoref{fig:alphas-models-14}
clearly indicates that the predictions of the four models $m_1$, $m_2$, $m_3$, and $m_4$
are often inconsistent with each other.
\autoref{tab:betterworse-models-14} summarizes the predictions of each model:
in particular, models $m_1$, $m_2$, and $m_3$'s estimates
are nearly the opposite of model $m_4$'s.
Since we have shown that $m_4$ outperforms the other three models in terms of predictive accuracy,
it makes sense to base the correlational analysis on $m_4$.

\paragraph{Robustness.}
When the qualitative results of a statistical analysis ``flip''
as we refine models
(such as when going from $m_3$ to $m_4$),
it may indicate that the associations we are measuring
are weak.
\autoref{sec:effects-comparison} will confirm this suspicion,
detailing how the specific contribution of \var{language}
is small compared to the other variables'.
From a different angle, this lack of robustness in the statistical analysis
also prompts us to think about possible \emph{causal} relations,
in order to go beyond mere predictive accuracy towards understanding
the genuine causes of the observed associations.

\subsection{Frequentist Statistics}
\label{sec:frequentist}

In related work~\cite{FFT-TSE19-Bayes2,TOSEM-Bayes-guidelines},
we argued extensively about the advantages of Bayesian statistics over the more traditional frequentist statistics.
This recommendation applies regardless of whether we are just interested in prediction or are looking for causal links.

Nevertheless, most of the observations and analyses we discuss in the present paper
would remain valid when using frequentist statistics.
To substantiate this claim, the replication package presents 
four statistical models $f_1, f_2, f_3, f_4$
that are the frequentist counterparts to the four Bayesian models $m_1, m_2, m_3, m_4$
discussed in the paper.
Except for a few technical differences---such as how we encode contrasts and binary indicator variables, and using flat rather than weakly informative priors---each frequentist model $f_k$ is a regression model using the same predictors and outcome variables,
the same negative binomial likelihood, and the same log-linear parameterization as the corresponding
Bayesian model $m_k$.
Fitting $f_k$ on the \cj dataset and analyzing it similarly to how we did for $m_k$
also leads to very similar results in terms of which programming languages
are associated with worse or better performance in the contest.
The frequentist analysis also finds 
that the predictions based on model $f_4$ (like $m_4$) are nearly opposite of those based on model $f_3$ (like $m_3$).

The frequentist approach also agrees with our Bayesian approach
regarding which model is ``best'' for predictions.
To this end, we apply a variable selection process that starts from
model $f_1$ (with a single predictor, like model $m_1$),
adds\footnote{A backward process has the same outcome: it suggests $f_4$ as the ``best'' model.}
one predictor at a time,
and checks whether it ``significantly'' improves model performance
(measured by $p$-values or by information criteria such as AIC).
This process continues until it selects model $f_4$, which is strictly better than all other simpler models
according to this criterion.

In summary, 
we can go back to our four Bayesian models 
for the rest of the analysis,
knowing that
the high-level contributions and results of the paper---most important, the causal analysis results---are largely independent of
whether one prefers to use Bayesian or frequentist
regression models.

\section{Causal Analysis of Programming Language Effects}
\label{sec:causation}

This section discusses how to build an analysis of the \cj dataset
that tries to identify true \emph{causal} effects
of the different programming languages;
and compares its results to the purely associational analysis presented in the previous section.

\subsection{From Correlation To Causation}
\label{sec:corr-doesnt-imply}

In general, one cannot identify causal relations by analyzing data alone,
as these relations depend also on how the data was \emph{generated}.\footnote{``In causal analysis we must incorporate some understanding of the process that produces the data, and then we get something that was not in the data to begin with.''~\cite[p.~85]{book-of-why}}
This implies that we cannot just lift the results of the statistical analysis
to infer causal effects without any assumptions on the data-generation process.

To illustrate this point, imagine that \cj were run as a fully controlled experiment
designed to find the impact of using different programming languages.
This means that all independent variables would be assigned randomly
within the population of interest:
random programmers (variable \var{nickname}) would be selected to join the competition;
they would be assigned a random problem (variable \var{challenge}),
which they would have to solve using a randomly chosen language (variable \var{language})
and producing a solution of a given size (variable \var{size}).
In this scenario, the randomization would cancel out
any indirect correlation between independent variables
(in other words, the groups of programmers using different programming languages would be homogeneous in all characteristics except for the used language);
thus, the statistical relation between \var{language} and \var{rank}
(measured, for instance, by a regression model)
would measure exclusively the direct causal effect of \var{language} on \var{rank}.

Obviously, this is not how \cj works.
Participants are free to join or leave the contest, can use the programming language they prefer,
and do not have any upfront constraints on the size of the program they submit.
Therefore, the statistical relation between \var{language} and \var{rank}
may actually measure an unknown mix of
direct causal effect of \var{language} on \var{rank}
and other confounding effects between independent variables.
For example, imagine that all best programmers like to use C++;
then, a statistical analysis would report a clear association between
submissions in C++ and worse ranks.
But this would mostly be an indirect effect of the best programmers' preference
for one language over the others;
if we asked them to only use Java, they would likely still get top ranks
just because of their superior programming skills.

Given that the \cj data is observational, and hence such confounding effects
are likely to take place, what can we do to isolate any causal effects?
While we cannot infer them from the data alone, we can:
\begin{enumerate*}[label=\emph{\roman*})]
\item design \emph{qualitative causal models} that capture
  plausible causal relations among variables;
\item partially \emph{validate} such models on the data,
  determining which models are more or less likely to be consistent with the data;
\item use a qualitative causal model that is consistent with the data
  as a guide to build a \emph{statistical model} that properly factors out any confounding effects.
\end{enumerate*}
If built following these guidelines,
the statistical model's association between \var{language} and \var{rank}
should measure the true causal effect between the two variables---at least, within the uncertainty given by the data, and assuming the
qualitative causal model is accurate.
The rest of this section goes through these steps for the \cj dataset.

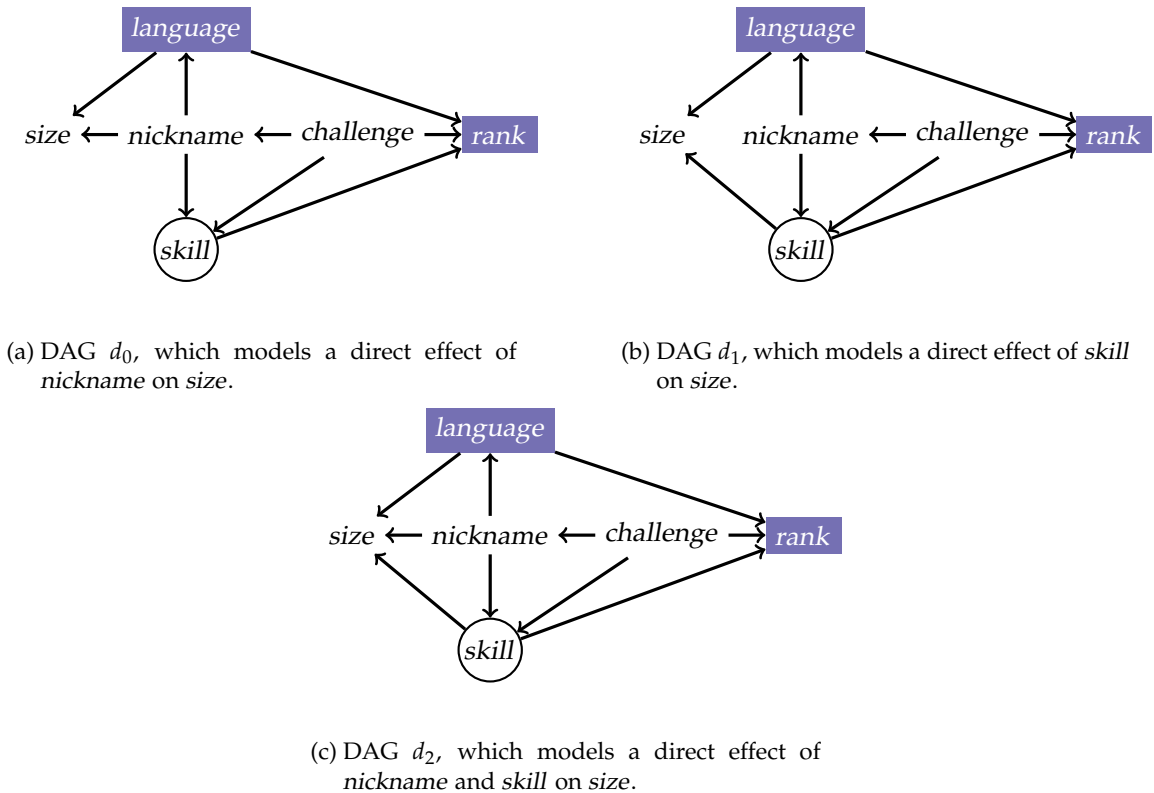
\begin{figure}[!tb]
  \centering
  \begin{subfigure}{0.4\textwidth}
    \begin{tikzpicture}
          \matrix[row sep=8mm, column sep=5mm] {
      & \node[fill=m3col,text=white] (language) {\var{language}}; 
      \\
      \node (size) {\var{size}};
      & \node (nickname) {\var{nickname}};
      & \node (challenge) {\var{challenge}};
      & \node[fill=m3col,text=white] (rank) {\var{rank}};
      \\
      & \node[unobserved] (skill) {\var{skill}};
      \\
    };
    \begin{scope}[very thick,->]
      \draw (language) to (rank);
      \draw (language) to (size);
      \draw (nickname) to (skill);
      \draw (nickname) to (language);
      \draw (skill) to (rank);
      \draw (challenge) to (skill);
      \draw (challenge) to (rank);
      \draw (challenge) to (nickname);
    \end{scope}

       \begin{scope}[very thick,->]
        \draw (nickname) to (size);
      \end{scope}
    \end{tikzpicture}
    \Description{The DAG includes arrows from \var{language} to \var{rank} and to \var{size}; from \var{nickname} to \var{language}, to \var{size}, and to \var{skill}; from \var{challenge} to \var{rank}, to \var{nickname}, and to \var{skill}; from \var{skill} to \var{rank}.}
    \caption{DAG $d_0$, which models a direct effect
      of \var{nickname} on \var{size}.}
    \label{fig:dag-d0}
  \end{subfigure}
  \hspace{12mm}
  \begin{subfigure}{0.4\textwidth}
    \begin{tikzpicture}
          \matrix[row sep=8mm, column sep=5mm] {
      & \node[fill=m3col,text=white] (language) {\var{language}}; 
      \\
      \node (size) {\var{size}};
      & \node (nickname) {\var{nickname}};
      & \node (challenge) {\var{challenge}};
      & \node[fill=m3col,text=white] (rank) {\var{rank}};
      \\
      & \node[unobserved] (skill) {\var{skill}};
      \\
    };
    \begin{scope}[very thick,->]
      \draw (language) to (rank);
      \draw (language) to (size);
      \draw (nickname) to (skill);
      \draw (nickname) to (language);
      \draw (skill) to (rank);
      \draw (challenge) to (skill);
      \draw (challenge) to (rank);
      \draw (challenge) to (nickname);
    \end{scope}

       \begin{scope}[very thick,->]
        \draw (skill) to (size);
      \end{scope}
    \end{tikzpicture}
    \Description{The DAG includes arrows from \var{language} to \var{rank} and to \var{size}; from \var{nickname} to \var{language} and to \var{skill}; from \var{challenge} to \var{rank}, to \var{nickname}, and to \var{skill}; from \var{skill} to \var{rank} and to \var{size}.}
    \caption{DAG $d_1$, which models a direct effect
      of \var{skill} on \var{size}.}
    \label{fig:dag-d1}
  \end{subfigure}
  \\
  \begin{subfigure}{0.4\textwidth}
    \begin{tikzpicture}
          \matrix[row sep=8mm, column sep=5mm] {
      & \node[fill=m3col,text=white] (language) {\var{language}}; 
      \\
      \node (size) {\var{size}};
      & \node (nickname) {\var{nickname}};
      & \node (challenge) {\var{challenge}};
      & \node[fill=m3col,text=white] (rank) {\var{rank}};
      \\
      & \node[unobserved] (skill) {\var{skill}};
      \\
    };
    \begin{scope}[very thick,->]
      \draw (language) to (rank);
      \draw (language) to (size);
      \draw (nickname) to (skill);
      \draw (nickname) to (language);
      \draw (skill) to (rank);
      \draw (challenge) to (skill);
      \draw (challenge) to (rank);
      \draw (challenge) to (nickname);
    \end{scope}

       \begin{scope}[very thick,->]
        \draw (nickname) to (size);
        \draw (skill) to (size);
      \end{scope}
    \end{tikzpicture}
    \Description{The DAG includes arrows from \var{language} to \var{rank} and to \var{size}; from \var{nickname} to \var{language}, to \var{skill}, and to \var{size}; from \var{challenge} to \var{rank}, to \var{nickname}, and to \var{skill}; from \var{skill} to \var{rank} and to \var{size}.}
    \caption{DAG $d_2$, which models a direct effect
      of \var{nickname} and \var{skill} on \var{size}.}
    \label{fig:dag-d2}
  \end{subfigure}
  \Description{Three graphs with seven nodes \var{rank}, \var{language}, \var{size}, \var{nickname}, \var{challenge}, and \var{skill}.}
  \caption{Three DAGs encoding possible causal relations
    among variables in the \cj dataset.
    Variables \var{rank} and \var{language} are colored because their relation is
    the primary target of the causal analysis.
    Variable \var{skill} is circled to denote that it is unobserved, that is not actually measured.
    All three DAGs have the same nodes (variables) but differ in the
    arrows (causal relations) that enter \var{size}.}
  \label{fig:dag-cj}
\end{figure}

\subsection{Qualitative Causal Models: DAGs}
\label{sec:dags}

Causal DAGs (Directed Acyclic Graphs)
are intuitive graphical models
to express qualitative causal relations between variables.
Let's demonstrate how they work by designing qualitative causal models for the \cj data.

A DAG
consists of nodes (also called vertices) and arrows (also called arcs, edges, or links)
connecting them.
In a causal DAG, \emph{nodes} correspond to \emph{variables}.
For the \cj data, these are 
\var{language}, \var{rank}, \var{size}, and \var{nickname}.
A DAG may also include \emph{unobserved} variables,
which are in a cause-and-effect relations with other variables, 
but are not directly available in the dataset.
For the \cj data, we introduce unobserved variable\footnote{Unobserved variables are also called ``latent'' variables in frequentist terminology.} \var{skill}
as a placeholder for the programming skills of participants
that are relevant for the contest.
\autoref{fig:dag-cj} displays all nodes---observed and unobserved---for the \cj contest,
marking unobserved variables with a circle,
and highlighting the two variables \var{language} and \var{rank} whose causal relation
our analysis focuses on.

A causal DAG's \emph{arrows}
denote causal relations:
an arrow $A \to B$ from node $A$ to node $B$
means that there is a direct causal effect of $A$ on $B$
(in other words, changing $A$ causes $B$ to change as well).
The model is qualitative in that an arrow's causal relation may be weak or strong, 
and is in general probabilistic 
(but it should be detectable within the precision of the data measurements).
For the \cj data, we design the DAGs in \autoref{fig:dag-cj} based on the following observations:

\begin{itemize}
\item Measuring the causal relation between \var{language} and \var{rank} is the main focus
  of our analysis; thus,
  we include an arrow $\var{language} \to \var{rank}$.\footnote{The main analysis goal is quantitatively estimating the causal effect
    of \var{language} on \var{rank}, but this effect may not exist.
    Thus, the rest of the analysis builds on the tentative assumption
    that \var{language} may directly affect \var{rank}, and 
    tries to remove sources of bias in the estimate of this causal effect
    as much as possible given the available information and any causal assumptions.
    In the end, the estimate itself may confirm a significant
    effect or indicate that the effect is negligible.
    Clearly, a causal link in the opposite direction
    (from \var{rank} to \var{language}) is impossible because
    a submission's \var{rank} is determined \emph{after} its \var{language}
    has been chosen.
}
  
\item Different programmers prefer to use some languages over others;
  thus, we include an arrow $\var{nickname} \to \var{language}$.

\item A program's size depends also on the programming language it is written in;\footnote{For example, Python programs are usually more concise than Java programs for the same task~\cite{NF-ICSE15}.}
  thus, we include an arrow $\var{language} \to \var{size}$.

\item Programmers differ in their coding skills, which obviously have an impact on their success
  in the contest; thus, we include arrows $\var{nickname} \to \var{skill}$ and $\var{skill} \to \var{rank}$.

\item Not all challenges in \cj have the same intrinsic difficulty;
  hence, the range of ranks that are assigned varies from challenge to challenge  (for example, because fewer participants submit to harder challenges).
  Thus, we include an arrow $\var{challenge} \to \var{rank}$.\footnote{Nevertheless, omitting the arrow $\var{challenge} \to \var{rank}$
    does not affect the conclusions of the ensuing causal analysis.
  }

\item A programmer's skills are usually problem-dependent:
  different programmers may be more familiar with certain kinds of problems;
  thus, we include an arrow $\var{challenge} \to \var{skill}$.

\item Similarly, a programmer may selectively decide to submit solutions
  only for challenges that they find congenial;
  thus, we include an arrow $\var{challenge} \to \var{nickname}$.\footnote{This does not mean that a challenge can \emph{change} a participant's nickname;
    it means that the value of \var{challenge} can affect which participants (identified by their nicknames) \emph{appear} in the data for that particular challenge.
  }

\item Connecting nodes \var{size} and \var{rank} would be inconsistent with how the \cj contest works:
  the contest rules (\autoref{sec:code-jam-data})
  do not take a submission's size into account to rank it (thus, no arrow from \var{size} to \var{rank});
  and the rank is determined \emph{after} submission, and hence changing
  a submission's rank cannot affect the submission's size (thus, no arrow from \var{rank} to \var{size}).
\end{itemize}

Finally, we may consider different plausible relations between \var{nickname} and \var{size}.
It's clear that programmers often have more concise or verbose programming styles,
but we can express this relation in a DAG in different ways.

\begin{description}
\item[DAG $d_0$.] If we consider a programmer's conciseness a form of personal preference,
  we include an arrow $\var{nickname} \to \var{size}$.

\item[DAG $d_1$.] If we consider conciseness a direct result of a programmer's skills,
  we include an arrow $\var{skill} \to \var{size}$.

\item[DAG $d_2$.] If we consider conciseness a combination of personal preference and skills,
  we include both arrows $\var{nickname} \to \var{size}$ and $\var{skill} \to \var{size}$.
\end{description}
\autoref{fig:dag-cj} shows the resulting DAGs $d_0$ (\autoref{fig:dag-d0}),
$d_1$ (\autoref{fig:dag-d1}),
and $d_2$ (\autoref{fig:dag-d2}).

\subsection{Validating DAGs on Empirical Data}
\label{sec:dag-validation}

Which causal DAG among $d_0, d_1, d_2$
should we adopt as qualitative causal model of the \cj data?
We can perform a kind of \emph{validation}
where we test whether some of the DAG's relations are consistent with the data.
This validation is necessarily \emph{partial}:
as we discussed previously,
we cannot derive causal relations from observational data in general (and exhaustively);
this limitation also applies to a posteriori validation.
However, even partial validation can be useful to choose among different plausible DAGs,
and to corroborate the intuition that led us to construct the DAGs in the first place.

To validate a DAG, we first derive its \emph{implied conditional independencies}:
these are relations of statistical independence among DAG variables
that can be derived from the DAG's topology.
A statistical independence relation
is written as $X \bigCI Y \mid Z$
and means that variable $X$ is independent of variable $Y$ given variable $Z$;
intuitively, it captures the idea that there is no meaningful association between
$X$ and $Y$ after we condition on (know the value of) $Z$.

To see an example of conditional independence implied by all DAGs $d_0, d_1, d_2$,
consider the path $\var{language} \leftarrow \var{nickname} \leftarrow \var{challenge}$.
Variable \var{challenge} affects \var{language} only indirectly through variable \var{nickname};
if we condition on \var{nickname}
(that is, we single it out explicitly),
we should then observe no remaining meaningful correlation
between \var{language} and \var{challenge}.
Therefore, the three DAGs imply the conditional independence:
\begin{gather}
  \begin{array}{ccccl}
    \var{language} & \bigCI & \var{challenge} & \mid & \var{nickname}
  \end{array}
  \label{eq:id-lang-chal}
\end{gather}
Intuitively, this conditional independence is very plausible
if we expect that participants tend to use the same programming language in all challenges they take part in;\footnote{In our dataset, more than 80\% of all participants used the same language in all their \cj submissions.}
then, after we fix \var{nickname},
we can already reliably predict \var{language}---regardless of \var{challenge}.

More precisely,
\eqref{eq:id-lang-chal} is the \emph{only} conditional independence implied by DAGs $d_1$ and $d_2$.\footnote{More precisely, it is the only \emph{testable} conditional independence: conditional independencies that involve unobserved variables (such as \var{skill}) cannot be tested since they are not available in the data.}
In contrast, DAG $d_0$ includes subgraphs $\var{size} \leftarrow \var{nickname} \leftarrow \var{challenge}$
and $\var{size} \leftarrow \var{nickname} \to \var{language} \to \var{rank}$,
but no path $\var{nickname} \to \var{skill} \to \var{size}$,
which imply two additional conditional independencies:
\begin{alignat}{4}
    \var{size} & \bigCI &&\ \var{challenge} &\ \mid &&&\ \var{nickname}
   \label{eq:id-size-challenge}
    \\
    \var{rank} & \bigCI &&\ \var{size} &\ \mid &&&\ \var{nickname}, \var{language}
   \label{eq:id-rank-size}
\end{alignat}

How do we \emph{test} a DAG's implied conditional independencies?
A simple way uses regression models to
measure the association between allegedly independent variables.
Given $X \bigCI Y \mid Z$,
we build a generalized linear model
with $X$ as outcome, and $Y$ and $Z$ as predictors;\footnote{Since $X \bigCI Y \mid Z$ is equivalent to $Y \bigCI X \mid Z$, we could equivalently use $Y$ as outcome,
  and $X$ and $Z$ as predictors.}
we fit the model on the data, and check whether $Y$'s fitted parameters
are consistently different from zero.
If they are, it means that there is a consistent association
between $X$ and $Y$, which invalidates the conditional independence;
in this case, validation \emph{fails}, as it seems the data exhibits correlations
that should be negligible according to the causal DAG.
Conversely, if $Y$'s parameters are indistinguishable from zero,
validation is \emph{successful},
in that the data agrees with the causal DAG on this aspect.
In practice, like every statistical analysis,
we should not apply this kind of validation
as a strictly binary (yes\slash no) check,
but rather evaluate the parameters' uncertainty in context
and possibly in comparison with other model coefficients.

\begin{figure}[!tb]
  \centering
  \begin{subfigure}[t]{0.47\linewidth}
    \begin{equation*}
      \begin{aligned}
        \var{language}_i &\sim\ \dist{NegativeBinomial}(\lambda_i, \phi) 
        \\
        \log(\lambda_i) &=\ \alpha + \alpha_{\var{challenge}[i]} + \alpha_{\var{nickname}[i]}
      \end{aligned}
    \end{equation*}
    \Description{Equations defining the likelihood of model $\mathit{mid}_1$: negative binomial distribution, logarithmic link function, \var{language} as outcome, and \var{challenge} and \var{nickname} as predictors.}
    \caption{Likelihood of model \idm{1} for testing conditional independence \eqref{eq:id-lang-chal}.}
    \label{eq:mid1}
  \end{subfigure}
  \hspace{2mm}
  \begin{subfigure}[t]{0.47\linewidth}
    \begin{equation*}
      \begin{aligned}
        \log(\var{size}_i) &\sim\ \dist{Normal}(\mu_i, \sigma) 
        \\
        \mu_i &=\ \alpha + \alpha_{\var{challenge}[i]} + \alpha_{\var{nickname}[i]}
      \end{aligned}
    \end{equation*}
    \Description{Equations defining the likelihood of model $\mathit{mid}_2$: normal distribution, identity link function, $\log(\var{size})$ as outcome, and \var{challenge} and \var{nickname} as predictors.}
    \caption{Likelihood of model \idm{2} for testing conditional independence \eqref{eq:id-size-challenge}.}
    \label{eq:mid2}
  \end{subfigure}
  \\
  \begin{subfigure}[t]{0.47\linewidth}
    \begin{equation*}
      \begin{aligned}
        \var{rank}_i &\sim\ \dist{NegativeBinomial}(\lambda_i, \phi) 
        \\
        \log(\lambda_i) &=\  \alpha + \beta \cdot \log(\var{size}_i) + \alpha_{\var{nickname}[i]} + \alpha_{\var{language}[i]}
      \end{aligned}
    \end{equation*}
    \Description{Equations defining the likelihood of model $\mathit{mid}_3$: negative binomial distribution, logarithmic link function, \var{rank} as outcome, and $\log(\var{size})$, \var{nickname} and \var{language} as predictors.}
    \caption{Likelihood of model \idm{3} for testing conditional independence \eqref{eq:id-rank-size}.}
    \label{eq:mid3}
  \end{subfigure}
  \Description{This figure consists of three subfigures with the equations of statistical models $\mathit{mid}_1$, $\mathit{mid}_2$, and $\mathit{mid}_3$.}
  \caption{Definitions of regression models \idm{1}, \idm{2}, and \idm{3}
    for testing the implied conditional independencies of DAGs $d_0$, $d_1$, and $d_2$.}
  \label{fig:models-mid123}
\end{figure}

\autoref{fig:models-mid123} shows the definitions of models
\idm{1}, \idm{2}, and \idm{3},
which test implied conditional independencies
\eqref{eq:id-lang-chal}, \eqref{eq:id-size-challenge}, and \eqref{eq:id-rank-size} respectively.
For brevity, we do not discuss choosing suitable priors for these models but only show the likelihoods.
All models introduce a parameter $\alpha$ that captures a population-level mean;
this way, all other parameters are differences over $\alpha$. Model \idm{1} uses a negative binomial probability distribution,
since outcome variable \var{language} is an integer identifier
and its variance is one order of magnitude bigger than its mean.
For similar reasons, model \idm{3} also uses a negative binomial distribution.
Model \idm{2} uses instead a normal distribution, since its outcome variable varies on a continuous scale.
Then, parameters $\alpha_{\var{challenge}}$, $\alpha_{\var{nickname}}$, and $\alpha_{\var{language}}$
are index variables---one for each challenge, nickname, and language---used as intercepts;
parameter $\beta$ is a slope associated with the (logarithm of) a submission's size.

\begin{figure}[!tb]
  \centering
  \includegraphics[width=\textwidth]{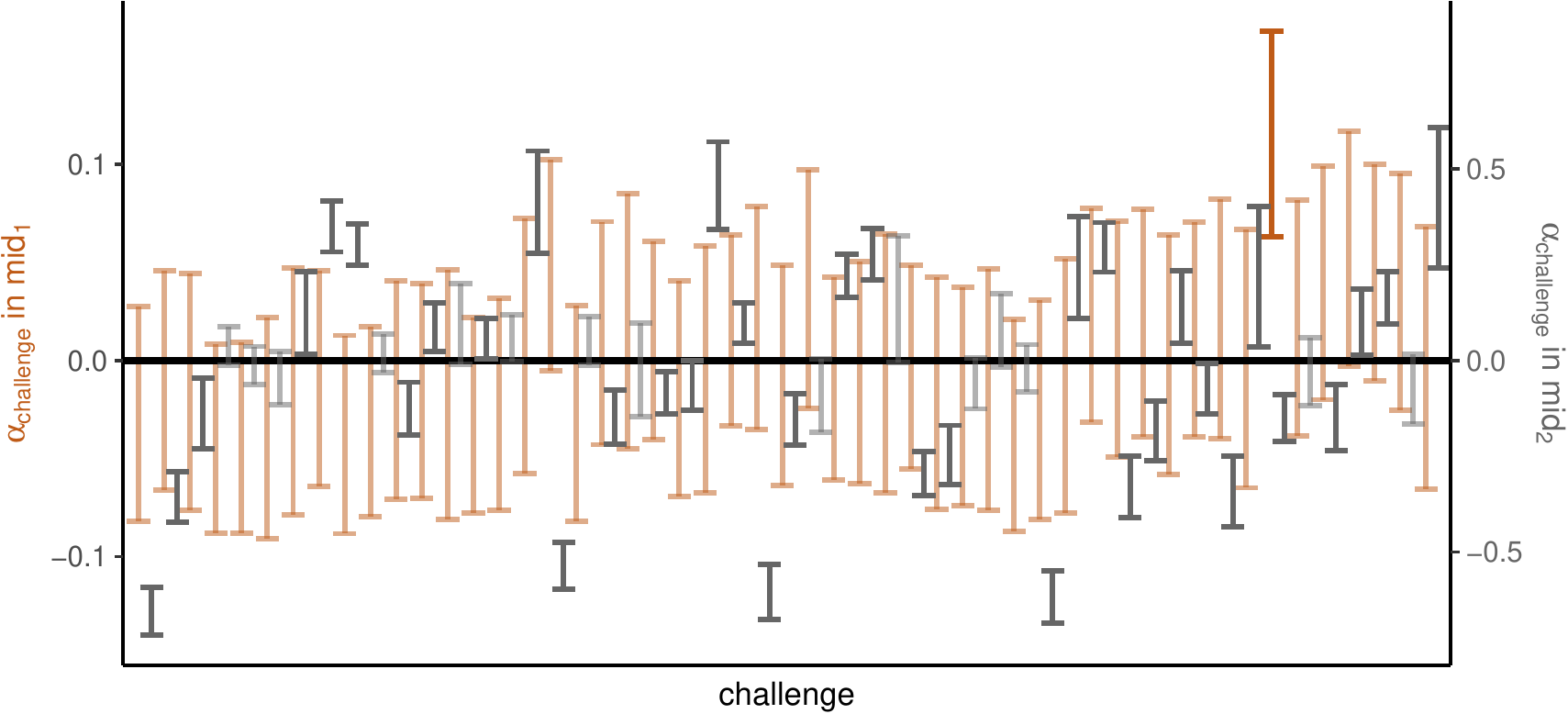}
  \Description{There are 51 intervals corresponding to the 50\% probability estimates of parameter $\alpha_c$ in model $\mathit{mid}_1$ for each challenge $c$; 50 of these intervals include zero (and many of them are nearly centered around it). There are another 51 intervals corresponding to the 50\% probability estimates of parameter $\alpha_c$ in model $\mathit{mid}_2$ for each challenge $c$; only 15 of these intervals include zero, the others are either completely above or completely below it.}
  \caption{For each \cj challenge $c$,
    the 50\% probability intervals of the estimates of parameters 
    $\alpha_c$ in models \idm{1} (brown, left scale)
    and \idm{2} (gray, right scale).
    Intervals that do not include zero are drawn in a darker color.
  }
  \label{fig:ci-analysis}
\end{figure}

\autoref{fig:ci-analysis} plots the 50\% probability intervals\footnote{Analyzing higher probability intervals leads to similar overall conclusions.}
of parameters $\alpha_c$, for each challenge $c$, in model \idm{1} (in brown color)
and in model \idm{2} (in gray color).
Only \n[0]{mid1 alpha perc}[100]| of all \n{mid1 alpha length} coefficients $\alpha_c$
are consistently different from zero in \idm{1};
in contrast, \n[0]{mid2 alpha perc}[100]| of them are in \idm{2}.
As for $\idm{3}$, coefficient $\beta$'s 50\% probability interval is
$(\n[2]{m3 beta low 50}, \n[2]{m3 beta high 50})$, which is quite clearly negative.
This analysis suggests that:
\begin{enumerate}
\item Variables \var{challenge} and \var{language} are negligibly associated in \idm{1};
  thus, the data confirms the conditional independence \eqref{eq:id-lang-chal}.
\item Variables \var{challenge} and \var{size} are consistently associated in \idm{2};
  thus, the data invalidates the conditional independence \eqref{eq:id-size-challenge}.
\item Variables \var{size} and \var{rank} are consistently associated in \idm{3};
  thus, the data invalidates the conditional independence \eqref{eq:id-rank-size}.
\end{enumerate}

We conclude that DAGs $d_1$ and $d_2$,
which imply the same conditional independence~\eqref{eq:id-lang-chal},
pass validation, whereas DAG $d_0$ fails it.
Thus, we use DAGs $d_1$ or $d_2$ as the basis of our causal analysis
of the effects of programming languages.

\subsection{Choosing Predictors to Measure Causality}
\label{sec:adjustment-sets}

After validating it,
a causal DAG can guide the choice of which predictors to include and which to exclude
in a regression model in order to estimate the direct causal effects of one variable on another one.
We'll demonstrate this on DAG $d_2$ in \autoref{fig:dag-cj};
using DAG $d_1$ leads to the same conclusions.

\subsubsection{Causal and Non-Causal Paths}
A (simple) \emph{path} on a DAG is a sequence of nodes connected by arrows
(ignoring the arrow direction), where each node appears at most once.
There are 14~paths in DAG $d_2$
that go from \var{language} to \var{rank}:
\begin{enumerate}[label=\emph{\Alph*}.,ref=\emph{\Alph*}]
\item \label{p:direct} $\var{language} \to \var{rank}$
\item $\var{language} \to \var{size} \leftarrow \var{nickname} \to \var{skill} \to \var{rank}$
\item $\var{language} \to \var{size} \leftarrow \var{nickname} \to \var{skill} \leftarrow \var{challenge} \to \var{rank}$
\item $\var{language} \to \var{size} \leftarrow \var{nickname} \leftarrow \var{challenge} \to \var{rank}$
\item $\var{language} \to \var{size} \leftarrow \var{nickname} \leftarrow \var{challenge} \to \var{skill} \to \var{rank}$
\item $\var{language} \to \var{size} \leftarrow \var{skill} \to \var{rank}$
\item $\var{language} \to \var{size} \leftarrow \var{skill} \leftarrow \var{challenge} \to \var{rank}$
\item $\var{language} \to \var{size} \leftarrow \var{skill} \leftarrow \var{nickname} \leftarrow \var{challenge} \to \var{rank}$
\item \label{p:closed-1} \label{p:backdoor-first} $\var{language} \leftarrow \var{nickname} \to \var{size} \leftarrow \var{skill} \to \var{rank}$
\item \label{p:closed-2} $\var{language} \leftarrow \var{nickname} \to \var{size} \leftarrow \var{skill} \leftarrow \var{challenge} \to \var{rank}$
\item \label{p:open-2} $\var{language} \leftarrow \var{nickname} \to \var{skill} \to \var{rank}$
\item \label{p:closed-3} $\var{language} \leftarrow \var{nickname} \to \var{skill} \leftarrow \var{challenge} \to \var{rank}$
\item \label{p:open} \label{p:backdoor} $\var{language} \leftarrow \var{nickname} \leftarrow \var{challenge} \to \var{rank}$
\item \label{p:open-3} \label{p:backdoor-last} $\var{language} \leftarrow \var{nickname} \leftarrow \var{challenge} \to \var{skill} \to \var{rank}$
\end{enumerate}

Since each pair of consecutive nodes in a path are correlated
variables, a path from $X$ to $Y$ represents one contribution 
to the overall correlation that is observed between $X$ and $Y$.
Some paths correspond to \emph{causal} correlations between $X$ and $Y$;
others represent \emph{spurious} correlations
that do not correspond to any causal effects.
For example, $d_2$'s path~\ref{p:direct} is clearly causal;
in contrast, path~\ref{p:backdoor} is not, since the resulting correlation
between \var{language} and \var{rank}
is due to the effect of \var{challenge}
on both \var{rank} and \var{language}:
changing \var{challenge} simultaneously affects \var{rank} and \var{language},
and a spurious correlation emerges.
In this scenario, \var{challenge} is called a \emph{confounder},
as it biases the net causal effect of \var{language} on \var{rank}.
More generally, every path from $X$ to $Y$ with an arrow entering $X$
is a so-called \emph{backdoor} path that biases the estimate of the
true causal effect of $X$ on $Y$.
In $d_2$, paths \ref{p:backdoor-first} through \ref{p:backdoor-last} are
backdoor paths.

\subsubsection{Open and Closed Paths}
Our goal is removing all backdoor paths,
so that the correlation we observe between $X$ and $Y$ is
\emph{unbiased}---that is, it is exclusively due to causal effects.
Obviously, we cannot do that by simply removing arrows from a DAG:
the arrows are supposed to represent direct causal relations
in the process that determined the data we are analyzing.
Instead, we can select which variables (nodes)
to include (``control for'')
in a regression model that estimates the correlation
between $X$ and $Y$.

Intuitively, we select variables to use as predictors
so as to \emph{close} all backdoor paths.
Closing a path means that the path's spurious correlation
cancels out in the estimate of the regression model.
Consider again path~\ref{p:open} in $d_2$;
if we include \var{nickname} among the regression variables,
the path becomes closed,
and it will not contribute any spurious correlation
between \var{language} and \var{rank}.
In fact, including a variable in a regression model
means \emph{conditioning} on the variable.
Conditioning on \var{nickname} means
that the model can specifically estimate the correlation of \var{nickname}
on the other variables,
which blocks the spurious correlation between \var{language} and \var{rank}
otherwise induced by the path through \var{nickname}.

Unfortunately, adding variables as predictors may also backfire,
as including some variables can \emph{open} backdoor paths
that would otherwise be closed.
For example, take path~\ref{p:closed-1} in $d_2$:
as long as we do \emph{not} include \var{size} among the regression variables,
the path remains closed, and it will not contribute any spurious correlation
between \var{language} and \var{rank};
but conditioning on \var{size} ``opens the path''
by introducing a correlation from \var{language} that goes
all the way through the path to \var{rank}.

Here is an intuitive explanation of why conditioning
on \var{size} may \emph{bias}
the estimate of the effect of \var{language} on \var{rank}.
A submission's \var{size}
is affected by its \var{language}
(some languages
are generally more concise~\cite{Prechelt:2000:ECS:619056.621567,NF-ICSE15})
and by the participant's \var{skill}
(more skilled programmers presumably can produce
more concise code).
This implies that there are two main ways for a submission
to be concise:
either it's written by a skilled programmer,
or it's written using a concise language.
In other words, if we fix a submission's \var{size},
it would appear as if \var{language} and \var{skill}
correlate:
thanks to their higher \var{skill}, 
the better programmers
using verbose languages
would still be able to produce submissions
of \var{size} comparable to
those of the worse programmers (lower \var{skill})
using concise languages.
This correlation, however, is merely a result of
self selection, and does not reflect any
genuine causal relation between \var{skill} and \var{language}.\footnote{Here is an analogous example of selection bias that is perhaps
  more intuitive to understand~\cite{collider-bias,collider-bias-2}.
  An actor's \var{success} is mainly determined by their acting \var{talent}
  and by their good \var{looks}
  ($\var{talent} \to \var{success} \leftarrow \var{looks}$).
  Thus, among successful actors, the more talented ones will be
  worse looking, and the less talented ones will be better looking.
  This inverse relation between \var{talent} and \var{looks} is spurious though:
  an actor's talent and good looks are presumably independent,
  but conditioning on \var{success} measures a correlation that is
  simply a figment of the way in which the data is analyzed.
  }
On top of this, \var{skill} and \var{language}
both directly affect \var{rank};
therefore, the spurious correlation introduced
by conditioning on \var{size}
conflates these two effects---ultimately biasing our estimate
of the true causal effect of \var{language} on \var{rank}.

Systematically, a path is \emph{open} unless
it includes a \emph{collider}: a node $Z$
where both arrows connecting it to the rest of the path enter it
(as in $X \cdots \to Z \leftarrow \cdots Y$).
In the previous example of path~\ref{p:closed-1},
\var{size} is a collider, which we should \emph{exclude}
from the regression variables lest backdoor path~\ref{p:closed-1}
becomes open.

\subsubsection{Adjustment Sets}
Finally, we can formulate a general recipe to
select which variables to include as regression variables
in a model to estimate the genuine causal effect of $X$ on $Y$
given a DAG
capturing the causal relations among variables~\cite{good-bad-controls}.
Consider all \emph{backdoor} paths from $X$ to $Y$ in the DAG;
for each open backdoor path,
pick a variable $Z$ within the path such that conditioning on $Z$
would close the path;
while doing so, make sure that you do not select any variable
that is also a collider in some closed backdoor path.
A set of variables that satisfy these constraints 
is called an \emph{adjustment set};
using them as regression variables
would close all open backdoor paths
without opening any closed backdoor paths.
Thus, a regression model that conditions on $X$ as well as on all variables
in an adjustment set
provides an \emph{unbiased} estimate of the correlation
between $X$ and $Y$ that is solely due to the causal link between them.

The constraints on which variables to include or exclude from
an adjustment set may be unsatisfiable---if conditioning on a collider is
needed to close another backdoor path.
Conversely, multiple valid adjustment sets
may exist for the same DAG.

Adding predictors to a regression model is a common practice to
address so-called ``omitted variable bias'',
which occurs when a causal effect's estimate
is biased because we did not correct for possible confounders.
As we are demonstrating, causal DAGs are useful to rigorously find confounders and filter them out
from the causal estimate.
The flip side is that there may also be a risk of ``included variable bias'':
adding the wrong predictor to a regression may \emph{introduce} a confounding effect
and spoil the estimate of a causal relation.
DAGs are also useful to detect and address this kind of confounding
that occurs when we include variables that should not be included.

\subsubsection{Unbiased Estimation for \cj Data}
Analyzing the paths from \var{language} to \var{rank}
in \autoref{fig:dag-cj}'s DAG $d_2$
indicates two adjustment sets\footnote{There exist several other adjustment sets, but they all include \var{skill}, which is unobserved, and hence cannot be used as predictor.}
to accurately estimate the causal effect of \var{language} on \var{rank}:
$A_1 = \{ \var{nickname} \}$
and $A_2 = \{ \var{nickname}, \var{challenge} \}$.
$A_1$ is the \emph{minimal} adjustment set, as conditioning on \var{nickname} is required to
close any backdoors.
$A_2$ is an alternative adjustment set, which indicates that conditioning on \var{challenge} is neither
needed nor harmful to filter out spurious association.
In contrast, \var{size} does not appear in either adjustment sets,
which means that we should \emph{not} condition on it.

This analysis indicates that models $m_2$ and $m_3$ in \autoref{fig:models-m14}
are suitable to reliably estimate the causal effect of \var{language} on \var{rank}.
In contrast, we should not use models $m_1$ or $m_4$ for this:
model $m_1$ uses only \var{language} as predictor (and none of the adjustment variables), 
whereas model $m_4$ also uses \var{size} as predictor (which adds a confounding effect).
This conclusion holds quite robustly independent of the details of the causal DAG that
we assume; in particular, DAGs $d_1$ and even $d_0$ have the same adjustment sets as $d_2$.

While including \var{challenge} as predictor
is neutral as far as removing confounding bias is concerned,
it may help improve the \emph{precision} of the estimate of the effect of \var{language} on outcome \var{rank}.
As a rule of thumb,
adding predictors ``close'' to the outcome helps reduce variance, since it may filter out
the effect of other unknown dependencies or non-linearities among variables~\cite{graphical-criteria-estimation}.
Since model $m_3$ also outperforms model $m_2$ in terms of predictive accuracy (see \autoref{tab:comparison-m14}),
we will use the former as the basis of our causal analysis.

\subsection{Analysis: Programming Language Effects}
\label{sec:analysis-effects}

The qualitative causal analysis based on DAGs
indicates that we can use model $m_2$ or $m_3$,
but should prefer $m_3$, 
to estimate the causal effect of \var{language} on \var{rank}.
Equipped with this knowledge, we can revisit the analysis of \autoref{sec:analysis-correlation}
by inspecting model $m_3$'s 50\% highest-posterior density intervals
of the distribution of $\effect[\ell]$
(the difference between language $\ell$'s contribution and the average language contribution)
for each language~$\ell$.
\autoref{fig:alphas-models-14} displays these intervals.

\paragraph{Results: model $m_3$.}
According to model $m_3$ (which causal analysis identified as the one
capturing most accurately the direct causal effects of languages on \cj results),
at the 50\% probability level:

\begin{itemize}
\item C++ is associated with smaller-than-average rank ordinals, that is \emph{better} contest results;
\item Python is associated with larger-than-average rank ordinals, that is \emph{worse} contest results;
\item there is no consistent association for Java, although its 50\% interval mostly covers better-than-average ranks.
\end{itemize}

Based on this, we can answer the paper's main RQ under the causal interpretation:

\begin{center}
\begin{boxedminipage}{0.9\columnwidth}
  \centering
  \textbf{AQ (causal)}: For experienced participants to the \cj contest,\\
  using C++ leads to better contest results,\\
  using Python leads to worse contest results,\\
  and using Java has no consistent effect (but leans towards better results).
\end{boxedminipage}
\end{center}

The answer to the paper's RQ has changed completely
after shifting our focus on causal effects,
as opposed to purely correlational associations!

\subsection{Correlation vs.\ Causation}
\label{sec:corr-vs-caus}

The difference between the correlational and causal answers
to the question of the relation between programming languages and \cj contest results
is striking,
as they are nearly each other's mirror image.

The contrast is amplified by other features of the analysis.
First, there is a remarkable agreement between the two ``causally consistent''
models $m_2$ and $m_3$, whose predictions are essentially the opposite of model $m_4$.
Second, model $m_4$ obliterates the other models in terms of predictive capabilities (see \autoref{sec:model-comparison});
thus, there is no reason to use models $m_3$ or $m_2$ according to purely predictive considerations.
Third, even though the simple model $m_1$ happens to agree with models $m_2$ and $m_3$'s predictions
despite not controlling for confounders,
it is dead last in \autoref{sec:model-comparison}'s model comparison;
thus, once again, model $m_4$ is indisputably the one to prefer from a purely predictive point of view.

This confirms the intuition that the \cj
contest is not run as a randomized controlled experiment.
Therefore, not all associations that we observe in the data
reflect actual causal effects.
If we want to estimate the latter, as opposed to the former,
we have to carefully select which information to include and which to exclude
from our statistical model.

\begin{figure}[!bt]
  \centering
  \includegraphics[width=\textwidth]{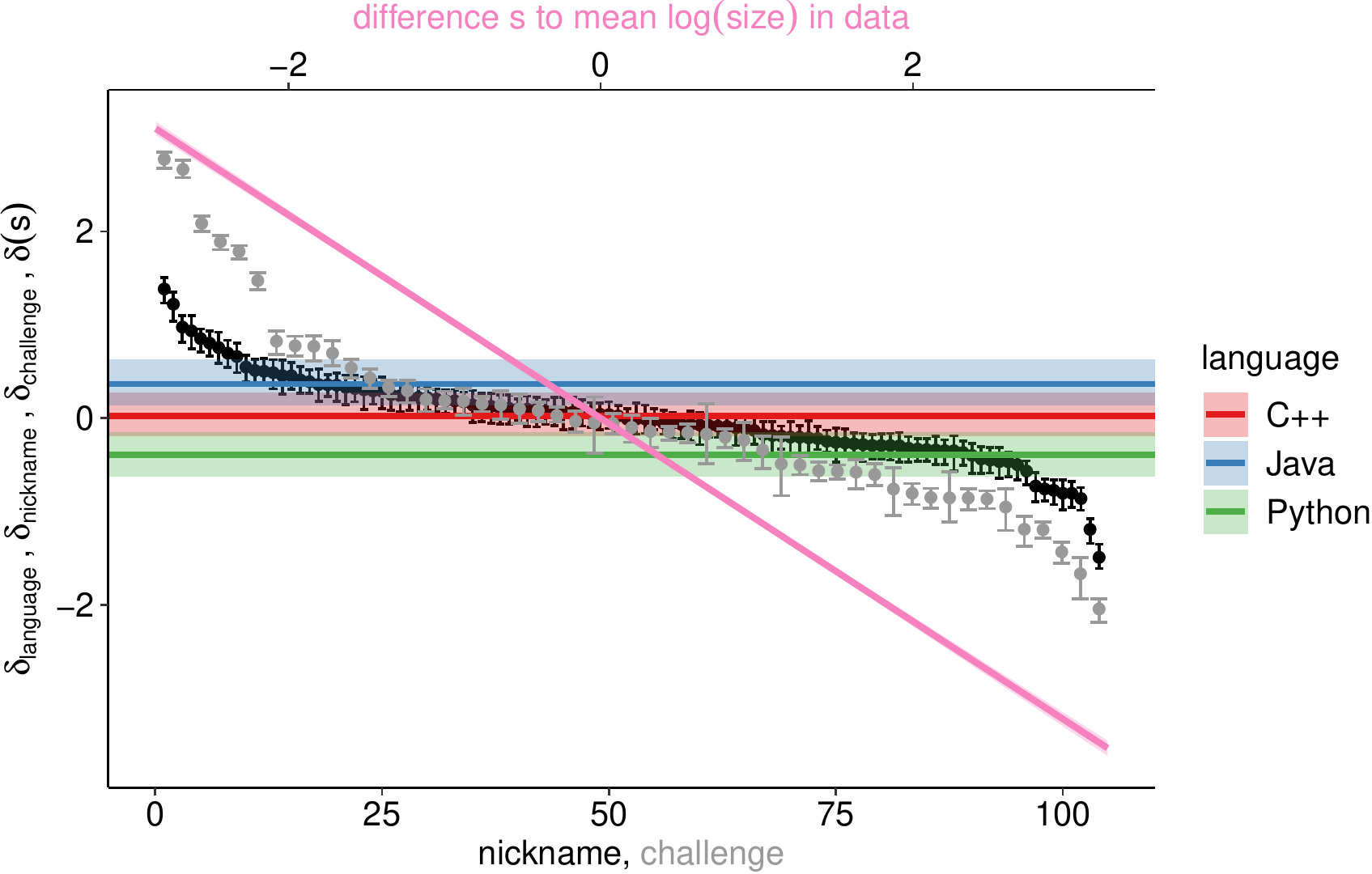}
  \Description{Three intervals show the centered effects of C++, Java, and Python; these three intervals overlap significantly, and overall span the range from $-0.6$ (Python's lower bound) to $+0.6$ (Java's upper bound).
    105 intervals show the centered effects associated with each \var{nickname}; each of these intervals is small (the widest is around $0.37$ units wide),
    but overall they span the range from $-1.6$ to $+1.5$.
    51 intervals show the centered effects associated with each \var{nickname};
    each of these intervals is fairly small (the widest is around $0.6$ units wide),
    but overall they span the range from $-2.2$ to $+2.8$.
    The regression line associated spans a wider range still---roughly from $+3.2$ down to $-3.4$.}
  \caption{Estimates and 50\% probability intervals of the centered effects:
    $\effect[\ell]$, for every language $\ell$;
    \effect[n], for every nickname $n$;
    \effect[c], for every challenge $c$;
    and \effect*[s] for every difference $s$ to mean $\log(\var{size})$.
  }
  \label{fig:releffs}
\end{figure}

\section{The Modest Impact of Languages}
\label{sec:effects-comparison}

How can we explain the striking difference between the
correlational and causal analyses of the same \cj data?
In other words, how is it possible that adding a single predictor
to model $m_3$ leads to nearly opposite predictions
about the relative impact of different programming languages?

In order to shed some light on the matter,
we break down model $m_4$ to compare the relative effect of
each predictor variable it uses.
We consider $m_4$ because it is the one that delivers the most
reliable predictions according to the model comparison of \autoref{sec:model-comparison}.
Furthermore, it includes all available data, so that we can compare \emph{all}
variables to each other.

The following discussion hinges on \autoref{fig:releffs},
which plots the estimates and 50\% probability intervals of
the effects of variables
\var{nickname}, \var{challenge}, \var{language}, and \var{size}
in model $m_4$.
As detailed in \autoref{sec:centered-model-m4},
the effects are centered so that we can compare their
absolute values on a consistent scale.
Then,
\autoref{sec:comparison-centered} and \autoref{sec:language-effects-vs-others}
analyze the plot in detail to
appreciate the different relative impact
of each variable on a submission's \var{rank}.

\subsection{Centered Effects in Model $m_4$}
\label{sec:centered-model-m4}

As shown in \autoref{eq:likelihoods-m14},
model $m_4$'s main term is the sum of four components,
each capturing the contribution of a different predictor
among \var{language}, \var{nickname}, \var{challenge}, and \var{size}.
Just like we did for \var{language} in the main analysis of the paper,
we introduce a derived variable $\delta_x$ for each nickname, challenge, and size;
in a nutshell, $\delta_x$ is a \emph{centered} version of $\alpha_x$, which
measures the differential contribution of $x$ relative to the average contribution of
other $x$s in the same group.

\begin{figure}[!hbt]
\begin{equation*}
  \begin{aligned}
    \effect[\ell] & = \alpha_\ell - \frac{\mean{\Sigma_{x \in L}\alpha_{x}}}{|L|}
    & \text{for every }\var{language}\ \ell\in L \\
    \effect[n] & = \alpha_n - \frac{\mean{\Sigma_{x \in N}\alpha_{x}}}{|N|}
    & \text{for every }\var{nickname}\ n\in N \\
    \effect[c] & = \alpha_c - \frac{\mean{\Sigma_{x \in C}\alpha_{x}}}{|C|}
    & \text{for every }\var{challenge}\ c\in C \\
    \effect*[s] & = \beta \cdot s
        & \text{for any value }s\text{ of }\log(\var{size})
  \end{aligned}
\end{equation*}
\Description{Equations describing the centered random variables of the effects associated with languages, nicknames, challenges, and logarithm of size.}
\caption{Definitions of centered random variables \effect[\ell], \effect[n], \effect[c], and \effect*[s].
\mean{x} denotes the expected value (mean) of $x$; $L$, $N$, and $C$ are the sets of all languages, nicknames, and challenges, respectively.}
\label{fig:centered-vars}
\end{figure}

More precisely, consider the definitions in \autoref{fig:centered-vars}.
Each \effect[\ell], \effect[n], \effect[c], and \effect*[s]
is a random variable derived from one of $m_4$'s parameters.
For example, for any language $\ell$ from the set $L = \{\text{C++}, \text{Java}, \text{Python}\}$ of all languages,
\effect[\ell] is the difference between $\alpha_\ell$ and the mean (expected value)
of all language-specific intercepts $\alpha_{\text{C++}}$, $\alpha_{\text{Java}}$, and $\alpha_{\text{Python}}$.
Since $\alpha_\ell$ is a random variable, whereas the mean is a constant,
\effect[\ell] is itself a (shifted) random variable,
with its own distribution that we can summarize with the usual statistics (mean, probability intervals, and so on).
In Bayesian analysis, \effect[\ell]'s
distribution is approximated by a collection of \emph{posterior samples}, on which we measure the statistics.

Similarly, \effect[n] is the centered version of $\alpha_n$ for any nickname $n$ among all nicknames $N$
in the \cj dataset;
and \effect[c] is the centered version of $\alpha_c$ for any challenge $c$ among all challenges $C$ in the dataset.
The definition of \effect*[s] is analogous but for the slope $\beta$ in model $m_4$,
which multiplies the logarithm of continuous variable \var{size}.
Consistently with the other centered variables,
we want \effect*[s] to measure \emph{difference}
over the ``average'' contribution associated with \var{size};
to this end, $s$ ranges over differences $x - \overline{\log(\var{size})}$,
where $\overline{\log(\var{size})}$ is the average (mean) logarithmic size of a submission in the \cj dataset.
For example, $s=0$ denotes a submission whose logarithmic size is the same as the average submission size.

With these definitions in place,
we can plot the estimates,
as well as the 50\% probability intervals of \effect[\ell], \effect[n], \effect[c], and \effect*[s]
on the same figure since they all range over the same scale of ``differences on top of the overall population mean''.
This also means that we can meaningfully \emph{compare} the relative contributions of each term;
according to the model's predictions,
those that are larger have a larger overall effect on the \var{rank} of a submission,
and hence are the main determinants of the overall results.
The resulting plot is shown in \autoref{fig:releffs}.

\subsection{Comparison of Centered Effects}
\label{sec:comparison-centered}

First, look at the contribution of \var{language},
which \autoref{fig:releffs} pictures by three colored lines with a shaded band, one per language $\ell$, 
that represent the estimate of \effect[\ell] and the 50\% probability interval of that estimate.
As you can see, \var{language}'s effects span a narrow range, roughly centered around zero,
and visibly overlap.
This means that the overall impact, on the predicted results,
of using one or the other programming language
is modest and hardly definitive.
The range between the upper endpoint of Java's 50\% probability interval
and the lower endpoint of Python's is just \n[2]{max diff/language:diff};
this translates to $\exp(\n[2]{max diff/language:diff}) \simeq \n[1]{max diff/language:exp}$
on \var{rank}'s outcome scale,\footnote{Taking the exponential inverts the logarithmic link function of model $m_4$'s linear relation.}
which means that using one language over another is usually
associated with only a few position's differences in rank.

Now, look at the contribution of \var{nickname},
which \autoref{fig:releffs} pictures as black dots and intervals (for mean and 50\% probability interval of the estimate);
for readability, all values $n$ of \var{nickname} are sorted by decreasing values of \effect[n] and given
a progressive numerical identifier (on the bottom horizontal axis).
As you can see, \var{nickname}'s effects span a wider range than the language ones:
the difference between the upper endpoint of the leftmost interval and the lower endpoint of the rightmost interval
is \n[1]{max diff/nickname:diff}, or $\n[1]{max diff/nickname:exp}$ on the outcome scale.
This value is one order of magnitude larger than the corresponding value for \var{language};
thus, the overall difference in contributions between different participants
is more conspicuous and consequential.
In other words,
if we fix a participant and change the language\footnote{
  This is a statistical prediction of the model;
  in reality, not all participants may be fluent in different languages
  or comfortable switching.
}
they use from Python to Java,
we still can only expect a modest change in rank;
but if we let the best and worst participant
use even the least effective language,
their results in the contest will remain substantially different.

We now repeat the analysis for \var{challenge},
whose contribution
\autoref{fig:releffs} pictures as gray dots and intervals (for mean and 50\% probability interval of the estimate);
as for \var{nickname}, all values $c$ of \var{challenge} are sorted by decreasing values of \effect[c] and given
a progressive numerical identifier (on the bottom horizontal axis).
The intervals associated with the different \var{challenge}s
span a wider range yet:
the difference between the upper endpoint of the leftmost interval and the lower endpoint of the rightmost interval
is \n[1]{max diff/challenge:diff}, or $\n[1]{max diff/challenge:exp}$ on the outcome scale.

Finally, the pink line in \autoref{fig:releffs}
diagrams the function $y = \effect*[x]$,
where $x$ measures a difference between a submission's logarithmic \var{size}
and the mean logarithmic size of submissions to \cj.
Around the line is a shaded ribbon, corresponding to the 50\% probability interval of the estimate of $\beta$;
the ribbon is barely visible because it is narrow, as there is little uncertainty in the estimate.
The horizontal range on the top horizontal axis spans
the actual range of differences to mean logarithmic size that we observed in the \cj dataset.
Within this range, \var{size}'s effects on \var{rank} are the biggest yet:
the difference between the vertical coordinates of the top-left and bottom-right ends of the pink line
is \n[1]{max diff/size:diff}, or $\n[1]{max diff/size:exp}$ on the outcome scale.
In part, the large effect attributed to \var{size} is a result of
measuring the \emph{logarithm} of a submission's size.
A submission that is larger than another submission by $s$ units on this scale
is actually $\exp(s)$ larger in bytes;
thus, the effects of \var{size}
emerge only when we observe substantial differences in actual size.

The relatively large effect of \var{size} also demonstrates
a spurious (i.e., non-causal) correlation:
clearly, a submission's size cannot be a direct cause of the submission's rank,
since \cj rules do not take size into account.
Similarly, if we take a poorly-ranked submission and artificially increase its size
(for example, by adding dead code), its rank won't improve.
Therefore, \var{size} is an effective predictor
(probably because it serves as a proxy for the expected effort required to meet a submission's requirements),
but does not immediately suggest any practical strategy to
improve success at \cj contests.

\subsection{Language vs.\ Other Effects}
\label{sec:language-effects-vs-others}

Remember that model $m_4$ makes better predictions than the other models.
Thus, the above analysis suggests that knowing a submission's \var{language}
sways only modestly the prediction of the submission's \var{rank}.
In contrast, knowing who wrote the submission (\var{nickname}),
how large the submission is (\var{size}), 
and which specific \var{challenge} the submission is for,
all affect predictions more strongly.

In fact, this is not a feature of model $m_4$ specifically:
a plot similar to the one in \autoref{fig:releffs} but for $m_3$\footnote{This plot is available in the replication package.}
would look very similar---except for the absence of \var{size}, which is not used in $m_3$---and confirm that the language-specific effect is modest compared to
the other predictors' effects.

One takeaway message is simply that programming language effects---whatever they are---are
\emph{modest} compared to others.
Thus, they may be tricky to detect reliably,
as they are easily confounded by other dominant factors.
As we discuss in \autoref{sec:related-work},
this observation is consistent with several other empirical studies investigating
the impact of programming languages.
Another way of looking at this 
is that, since the effects of programming languages are easily confounded,
it is very hard for an empirical study to collect
all the necessary data in an accurate enough way;
in other words, studies are easily underpowered~\cite{bsp}.

How do these findings relate to the main focus of the paper---demonstrating
how causal analysis techniques
can help improve how we answer software engineering questions empirically?
In a way, this section's analysis
helps validate the causal analysis,
in that it confirms that the impact of programming languages
is limited in our (as in other studies') data,
and hence not taking causality into account
can easily distort their actual effects.
Conversely, explicitly considering and modeling causal effects
was instrumental in obtaining a precise understanding of
the role of programming languages
and their relations to other factors.

\begin{table}[!bt]
  \centering
  \footnotesize
  \begin{tabular}{rrrrr}
  \toprule
  \multicolumn{1}{c}{\textsc{index}}
  & \multicolumn{1}{c}{\textsc{years}}
  & \multicolumn{1}{c}{\textsc{rounds}}
  & \multicolumn{1}{c}{\textsc{datapoints}}
  & \multicolumn{1}{c}{\textsc{participants}}
  \\
  \midrule
  \n{other datasets/1:size} & \n{other datasets/1:num.years} & \n{other datasets/1:num.rounds} & \n{other datasets/1:geq.years.rounds} & \n{other datasets/1:n.nicknames}
  \\
  \n{other datasets/2:size} & \n{other datasets/2:num.years} & \n{other datasets/2:num.rounds} & \n{other datasets/2:geq.years.rounds} & \n{other datasets/2:n.nicknames}
  \\
  \n{other datasets/3:size} & \n{other datasets/3:num.years} & \n{other datasets/3:num.rounds} & \n{other datasets/3:geq.years.rounds} & \n{other datasets/3:n.nicknames}
  \\
  \n{other datasets/4:size} & \n{other datasets/4:num.years} & \n{other datasets/4:num.rounds} & \n{other datasets/4:geq.years.rounds} & \n{other datasets/4:n.nicknames}
  \\
  \n{other datasets/5:size} & \n{other datasets/5:num.years} & \n{other datasets/5:num.rounds} & \n{other datasets/5:geq.years.rounds} & \n{other datasets/5:n.nicknames}
  \\
  \n{other datasets/6:size} & \n{other datasets/6:num.years} & \n{other datasets/6:num.rounds} & \n{other datasets/6:geq.years.rounds} & \n{other datasets/6:n.nicknames}
  \\
  \n{other datasets/7:size} & \n{other datasets/7:num.years} & \n{other datasets/7:num.rounds} & \n{other datasets/7:geq.years.rounds} & \n{other datasets/7:n.nicknames}
  \\
  \n{other datasets/8:size} & \n{other datasets/8:num.years} & \n{other datasets/8:num.rounds} & \n{other datasets/8:geq.years.rounds} & \n{other datasets/8:n.nicknames}
  \\
  \n{other datasets/9:size} & \n{other datasets/9:num.years} & \n{other datasets/9:num.rounds} & \n{other datasets/9:geq.years.rounds} & \n{other datasets/9:n.nicknames}
  \\
  \n{other datasets/10:size} & \n{other datasets/10:num.years} & \n{other datasets/10:num.rounds} & \n{other datasets/10:geq.years.rounds} & \n{other datasets/10:n.nicknames}
  \\
  \n{other datasets/11:size} & \n{other datasets/11:num.years} & \n{other datasets/11:num.rounds} & \n{other datasets/11:geq.years.rounds} & \n{other datasets/11:n.nicknames}
  \\
  \n{other datasets/12:size} & \n{other datasets/12:num.years} & \n{other datasets/12:num.rounds} & \n{other datasets/12:geq.years.rounds} & \n{other datasets/12:n.nicknames}
  \\
  \n{other datasets/13:size} & \n{other datasets/13:num.years} & \n{other datasets/13:num.rounds} & \n{other datasets/13:geq.years.rounds} & \n{other datasets/13:n.nicknames}
  \\
  \n{other datasets/14:size} & \n{other datasets/14:num.years} & \n{other datasets/14:num.rounds} & \n{other datasets/14:geq.years.rounds} & \n{other datasets/14:n.nicknames}
  \\
  \n{other datasets/15:size} & \n{other datasets/15:num.years} & \n{other datasets/15:num.rounds} & \n{other datasets/15:geq.years.rounds} & \n{other datasets/15:n.nicknames}
  \\
  \n{other datasets/16:size} & \n{other datasets/16:num.years} & \n{other datasets/16:num.rounds} & \n{other datasets/16:geq.years.rounds} & \n{other datasets/16:n.nicknames}
  \\
  \n{other datasets/17:size} & \n{other datasets/17:num.years} & \n{other datasets/17:num.rounds} & \n{other datasets/17:geq.years.rounds} & \n{other datasets/17:n.nicknames}
  \\
  \n{other datasets/18:size} & \n{other datasets/18:num.years} & \n{other datasets/18:num.rounds} & \n{other datasets/18:geq.years.rounds} & \n{other datasets/18:n.nicknames}
  \\
  \n{other datasets/19:size} & \n{other datasets/19:num.years} & \n{other datasets/19:num.rounds} & \n{other datasets/19:geq.years.rounds} & \n{other datasets/19:n.nicknames}
  \\
  \n{other datasets/20:size} & \n{other datasets/20:num.years} & \n{other datasets/20:num.rounds} & \n{other datasets/20:geq.years.rounds} & \n{other datasets/20:n.nicknames}
  \\
  \n{other datasets/21:size} & \n{other datasets/21:num.years} & \n{other datasets/21:num.rounds} & \n{other datasets/21:geq.years.rounds} & \n{other datasets/21:n.nicknames}
  \\
  \bottomrule
\end{tabular}
\caption{Different way of selecting the \cj data: each row identifies a dataset with \textsc{datapoints} about submissions made by a number of \textsc{participants}. These are the participants who entered at least \textsc{years} yearly edition of the \cj contest, at least \textsc{rounds} consecutive rounds in any one year, and who used the C++, Java, or Python programming languages. The rows are sorted by number of datapoints and numbered with an ordinal \textsc{index} of size. The paper's main analyses target the dataset with \textsc{index} \n{main size index}.}
\label{tab:other-datasets-size}
\end{table}

\section{Robustness and Threats to Validity}
\label{sec:robustness-and-threats}

Before discussing (in \autoref{sec:threats})
any threats to the validity of our results,
we outline (in \autoref{sec:robustness})
some additional analysis that we performed
to understand the robustness and generalizability of our main results.

\begin{figure}[!tb]
  \centering
  \includegraphics[width=\textwidth]{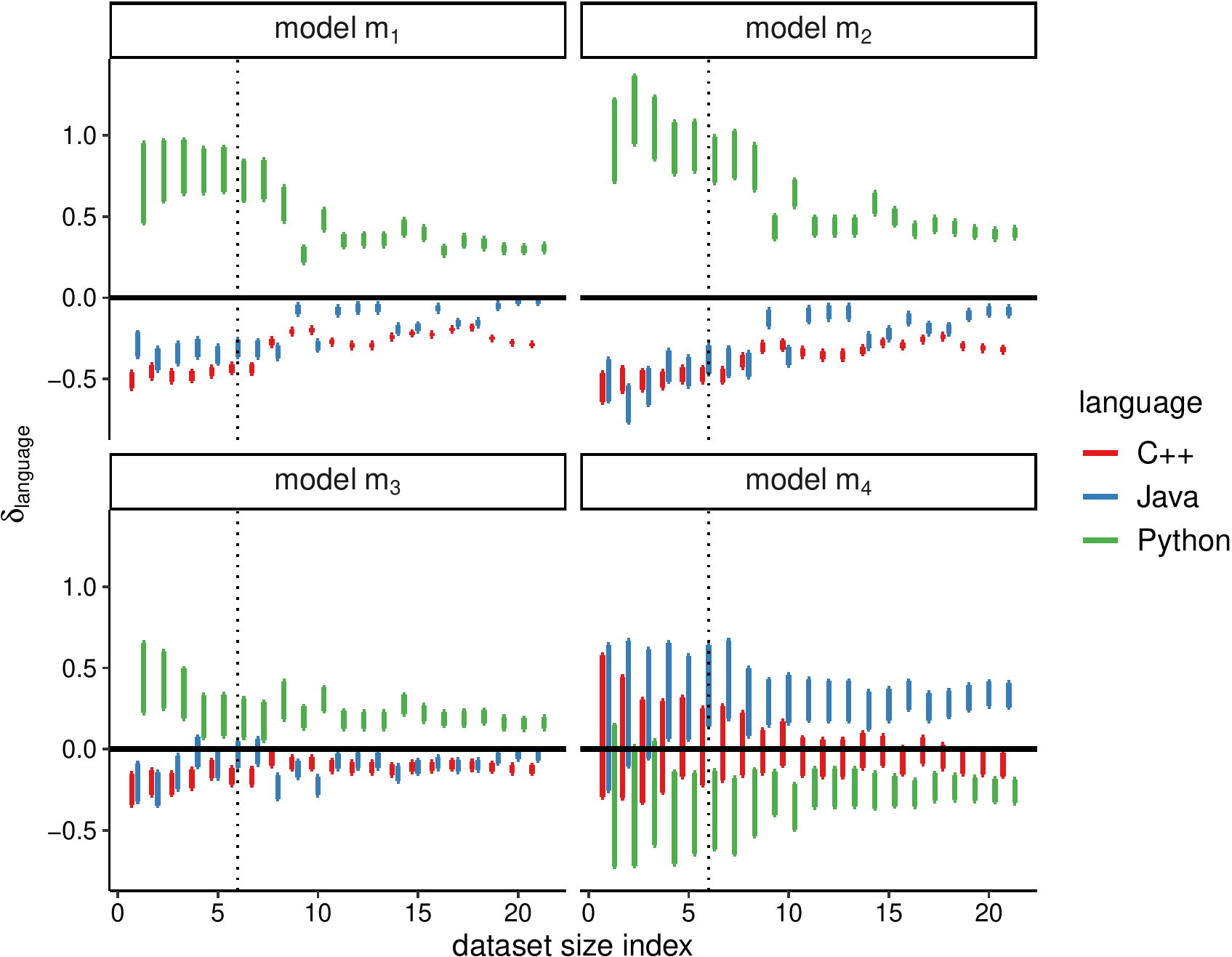}
  \Description{In each quadrant corresponding to one of the models $m_1, m_2, m_3, m_4$, there are 21 intervals for each language C++, Java, Python (a total of 63 intervals per model). Each interval is the 50\% probability interval of the coefficient associated with that language in that model fitted over one of the 21 datasets. The intervals move around depending on the dataset, but (with few exceptions), given a model and a language, whether the interval is above zero, below zero, or overlaps zero does not depend on the dataset.}
  \caption{For each language $\ell$, in each model $m_1, m_2, m_3, m_4$,
    the 50\% highest posterior density intervals of the distribution
    of $\effect[\ell]$ as we select larger datasets from all \cj data.
    The ``dataset size index'' corresponds to column \textsc{index} in \autoref{tab:other-datasets-size}:
    the larger, the more participants and datapoints are included in the dataset.
    The dotted lines mark the dataset used in the main paper analyses.
  }
  \label{fig:scal-analysis}
\end{figure}

\subsection{Analysis Robustness}
\label{sec:robustness}

In \autoref{sec:expert-selection} we explained how we selected a subset of all available \cj data
to compile a dataset that
is of reasonable size (neither unrealistically small nor impractically large)
and captures the performance of \emph{experienced} participants (as opposed to any casual participants).
To this end, we only considered data of programmers who:
\begin{enumerate*}[label=\emph{\roman*})]
\item took part in at least \n.{cutoffs/num.years}! yearly editions of \cj;
\item in any one edition, entered at least \n.{cutoffs/num.rounds}! rounds.
\end{enumerate*}
In addition, we only considered submissions using the most popular programming languages C++, Java, and Python.
Do the main results of our analysis change if we select a subset of data differently?

To answer this question, we consider different subsets of the \cj data built as follows.
A pair $Y, R$ 
identifies the subset of the \cj data including all submissions
using languages C++, Java, and Python by programmers who:
\begin{enumerate*}[label=\emph{\roman*})]
\item took part in at least $Y$ yearly editions of \cj;
\item in any one edition, entered at least $R$ rounds.
\end{enumerate*}
\autoref{tab:other-datasets-size}
lists 21 datasets, each corresponding to a pair $Y, R = \textsc{years}, \textsc{rounds}$---where $1 \leq Y \leq 7$ (since our data spans 7 editions of \cj)
and $1 \leq R \leq 6$ (since no edition has more than 6 rounds).
More precisely, \autoref{tab:other-datasets-size}
enumerates the first 21 datasets in increasing order of size (column \textsc{datapoints}).
We use column \textsc{index} to identify each dataset
according to its size; the paper's main analysis thus corresponds to the dataset with index \n{main size index}.

We fitted our four models 
on each of these 21 datasets.
\autoref{fig:scal-analysis}
pictures the 50\% probability intervals of the language-specific \effect[\ell].
While the intervals change in size and position relative to zero as we include more submissions,
the high-level picture remains generally consistent.
In particular, in the ``causal'' models $m_1, m_2, m_3$,
C++ and Java are usually associated with negative \effect[\ell] (better-than-average ranks),
whereas Python is usually associated with positive \effect[\ell] (worse-than-average ranks);
the situation is opposite in model $m_4$,
where Python is usually associated with negative \effect[\ell],
Java is usually associated with positive \effect[\ell],
and C++ usually has no consistent associations. 

There are a few exceptions to these trends:
with the smallest datasets,
there is more uncertainty (i.e., the intervals are wider)
because fewer data is used to fit the models;
with the largest datasets,
some associations shift
because more heterogeneous participants are included.
However, by and large, the contrast in predictions
between models $m_1, m_2, m_3$ and, on the other hand, model $m_4$
remains.

Similarly to \autoref{sec:effects-comparison}'s
analysis, this section's robustness analysis
further validates the causal analysis
by confirming that its results
are not merely a product of how the data was selected.
While we provided a justification for selecting
the original dataset---based on the intuitive notion of ``experienced'' participant---ours was the first detailed analysis of \cj data,
and hence validating our data selection process was important.
In contrast, other studies, targeting more widely used and accepted
curated datasets, may forgo such an exploration
of the analysis robustness without losing much in terms
of generalizability.

\subsection{Threats to Validity and Analytics Smells}
\label{sec:threats}

We scrutinize several aspects of the paper's analysis,
and connect them both to the traditional ``threats to validity''
categories (internal, external, construct, and conclusion)
and to Menzies and Shepperd's ``Bad smells'' in analytics~\cite{SmellsAnalytics}.
\autoref{tab:smells-summary} gives an overview of the analytics smells that
are covered in this section.

\begin{table}[!bt]
  \centering
  \begin{tabular}{lrr}
    \toprule
    \textsc{analytics smell} & \textsc{discussed in} & \textsc{addressed in} \\
    \midrule
    Using deprecated data & \ref{sec:threats-data} & \ref{sec:robustness} \\
    Not interesting & \ref{sec:threats-data} & \ref{sec:rw:pl-studies} \\
    Not exploring simplicity & \ref{sec:threats-modeling} & \ref{sec:correlation} \\
    Assumptions in statistical analysis & \ref{sec:threats-modeling} & \ref{sec:correlation}  \\
    $p < 0.05$ and all that! & \ref{sec:threats-modeling} & \ref{sec:correlation} \\
    Not exploring stability & \ref{sec:threats-analysis} & \ref{sec:robustness} \\
    Inadequate reporting & \ref{sec:threats-others} & \cite{replication-package} \\
    No data visualization & \ref{sec:threats-others} &  \ref{sec:correlation}, \ref{sec:causation}, \ref{sec:effects-comparison}, \ref{sec:robustness}\\
    Not using related work & \ref{sec:threats-others} & \ref{sec:related-work} \\
    \bottomrule
  \end{tabular}
  \caption{A summary of the analytics smells~\cite{SmellsAnalytics} relevant for the paper. For each \textsc{smell}, the subsection where it is \textsc{discussed}, and the section(s) or reference where it is \textsc{addressed}.}
  \label{tab:smells-summary}
\end{table}

\subsubsection{Data}
\label{sec:threats-data}

The operationalizations in the \cj data are generally straightforward,
as they consist of standard attributes (participant's nickname, size, and so on)
that should be unproblematic to obtain.
In particular, variable \var{rank} is computed automatically by the \cj organizers
according to predefined rules.

Determining a submission's programming language may be tricky in general,
but should not be an issue for \cj data,
since ranking a submission requires to run it on the contest's predefined environment,
and hence it must use one of the available language compilers.
Besides, we only used the three most widely used programming languages,
which helps avoid odd corner cases.

We mostly used the data as is,
except for aggregating multiple submissions by participant in a round,
and for taking the logarithm of the size.
The former is consistent with how the rank is determined by \emph{all} participant submissions;
the latter is customary when a quantity spans several order of magnitude.

The above observations indicate how threats to \emph{construct validity} are mitigated.

We could not use \emph{all} available \cj data both because it is impractically too much,
and because it is too heterogeneous in terms of participants' skills.
\autoref{sec:code-jam-data}
explains how we aimed for a reasonable selection that should be representative of
``experienced'' participants;
but we also performed additional analysis on larger and smaller data selections
to investigate robustness of the main results (\autoref{sec:robustness}).
While this selection might somewhat restrict the validity of our results to
a certain category of participants, 
it certainly helps address bad smell \absmell{using deprecated data}, as well as threats to \emph{internal validity}.
The latter, in particular, has to do with whether the paper's statistical analysis
is capable of identifying cause-effect relations by addressing biases and confounding effects;
this is the crux of the paper's contributions.

The advantages brought by using different programming languages is
a popular question for both researchers and practitioners
(see \autoref{sec:rw:pl-studies});
thus, the bad smell \absmell{not interesting} is not a risk for our research.

\subsubsection{Modeling}
\label{sec:threats-modeling}

Statistical modeling in the paper (\autoref{sec:correlation}) is based on 
widely used generalized regression models,
starting from a minimal model $m_1$ and extending it with other predictors that are available in the data.
We then argued---based on statistical model comparison techniques, as well as on causal analysis considerations---about the advantages and disadvantages of each model over the others.
This process helps avoid the bad smell \absmell{not exploring simplicity.}

Conversely,
no analysis can be truly exhaustive
and consider all possible models---nor can it claim that the models that it considered are \emph{definitive}.
Thus, it is always possible
that more refined, complex, or extended models
may reveal more nuanced trends in the data.
As we briefly mentioned earlier in the paper,
the replication package includes two more complex variants
of model $m_3$, which however do not lead to qualitatively different
conclusions about the impact of languages in \cj.
As we have seen in \autoref{sec:causation},
the causal models we consider in the paper are also somewhat robust,
in that adding or removing some assumed causal links
does not affect the model's implications.

As we argued in previous work~\cite{FFT-TSE19-Bayes2,TFFGGLE-TSE20-Bayes-practical},
using Bayesian modeling techniques
offers distinct advantages over the frequentist approaches that were dominant in the past.
One of them is that there exist
flexible \emph{validation} techniques~\cite{TOSEM-Bayes-guidelines};
these techniques 
buttress the choice of statistical models by
helping detect flawed models that are unsuitable to capture the analyzed data,
as well as other problems related to unwarranted statistical \absmell{assumptions} (another bad smell~\cite{SmellsAnalytics}).
For brevity, we did not discuss in detail the validation process in the paper,
but the replication package documents how we followed it scrupulously,
so as to mitigate any threats to \emph{conclusion validity}.

Using Bayesian data analysis techniques
also simplifies moving away from
dichotomous (binary) views of ``statistical significance'',
whose skewing effects and questionable theoretical justifications
have been repeatedly criticized~\cite{pvalue-cohen,pvalue-psychology,pvalue-medicine,ASA-statement,gelman-pvalues,abandon-significance,riseup,down-to-005}
(also as the related bad smell \absmell{$p < 0.05$ and all that!}).

\subsubsection{Analysis}
\label{sec:threats-analysis}

We observed how the main object of our analysis---the effect of programming languages---is elusive as any programming-specific effects tend to me small in
comparison to other factors.
This necessarily limits any strong claims of \emph{generalizability}---which is what \emph{external validity} is mostly concerned with.

We took some measures to mitigate these issues as best as possible.
First, we mainly focused on 50\% probability intervals instead of the stricter 95\% that are customarily used.
This is a way of avoiding ending up with a greatly \emph{underpowered study} (another bad smell of \cite{SmellsAnalytics}).
Second, we explored how the results change as we change the subset of \cj data that we analyze (\autoref{sec:robustness}),
which is also a way of exploring robustness\slash stability (avoiding bad smell \absmell{not exploring stability}).

Finally, the replication package includes further analyses of sensitivity
and using more complex multi-level models~\cite{gelman2020regression},
which we do not present here for brevity.
The sensitivity analysis confirms the modest relative effect of the programming language.
The analysis results of using more complex models corroborate 
those obtained by using the paper's plain regression models---which provides a further form of validation.

\subsubsection{Other Issues}
\label{sec:threats-others}

Providing a detailed replication package adheres to the best practices,
and avoids the bad smell \absmell{inadequate reporting}.
The replication package uses data visualization extensively, and so does the paper
(addressing bad smell \absmell{no data visualization}).

As we discuss in \autoref{sec:related-work}
there is very little work on causal analysis for software engineering data;
thus, this paper avoids the bad smell \absmell{not using related work}.

\section{Discussion and Conclusions}
\label{sec:conclusions}

As a concluding discussion, let's review some of the high-level
lessons that emerged from this paper's case study.

\paragraph{Small means small.}
Regardless of the specific causal relations,
our analysis confirms the observation,
made by several other empirical studies (see \autoref{sec:related-work}),
that the effects of using different programming languages
tend to be small---especially relative to those of other variables.
When effects are small, the robustness of any findings is necessarily lessened.
However, causal analysis can still help to isolate general interactions 
whose overall impact may become more prominent in other scenarios;
in other words, thinking causally can still improve external validity.

\paragraph{When associations are enough.}
If your exclusive or primary goal is making accurate predictions,
a purely correlational model may be sufficient; in fact, it may even outperform
a causally consistent model. For example, if you just want to predict the winner of the next
Google \cj contest, you're probably better off basing your predictions on model $m_4$
(or an even more sophisticated model).
Nevertheless,
even in such situations,
thinking about causality can still be useful as a sanity check
and as a safeguard against misinterpreting or overgeneralizing
your analysis results.

\paragraph{Bayesian models remain more flexible.}
As we repeatedly observed in the paper,
the high-level results our analysis---in particular, the contrast between causal and non-causal statistical models---are largely independent of whether we deploy Bayesian or frequentist statistics.
More generally, a lot of the causal modeling techniques we discussed in the paper
can be applied to frequentist models as well.
Nevertheless, we maintain that Bayesian statistics are preferable~\cite{FFT-TSE19-Bayes2},
as they are flexible and natural to interpret.
For example, performing \autoref{sec:effects-comparison}'s analysis
on a frequentist model would be cumbersome (but still possible), whereas
it was straightforward on the full posterior probability distribution
that Bayesian models provide.

\paragraph{Correlation vs.\ causation, again.}
Thinking about causal relations is not the endgame of empirical data analysis.
First, we already noted that there are scenarios in which predictive accuracy
takes precedence.
Second, causal and predictive models
need not disagree---in fact, they are often consistent~\cite{Prediction-for-Explanation}.
Third, the hypothesis underlying a causal model may need their own validation
by other means, and may change as the data generating process changes or is better understood.
It remains that, if we do not take causal relations into account,
we may miss (important) parts of the picture.
Conversely, thinking about causality prods us into looking
at the data from a fresh perspective,
rigorously thinking about confounding factors,
and ultimately interpreting any potential findings in a more sound way.

\ifarxiv
\else
\fi

%% \bibliography{causality,rosetta}

\end{document}